\documentclass{umthesis}
\usepackage{amsmath,amssymb,amsfonts}
\usepackage{latexcad}
\usepackage{epsfig}

\begin{document}

\title{Calibrated brane solutions of M-theory}
\author{Moataz H. Emam}
\date{September 2004}
\copyrightyear{2004}

\bachelors{B.Sc.}{Cairo University} %
\masters{M.Sc.}{Cairo University}   %

\committeechair{David Kastor}       %
\firstreader{Eugene Golowich}       %
\secondreader{Jennie Traschen}      %
\thirdreader{Markos Katsoulakis}    %
\departmentchair{Jonathan Machta}   %
\departmentname{Physics}

\degree{Doctor of Philosophy}{Ph.D.}

\frontmatter                        %
\maketitle                          %
\copyrightpage                      %
\signaturepage

\begin{dedication}
  \begin{center}
    \emph{To Miriam and Manal \\ $\quad $ \\
          $\quad \quad \quad \quad \quad \quad \quad \quad \quad \quad \quad \quad$
          My love and devotion forever ...}
  \end{center}
\end{dedication}

\begin{dedication}
  \begin{center}
        \centerline{\epsfxsize=7.0cm\epsfysize=4.0cm\epsfbox{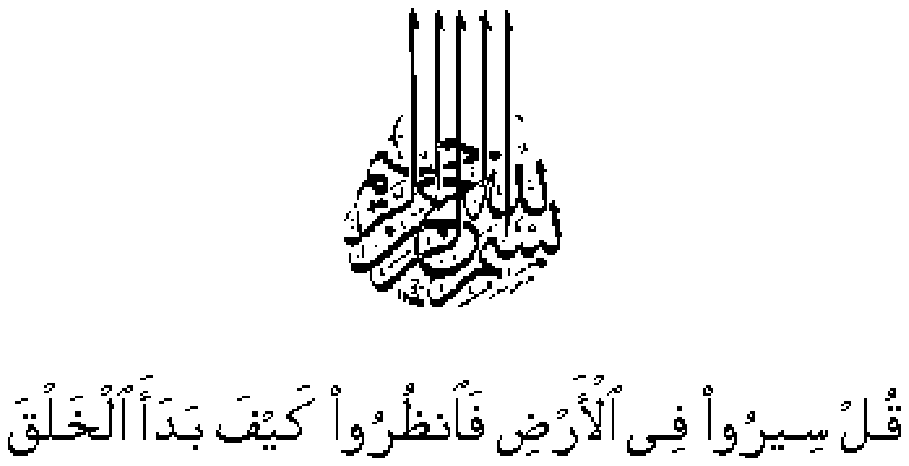}}\vskip5pt
        \centerline{``Tell them to walk the earth and discover how creation was started'' [Quran 29:20]}
        \centerline{\scriptsize{God commanding the prophet Mohammed to encourage humanity to explore and seek knowledge}}
  \end{center}
\end{dedication}

\chapter{Acknowledgments}

\begin{itemize}
    \item The theory formally known as string theory is, in fact, a vast
    field with a multitude of complex and intricate details. Without the
    guidance of a patient expert, understanding the
    big picture would seem to be a daunting, if not impossible,
    task to the beginner. My advisor, Dr. David Kastor, has proven to be
    such a guide. I am very grateful for his inspiration, without
    which, I would have drowned in the string sea long ago.
    \item Graduate school is more than just research and study. Part of
    the experience essential, in my view, for building up the
    scientific character of the student is the environment itself.
    I have enjoyed being a member of the community of the physics
    department at UMass Amherst. I feel indebted to the entire department
    for having made this experience such a great pleasure, as well as an
    excellent education. Special thanks to our teachers whose knowledge they
    have freely and generously handed down to us.
    \item The making of a scientist starts a lot earlier than school. I am greatly indebted to my father, who has
    always been an example that one aspires to, and to my mother,
    for being exactly the mother she is. Thank you for your love
    and devotion.
    \item Last, but certainly not least, I am grateful to my wife
    Manal for putting up with me all these years, which, as those who
    know me well can confirm, is a nontrivial task indeed.
\end{itemize}

\begin{abstract}

Close studies of the solitonic solutions of $D=11$ $\mathcal{N}=1$
supergravity theory provide a deeper understanding of the elusive
M-theory and constitute steps towards its final formulation. In
this work, we propose the use of calibration techniques to find
localized intersecting brane solutions of the theory. We test this
hypothesis by considering K\"{a}hler and special Lagrangian
calibrations. We also discuss the interpretation of some of these
results as branes wrapped or reduced over supersymmetric cycles of
Calabi-Yau manifolds and we find the corresponding solutions in
$D=5$ $\mathcal{N}=2$ supergravity.

\end{abstract}

\setcounter{tocdepth}{2}        %

\tableofcontents                %

\mainmatter                     %

\unnumberedchapter{Introduction}

The current understanding in fundamental physics is that there
exists a unique, nonperturbative quantum theory in eleven
spacetime dimensions, from which the known superstring theories in
ten dimensions arise as perturbative limits. The theory is known
as M-theory, and the search for its explicit form is one of
physics' greatest challenges of the twenty first
century\footnote{There is no general consensus as to what the `M'
stands for; Matrix, Membrane, Magic, Mystery, or even Mother (of
all theories), have been proposed. It is perhaps intentionally
left as ambiguous as the theory itself.}. The well known
$\mathcal{N}=1$ $D=11$ supergravity theory is postulated to be its
low energy limit. This means that its supersymmetric solitonic
solutions are also solutions of M-theory. There are exactly two
BPS brane solutions in eleven dimensions; these are the M2-brane
and its magnetic dual; the M5-brane. Other solutions may be
constructed by overlaps, intersections and wrappings of the
fundamental two.

The mathematical theory of calibrations is a technique developed
to find minimal manifolds that have the ability to sustain
constant spinors; a key ingredient in supersymmetric solutions. We
use it to find solutions representing localized M-brane
intersections, and show that their metrics have the general form
found earlier by Fayyazuddin and Smith. In addition, some of these
solutions are related to branes wrapped over supersymmetric cycles
of manifolds with restricted holonomy. The BPS solutions in the
reduced theory couple to fields that depend on the particular
topology of the compact space. This is what we set out to
demonstrate in this work. The dissertation is organized as
follows:

\begin{description}
    \item[Chapter One] We review the fundamentals of string theory
    and discuss the latest developments.
    \item[Chapter Two] We present a detailed review of brane physics with
    particular emphasis on intersections of M-branes. We discuss the $D=11$ solutions found by Fayyazuddin and Smith (FS) in 1999,
    which were the basis and motivation for this work.
    \item[Chapter Three] We present the theory of calibrations then proceed to generalize the FS result to include various K\"{a}hler
    calibrated cases and a certain special Lagrangian calibrated solution. Our
    discussion motivates the search for five dimensional solutions that represent the wrapping of M-branes over SLAG calibrated surfaces.
    \item[Chapter Four] We dimensionally reduce the theory over a Calabi-Yau
    manifold and present the result in the standard form of $D=5$
    supergravity theory with hypermultiplet fields. The calculation reveals the deeply rich structure of the lower dimensional theory.
    \item[Chapter Five] We find five dimensional 2-brane solutions
    and interpret them as M2 and M5-branes wrapped over a torus. We show that this last corresponds to the special case
    solution we found in eleven dimensions. We generalize this case to an arbitrary Calabi-Yau
    manifold. The general case is studied in detail and we briefly comment
    on a recently published $D=11$ solution that is supposed to
    reduce to it.
    \item[Conclusion] We summarize our results and point out
    various ways with which one can proceed in future research.
\end{description}

We set the notational convention as follows. Throughout, unless
otherwise mentioned, all Lorentzian metrics will have the
signature choice $\left( { - , + , \ldots , + } \right)$. The
volume form is related to the totally antisymmetric Levi-Civita
symbol by
\begin{equation}
    \varepsilon _{\mu _1 \mu _2  \cdots \mu _D}  = e\bar \varepsilon _{\mu
    _1 \mu _2  \cdots \mu _D},
\end{equation}
where $e = \sqrt {\det g_{\mu \nu } }$, $\bar \varepsilon _{012
\cdots }  =  + 1$ and
\begin{equation}\label{}
\bar \varepsilon _{\mu _1 \mu _2  \cdots }  = \left\{
{\begin{array}{*{20}c}
   0 & {{\rm any}\;{\rm two}\;{\rm indices}\;{\rm repeated}}  \\
   { + 1} & {{\rm even}\;{\rm permutaions}}  \\
   { - 1} & {{\rm odd}\;{\rm permutations}}  \\
\end{array}} \right.
\end{equation}

Other notations will be defined as the need arises.

\chapter{All roads lead to M-theory}

\begin{figure}
    \centering
    \input{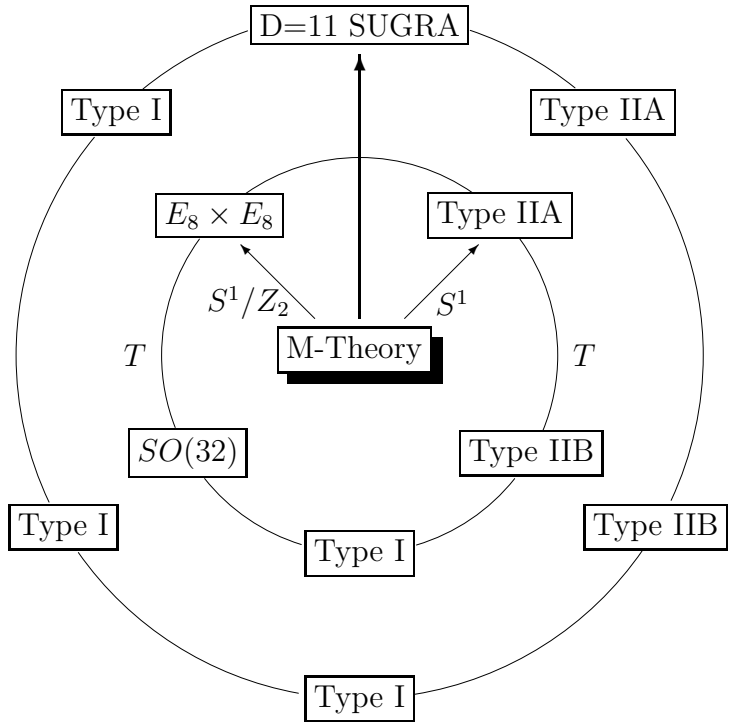}
    \caption{The current state of string theory: The inner circle gives the five
            known superstring theories in ten dimensions. The outer circle represents
            the three known supergravity theories in ten dimensions in addition to
            the eleven dimensional theory.}
    \label{mtheory}
\end{figure}

The search for a theory of everything (TOE) is certainly
theoretical physics' most exciting quest of the new century, if
not of the millennium. In this chapter, we will describe the
current status of this task in an attempt to motivate this work
and demonstrate how it fits in the big picture. A quick glance at
Fig.(\ref{mtheory}) summarizes the story. We will start with a
discussion of the inner circle in the figure, then expand and
contract in order to link the whole picture together.

Currently, our most successful understanding of the fundamental
workings of the universe is manifest in the collection of theories
known as the standard model (SM), which in turn is based on the
principles of quantum field theory. Absent in the model is any
characterization of the gravitational field. This last is
described classically by the general theory of relativity (GR),
but all attempts at a quantum description of gravity are, at best,
purely theoretical with no experimental backing.

In many physicists' view, the beauty of the standard model lies in
the fact that it has shown us that nature is based on simple
principles of symmetry. In addition, the standard model has
managed to provide us with incredibly accurate predictions of the
properties of elementary particles. Despite this success, many
feel that the standard model is lacking in several respects.
Foremost among those is its failure to account for the
gravitational field, in that quantizing general relativity yields
only an effective field theory over large length scales. Another
weakness is the arbitrariness of the model; in that it contains
far too many dimensionless parameters. In addition to that, the
appearance of ultraviolet divergences, albeit removable by
renormalization, signals our ignorance of an even deeper structure
of nature. The origin of these divergencies can be traced to the
fact that elementary particles are treated as point objects.
Attempts to remedy this situation have been with us as early as
the late nineteenth century via the works of Poincar\'{e} and
others on the problem in its classical form. But a truly
divergence-free quantum model of all interactions, including
gravity, has not been proposed until the discovery of string
theory in the 1970's. One of the theory's major attractive
features is that it is quite self-contained. Once a few initial
assumptions are made, the theory appears to require no further
adjustment, but rather spews out its own requirements. It even
tells us what number of spacetime dimensions it wishes to live in,
a feat unaccomplished by any previous theory.

We will proceed to tell the tale from the start. In addition to
cited sources, we will freely quote from the standard texts
\cite{GSW,Polchinski}. Other excellent reviews and lecture notes
include \cite{Das,Duff5,Forste,Sen,Pol2,Town,West}.

\section{Constructing a theory of strings}

We begin by assuming that all particles are described by
oscillating strings, freely propagating in a flat $D$-dimensional
spacetime background. In addition, we demand that the theory be
Poincar\'{e} invariant. Later, we will see that this is not
enough, a deeper symmetry is required.

The simplest Poincar\'{e} invariant string action is the
Nambu-Goto action:
\begin{eqnarray}\label{nambu}
        S_{NG} &=&  - \frac{1}{{2\pi \alpha '}}\int {d\tau d\sigma \sqrt { - \det h_{ab}}} \nonumber \\
               &=&  - \frac{1}{{2\pi \alpha '}}\int {d\tau d\sigma \sqrt { - \left| {\left( {\partial _a X^\mu } \right)\left( {\partial _b
                    X^\nu  } \right)\eta _{\mu \nu} } \right| } },
\end{eqnarray}
where $\tau$ is the proper time of the string's worldsheet, and
$\sigma$ is the proper length\footnote{In the more familiar case
of point particle theory, one number only; the proper time, is
sufficient to parameterize the particle's worldline.}. The indices
$a$ and $b$ run over $\left( {\sigma, \tau } \right)$, and
$h_{ab}$ is the induced metric of the target space, defining the
pullback of spacetime on the worldsheet. The spacetime coordinates
$X^\mu \left( {\sigma, \tau } \right)$ $\left( {\mu,\nu = 0,
\ldots D - 1} \right)$ describe the embedding of the worldsheet in
the flat background. The overall numerical factor is the tension
of the string, or its mass per unit length. For historical
reasons, the constant $\alpha '$ is known as the Regge slope, and
sets the size scale of the theory\footnote{Based on estimates of
graviton exchange between two particles one finds that the length
scale of the string has to be of the order of the Planck length
$l_P  = 1.6 \times 10^{ - 33}$cm. This translates into the Planck
mass $M_P  = 1.22 \times 10^{19}$GeV, which explains why the
stringy nature of particles (if at all correct) has not been
observed in present day particle accelerators.}.

Mathematically, what this action describes is simply the area of a
$\left( {1 + 1} \right)$ dimensional surface embedded in
$\mathbb{R}^{D - 1,1}$, and invariant under $SO
\left(D-1,1\right)$ rotations. Minimizing this action amounts to
minimizing the surface area of the worldsheet described by the
string as its sweeps its way through spacetime.

The square root in the Nambu-Goto action is a cumbersome object in
a quantum theory. It is avoided by defining an equivalent action,
in that it provides the same equations of motion, known as the
Polyakov action:
\begin{equation}\label{polyakov}
    S_P  =  - \frac{1}{{4\pi \alpha '}}\int {d\tau d\sigma \sqrt { -
    \gamma } \gamma ^{ab} \left( {\partial _a X^\mu } \right)\left(
    {\partial _b X^\nu } \right)\eta _{\mu \nu} },
\end{equation}
where $\gamma_{ab}$ is the worldsheet Lorentzian metric and
$\gamma = \det  {\gamma _{ab} } $. The action is mathematically
equivalent to a free scalar field theory in a two dimensional
background. In this interpretation, the index $\mu$ simply counts
the scalar fields $X$. For this reason, the spacetime coordinates
$X^\mu$ are sometimes called worldsheet fields. They are $D$
noninteracting massless scalar fields in a (1+1) dimensional world
with a metric $\gamma_{ab}$.

In addition to $D$-dimensional Poincar\'{e} invariance, we find
that the action is also diffeomorphism invariant under a change of
worldsheet coordinates and Weyl invariant under rescaling of the
worldsheet metric. In fact, Weyl invariance is only possible in
the case of a two dimensional worldsheet, making the string more
prominent among other extended objects such as membranes.

Varying the action with respect to $\gamma_{ab}$ shows that it is
proportional to the induced metric $h_{ab}$ and yields the
energy-momentum tensor on the worldsheet in the usual manner.
Plugging this in (\ref{polyakov}) retrieves (\ref{nambu}). Varying
with respect to $X^\mu$ yields the required string equations of
motion:
\begin{equation}\label{wave}
    \nabla ^2 X^\mu  = 0,
\end{equation}
where the Laplacian is with respect to the worldsheet coordinates.
This is a single string propagating without sources. Since the
canonical momentum to $X^\mu$ is $\left( {\partial _a X^\mu }
\right)$, we find that, for an open string, we are forced to
impose Neumann boundary conditions in order to avoid momentum flow
off the ends of the string. For a closed string we choose periodic
boundary condition.

Although the Nambu-Goto and Polyakov actions are the simplest one
can write, we demand that the theory be as general as possible
under its given symmetries. We find that we can add one more term
to the theory that preserves (diff$\times$Weyl) and Poincar\'{e}
invariances, while keeping the action polynomial in derivatives.
This term is as follows
\begin{equation}
    S_\chi   = \lambda \chi  = \lambda \left( {2 - 2g} \right) =
    \frac{\lambda }{{4\pi }}\int {d\tau d\sigma \sqrt { - \gamma }
    R},
\end{equation}
where $R$ is the Ricci scalar of the worldsheet, $\lambda$ is a
numerical factor, $g$ is the genus of string loop diagrams and
$\chi$ is the Euler characteristic of the worldsheet. Since this
term is topologically invariant, it does not alter the equations
of motion, but rather plays the role of the string coupling
constant in string interactions; $g_s \sim e^\lambda$.

\subsection{The string spectrum}

The string equation of motion (\ref{wave}) is obviously a wave
equation. For open strings, the most general form of its solution,
with Neumann boundary conditions, may be written as an expansion
in Fourier modes as follows:
\begin{equation}\label{spectrum}
    X^\mu \left( {\sigma, \tau } \right) = x^\mu  + 2\alpha ' p^\mu \tau  +
    i\sqrt {2\alpha '} \sum\limits_{\scriptstyle n =  - \infty  \hfill \atop
    \scriptstyle n \ne 0 \hfill}^{n = \infty } {\frac{1}{n}\alpha
    _n^\mu  e^{ - in\tau } \cos \left( {n\sigma } \right)},
\end{equation}
where $x^\mu$ and $p^\mu$ are the position and momentum vectors of
the center of mass of the string respectively, and $\alpha^\mu_n$
are the Fourier coefficients, which in the quantum theory are
interpreted as annihilation and creation operators in the Hilbert
space of the string's oscillations. For closed strings with
periodic boundary conditions, there are two solutions; one for
right moving waves and the other for left moving waves.
Corresponding to each direction of motion, there are two sets of
annihilation and creation operators; $\alpha$ and ${\tilde \alpha
}$. The Hilbert space of string oscillations naturally has number
operators for each case that can be used to find the string
spectrum in the usual manner familiar from field theory. The
general state for an open string moving in spacetime with center
of mass momentum $p$ can be constructed as follows:
\begin{equation}
    \left| {N;p} \right\rangle  = \left[ {\prod\limits_{\mu = 2}^{D - 1}
    {\prod\limits_{n = 1}^\infty  {\frac{{\left( {\alpha _{ - n}^\mu }
    \right)^{N_{in} } }}{{\left( {n^{N_{in} } N_{in} !} \right)^{{1
    \mathord{\left/
     {\vphantom {1 2}} \right.
     \kern-\nulldelimiterspace} 2}} }}} } } \right]\left| {0;p}
     \right\rangle,
\end{equation}
where the corresponding spectrum for the closed string states
contains two sets of $\alpha$'s and an extra ${\tilde N}$ for
waves moving in the opposite direction to those belonging to $N$.
The masses of open string states can be calculated via the formula
\begin{equation}
    m^2  = \frac{1}{{\alpha '}}\left( {N - 1} \right),
\end{equation}
where we have used the fact that this string theory lives in
$D=26$ spacetime dimensions as required by the cancellation of
quantum anomalies\footnote{Featured by the vanishing of negative
norm states, or ghosts.}. Obviously, the ground state of the
theory ($N=0$) is tachyonic. This signals the instability of the
theory and can only be dealt with by introducing an even more
general spacetime symmetry than Poincar\'{e}'s, as we will see
shortly.

For closed strings the formula is
\begin{equation}\label{mass}
    m^2  = \frac{2}{{\alpha '}}\left( {N + \tilde N - 2} \right).
\end{equation}

The string states presumed to yield the observed elementary
particles in lower dimensions are the ones corresponding to
$N={\tilde N}=1$, representing massless particles. An important
such state is a particle with spin two. This, naturally, is
interpreted as the graviton\footnote{A spin two boson is the
minimum possibility for a force-carrying particle to be attractive
only. For a review of this argument, we recommend Feynman's
excellent, albeit outdated, book \cite{Feynman}.}, confirming that
string theory necessarily contains gravity as a solution. Since
the harmonics on the string can go arbitrarily high, we find that
there is an infinite tower of massive states allowed on the
string. The masses of these `particles' are of the order of the
Planck mass, making them highly inaccessible to us. It is assumed
that the particles we observe today are the result of the massless
string states, acquiring mass via spontaneous symmetry breaking.

\subsection{\label{inter}Interactions and other issues}

In ordinary field theories, one normally adds new terms to the
action in order to account for interactions. String theory, as we
have noted before, is quite self-contained, in that one need add
no more terms to the `free' string action in order to introduce
interactions. The theory contains within itself all the necessary
information required to calculate multi-string collision cross
sections. Applying Feynman's path integral quantization method
starts by formulating the so called Polyakov path integral:
\begin{equation}\label{path}
    Z_P  = \int {\left[ {dXd\gamma } \right]e^{ -
    S\left[ {X,\gamma } \right]}},
\end{equation}
where from here on, $\gamma$ is redefined to be a Euclidean
worldsheet metric instead of a Lorentzian one as has been
previously assumed\footnote{This, of course, is clear from the
absence of a factor of $i$ in the exponent of (\ref{path}).}. The
action $S$ is the Polyakov action plus the Euler characteristic
term. The integral is performed over all string topologies
possible, where the lowest order diagram for the closed string is
the sphere, the first order diagram is the genus-one torus and so
on. This is where the Euler term in the action shows prominence.

Treated as a two dimensional conformal field theory, one can show
that string theory in a flat 26-dimensional background can be
reformulated as if it were in a curved background. Roughly
speaking, a theory of strings on a curved background is equivalent
to one embedded in a flat metric with a background teaming with
coherent gravitonic string states. This enables us to rewrite the
Polyakov action in terms of a curved background metric $G_{\mu
\nu} \left( X \right)$. We further generalize this by including
other massless background fields; namely a totally antisymmetric
tensor field $B_{\mu \nu} \left( X \right)$ and a scalar field
$\phi$ known as the dilaton. So, even though these are stringy
states, one can treat them as independent fields that couple to
the string. The Polyakov action then becomes
\begin{equation}
    S_\sigma   = \frac{1}{{4\pi \alpha '}}\int {d\tau d\sigma \sqrt
    \gamma  \left[ {\left( {\gamma ^{ab} G_{\mu \nu}  + i\varepsilon ^{ab}
    B_{\mu \nu} } \right)\left( {\partial _a X^\mu } \right)\left( {\partial
    _b X^\nu } \right) + \alpha 'R\phi } \right]}.
\end{equation}

For historical reasons, this action is known as the sigma model
\cite{Bagger}. In addition to these fields, we will see in the
next section that certain $p$-form gauge fields arise in
superstring theory. Again, these are really massless string states
which we organize in terms of classical background fields. In
other words, the spectrum of massless states in string theory is
identical to that of a free field theory with such fields.

An important thing to note about string theory is that there are
no free parameters. The Regge slope ${\alpha '}$ is dimensionful,
setting the scale of the theory, hence not a free parameter and
can simply be absorbed in the definition of $X^\mu$. What about
the coupling constant $g_s$ arising from the Euler characteristic
term? It turns out that this is dependent on the expectation value
of the dilaton field, hence not really a constant, but a parameter
that may, in principle, be determined from the dynamics.

\section{Superstrings}

The string theory briefly reviewed in the last section suffers
from some serious drawbacks. Firstly, its ground state is a
tachyon, signaling the instability of the vacuum. Secondly, it
contains only bosonic degrees of freedom, meaning that it can only
account for spacetime bosons. Thirdly, it is consistent and
anomaly free only in 26 spacetime dimensions. The first and second
problems can be fully solved, along with a partial solution to the
third, with a single stroke.

If the Polyakov action deals with the (commuting) bosonic degrees
of freedom $X^\mu$, then perhaps the solution to the absence of
fermions is to add more terms that would correspond to
(anticommuting) fermionic worldsheet fields. Does this mean that
we have to forgo our insistence that the theory is self-contained
within its symmetries and start adding terms by hand whenever
deemed convenient? Not necessarily. The keyword here is
`symmetries'. If one starts with Poincar\'{e} symmetry only, we
find the bosonic theory discussed before, with (diff$\times$Weyl)
invariance as a bonus. However, Poincar\'{e}'s is not the most
general spacetime symmetry possible. In fact, it has been argued
in the literature that the only possible generalization of
Poincar\'{e} invariance is a symmetry that ascribes both bosonic
and fermionic degrees of freedom to all particles (read: strings),
and demands the invariance of physics under rotations in this
generalized `superspace'. This is, of course, supersymmetry
(SUSY). One could have started this chapter by demanding that our
action be supersymmetric from the beginning, in which case the
Polyakov action would have been originally written with worldsheet
fermions included. In gauge fixed form, this is
\begin{equation}\label{super}
    S = \frac{1}{{4\pi }}\int {d^2 z\left[ {\frac{2}{{\alpha '}}\left(
    {\partial X^\mu } \right)\left( {\bar \partial X^\nu } \right) + \psi
    ^\mu \left( {\bar \partial \psi ^\nu } \right) + \tilde \psi ^\mu (
    {\partial \tilde \psi ^\nu } )} \right]\eta _{\mu \nu} },
\end{equation}
where the spinor $\psi$ and its adjoint $\tilde \psi$ are the
fermionic degrees of freedom (massless worldsheet fermions), with
Dirac $\Gamma$ matrices obeying the Clifford algebra $\left\{
{\Gamma _a ,\Gamma _b } \right\} = 2\eta _{ab} $ in the usual
manner. As is conventional, we have switched to the coordinate
redefinition
\begin{eqnarray}
        z &=& \sigma  + i\tau ,\quad \bar z = \sigma  - i\tau , \nonumber\\
        dz^2  &=& 2d\tau d\sigma.
\end{eqnarray}

In addition, Diff and Weyl symmetries allow us to choose the
worldsheet metric $\gamma_{ab}$ to be locally flat $\gamma _{ab}
\to \eta _{ab}$. This enables us to study the conformal invariance
of the two dimensional worldsheet theory and use its nature as a
conformal field theory (CFT) to calculate probability amplitudes.
Note that the equations of motion one gets by varying the action
with respect to the $\psi$'s are none other than the Dirac
equations for massless fermions in two dimensional flat spacetime.

Furthermore, as a supersymmetric theory, this action is invariant
under a set of infinitesimal superconformal transformations that
interchanges bosonic and fermionic degrees of freedom.

It is worthwhile emphasizing that supersymmetry is the maximal
spacetime symmetry\footnote{To be more precise, it is the most
general symmetry that mixes space and time with internal degrees
of freedom.}. No further generalizations are possible. Hence, with
the addition of the Euler term as usual, the action (\ref{super})
is the most symmetric string action one can write.

\subsection{Neveau-Schwarz and Ramond}

In the case of the closed string, one finds that a fermionic
oscillation, as it travels full circle around the string, has the
choice of returning back to either itself or negative itself.
These periodicity conditions are called:
\begin{eqnarray}
        {\rm Ramond\; (R)\;}&:&\quad \psi ^\mu \left( {z + 2\pi } \right) =  + \psi ^\mu \left( z \right), \\
        {\rm Neveau - Schwarz\; (NS)\;}&:&\quad \psi ^\mu \left( {z + 2\pi } \right) =  - \psi ^\mu \left( z \right),
\end{eqnarray}
for all values of $\mu$. The same conditions are imposed on
${\tilde \psi }$ also. To investigate the spectrum in a given
sector, we may expand the fermionic degrees of freedom similarly
to (\ref{spectrum}) as a Fourier series:
\begin{equation}
    \psi ^\mu  \left( z \right) = i^{ - \frac{1}{2}} \sum\limits_{r
    \in \mathbb{Z} + v} {\psi _r^\mu  e^{irz} },
\end{equation}
and similarly for $\tilde \psi$. The coefficients $\psi _r^\mu$
and ${\tilde \psi _r^\mu  }$ are the analogues to the bosonic
$\alpha_n ^\mu$ and $\tilde \alpha_n ^\mu$; they create and
annihilate worldsheet fermions.

The choice of periodicity conditions enables us to write:
\begin{equation}
    \begin{array}{l}
        \psi ^\mu \left( {z + 2\pi } \right) = e^{2\pi i v } \psi ^\mu \left( z \right) \\
        \tilde \psi ^\mu \left( {\bar z + 2\pi } \right) = e^{ - 2\pi i \tilde v } \tilde \psi ^\mu \left( {\bar z} \right) \\
    \end{array}
\end{equation}
where $v$ and $\tilde v$ take the values $0$ and $\frac{{\rm
1}}{{\rm 2}}$. So, for the four possible combinations of $v$ and
$\tilde v$, corresponding to periodic and antiperiodic boundary
conditions for either right or left moving fermionic waves, we get
four different Hilbert spaces; NS-NS, R-NS, NS-R and R-R.

The NS and R sectors are in the tensorial and spinorial
representations of $SO(9,1)$ Lorentz algebra respectively.
Obviously then, the NS-NS and R-R sectors contribute integer spin
states; the spacetime bosons of the theory (as opposed to
worldsheet bosons). Their spacetime fermionic partners come from
the R-NS and NS-R sectors. Presumably, upon dropping down to four
dimensions and SUSY breaking, these are the objects responsible
for what we observe as the particles of the standard model. The
NS-NS sector is the one responsible for the graviton $G_{\mu
\nu}$, the antisymmetric $B_{\mu \nu}$ field and the dilaton
$\phi$. It is sometimes called the universal sector since it is
present in bosonic string theory (as we saw explicitly in
\S\ref{inter}) as well as all superstring theories discussed in
the next subsection.

\subsection{Superstrings in ten dimensions}

The discovery of supersymmetry solved the absence of fermions
problem in string theory as we saw in the last subsection. In
addition to that, it turns out that the problem of the extra
dimensions is also partially solved. Supersymmetry puts additional
constraints on the theory in such a way that the cancellation of
quantum anomalies now requires the theory to live in `only' ten
spacetime dimensions. Furthermore, worldsheet supersymmetry has a
property known as modular invariance that allows us to define the
so called GSO projection, which projects out unphysical string
states\footnote{The action \ref{super} is worldsheet
supersymmetric, the GSO projection is required in order to realize
spacetime supersymmetry. The standard quantization approach known
as the Ramond-Neveau-Schwartz formalism uses the GSO projection.
Another approach known as the light-cone Green-Schwartz formalism
does not, and spacetime SUSY is explicit from the start. However,
this superstring theory cannot be quantized in a fully covariant
way, hence it is rarely used.}. Foremost among those is the
tachyonic ground state. So, supersymmetry also takes care of the
vacuum instability that has beset the bosonic theory.

Putting all of the SUSY restrictions together, we find that the
number of physically feasible superstring theories is exactly
five. These are:
\begin{description}
  \item[Type IIA string theory] This theory is ${\cal N}=2$
  supersymmetric, as denoted by the ``II'' in its name. The strings are closed and orientable. The
  gauge symmetry is $U(1)$, provided by a R-R 1-form field $A_\mu$.
  In addition, the theory contains a 3-form gauge field $C_{\mu \nu
  \rho}$. These R-R $p$-form fields will play an essential role
  in our discussion of the solitonic solutions of string theory,
  known as branes.
  \item[Type IIB string theory] A theory of closed orientable
  strings with ${\cal N}=2$ supersymmetry containing no gauge
  symmetry. In addition to the universal sector present in all
  five theories, the R-R sector contributes another scalar field $\ell$
  and a second antisymmetric tensor field $C_{\mu \nu }$
  as well as a 4-form $C_{\mu \nu \rho\sigma}$ with self-dual field strength.
  \item[Type I string theory] The only theory with both open and
  closed strings. It is ${\cal N}=1$ supersymmetric with unorientable strings. The theory has $SO(32)$ gauge symmetry.
  \item[$SO(32)$ heterotic string theory]A theory of closed
  orientable strings with ${\cal N}=1$ supersymmetry. The gauge group is, as implied by the name, $SO(32)$. A
  heterotic theory is constructed by considering the possibility
  that the number of left moving oscillations is not necessarily
  the same as the number of right movers.
  \item[$E_8\times E_8$ heterotic string theory] Also ${\cal N}=1$
  with orientable strings. The gauge group is $E_8\times E_8$.
  It turns out that this particular theory is the most likely
  candidate for a theory that produces the standard model. This can
  be seen by noting that the exceptional group $E_8$ contains
  within it $SU(3)\times SU(2)\times U(1)$ as a subgroup. This is of
  course the group structure of the standard model. The other
  $E_8$ predicts another ``parallel standard model'' that has not been observed
  in nature. For this reason, this second $E_8$ is said to be
  responsible for a ``shadow world'' that is closely related to
  ours but not as easily observable.
\end{description}

Although only five superstring theories are physically possible,
this still seemed like an embarrassment of riches, as one would
normally hope for a unique theory. As it turns out, certain
relations known as dualities have been found that connect all five
theories together, showing that they are not independent after
all, but rather descriptions of five different sectors of a single
truly unique theory.

\section{\label{SUGRA}Supergravity}

The set of theories known as supergravity (SUGRA) have been
constructed as early as the 1970's and developed quite separately
from string theory. These are what one gets by combining
Einstein's general relativity with supersymmetry in any arbitrary
number of dimensions\footnote{There are constraints on this as we
will note later.}. As it turns out, supergravity theories are in
fact low energy limits of string theory. Basically, one expands
the string action in $\alpha '$ and drops all terms of
$\mathcal{O}(\alpha '^2)$ and higher. This is of course not the
way supergravity theories were originally discovered.
Historically, they were constructed by supersymmetrizing the
Einstein-Hilbert action and hoping that it gave a theory of
everything unifying all interactions with gravity. This did not
work, mainly because on quantization the theories were found to be
as nonrenormalizable as general relativity itself. Hence they can
either be totally wrong or, at best, low energy effective limits
of a better idea, which is indeed what they turned out to be.

Another way to construct a supergravity theory is to begin with a
supersymmetric Yang-Mills theory without gravity and then insist
that the supersymmetry transformations act locally, i.e. become
spacetime dependent. Surprisingly, or perhaps not so much, this
automatically introduces gravity in the theory. This can roughly
be understood as follows: Treated as a Noether symmetry,
supersymmetry has a set of generators known as the supercharges
$Q$. The operation of those charges on either bosons or fermions
is what flips them to the other side. They are anticommuting
spinorial operators that obey algebras of the general form
(spinorial indices suppressed):
\begin{equation}
    \left\{ {Q,Q} \right\} \sim \Gamma ^\mu  P_\mu   +
    \sum\limits_{n} {\Gamma ^{\mu _1  \cdots \mu _n }
    Z_{\mu _1  \cdots \mu _n } } \label{SUSYalgebra}
\end{equation}
where the $\Gamma^{\mu _1  \cdots \mu _n}$ matrices are the
antisymmetrized products of the usual Dirac $\Gamma^\mu$ matrices.
The second term is a sum over all possible values of $n$ as
dictated by supersymmetry in the particular spacetime dimensions
we are interested in. The $Z$ operators are related to the gauge
potentials of the theory and when diagonalized yield the central
charges. We will have more to say about this term later. The
important parameter for our purposes here is the vector $P_\mu$.
This is the energy-momentum vector in $D$ dimensions and is, of
course, the generator of translations. Phenomenologically, we may
view the presence of this vector as follows: Suppose two bosonic
point particles are located in the same spacetime location. Now
consider a supersymmetric transformation that is induced on these
particles at some instant in time, changing them into fermions.
Fermions cannot, by virtue of Pauli's exclusion principle, stay at
the same point. We conclude that the two particles must move
apart, i.e. a translation has occurred.

The presence of the spacetime translation operator means spacetime
diffeomorphisms, and the invariant actions are generalizations of
the Einstein-Hilbert action, i.e. gravity arises naturally in a
locally supersymmetric theory. Our conclusion then is that
supersymmetry is not such an arbitrary restriction on nature, but
rather a maximal symmetry of spacetime with far reaching
implications.

Now, we will give a quick review of ten dimensional supergravity
theories and how they relate to the superstring theories.
Eventually, we will focus on two other SUGRA theories that will
become of interest to this work later on. The reader will note
that, as is traditional, only the Bosonic parts of the actions are
displayed. In addition to that, one also needs to construct the
supersymmetry transformations each (full) action is invariant
under.

\begin{description}
  \item[Type IIA string theory] reduces to type IIA SUGRA
  with the action:
    \begin{eqnarray}
        S_{IIA} &=&  \frac{1}{{2\kappa _{10}^2 }}\int {d^{10} x\sqrt { - G} e^{ - 2\phi } \left[ {R + 4\left( {\partial _\mu  \phi } \right)\left( {\partial ^\mu  \phi } \right) - \frac{1}{2}\left( {dB_2} \right)^2 } \right]} , \nonumber \\
                  &-& \frac{1}{{4\kappa _{10}^2 }}\int {d^{10} x\sqrt { - G} \left[ {(dA_1) ^2  + \left( {dC_3  - A_1  \wedge dB_2 } \right)^2 } \right]} , \nonumber \\
                  &-& \frac{1}{{4\kappa _{10}^2 }}\int {B_2  \wedge  dC_3  \wedge
                  dC_3}.
    \end{eqnarray}
  The last term is known as a Chern-Simons (CS) term. As we will see, CS terms are abundant in supergravity theories. They are invariant under the gauge transformation only up to a total derivative.
  From the NS-NS sector, we have the graviton $G$, the dilaton $\phi$ and the antisymmetric $B$ field as usual. The
   R-R sector gives the $U(1)$ $A_1$ field (the subscript giving the value of $p$ in $p$-form), and the 3-form $C_3$ field. As usual, `$d$' is the exterior derivative acting on
   $p$-forms. Finally, $\kappa$ is a constant relating Newton's gravitational constant in ten dimensions with the Planck length scale. For simplicity, we will refer to $\kappa$ as Newton's constant.
  \item[Type IIB string theory] reduces to type IIB SUGRA.
  However, due to the presence of the self dual field strength $dC_4 = \star dC_4$, where
  $\star$ is the Hodge dual operator, a covariant action of the theory is not known, but the
  following comes close:
    \begin{eqnarray}\label{IIB}
        S_{IIB}  &=& \frac{1}{{2\kappa _{10}^2 }}\int {d^{10} x\sqrt { - G} e^{ - 2\phi } \left[ {R + 4\left( {\partial _\mu  \phi } \right)\left( {\partial ^\mu  \phi } \right) - \frac{1}{2}\left( {dB_2} \right)^2 } \right]} , \nonumber \\
                  &-& \frac{1}{{4\kappa _{10}^2 }}\int {d^{10} x\sqrt { - G} \left[ {\left( {d\ell} \right)^2  + \tilde F_3^2  + \frac{1}{2}\tilde F_5^2 } \right]} , \nonumber \\
                  &-& \frac{1}{{4\kappa _{10}^2 }}\int {C_4  \wedge  dB_2  \wedge dC_2}
                  ,\\
                \tilde F_3  &=& dC_2  - \ell  \wedge dB_2 , \nonumber \\
        \tilde F_5  &=& dC_4  - \frac{1}{2}C_2  \wedge dB_2  + \frac{1}{2}B_2  \wedge
        dC_2.
    \end{eqnarray}

  In addition, we impose the self duality of $dC_4$ on the equations of motion.
  \item[Type I string theory]reduces to type I SUGRA with the
  following action:
  \begin{eqnarray}
    S_I  &=& \frac{1}{{2\kappa _{10}^2 }}\int {d^{10} x\sqrt { - G} e^{ - 2\phi } \left[ {R + 4\left( {\partial _\mu  \phi } \right)\left( {\partial ^\mu  \phi } \right) - \frac{1}{2}\tilde F_3 ^2 } \right]}  \nonumber \\
    &-& \frac{1}{{4{\kappa _{10}^2 } }}\int {e^{ - \phi } Tr\left( {dA_1 } \right)^2
    }, \\
    {\rm where}\;\; \tilde F_3  &=& dC_2  - \frac{1}{2}Tr\left( {A_1  \wedge dA_1  + \frac{2}{3}A_1  \wedge A_1  \wedge A_1 }
        \right), \label{tildeF}
  \end{eqnarray}
  and the trace is over the internal $SO(32)$ indices.
  \item[The heterotic string theories] both reduce to type I
  SUGRA since they have the same ${\cal N}=1$ supersymmetry,
  except that now the exponent of the dilaton is the same
  throughout, and the gauge fields will be either $SO(32)$ or $E_8\times
  E_8$:
  \begin{eqnarray}
    S_I  = \frac{1}{{2\kappa _{10}^2 }}\int {d^{10} x\sqrt { - G} e^{
    - 2\phi } \left[ {R + 4\left( {\partial _\mu  \phi } \right)\left(
    {\partial ^\mu  \phi } \right) - \frac{1}{2}\tilde F_3 ^2  +
    \frac{1}{2}Tr\left( {dA_1 } \right)^2 }
    \right]}
  \end{eqnarray}
  with $\tilde F_3$ defined by (\ref{tildeF}).
\end{description}

\subsection{${\cal N}=1$ $D=11$ supergravity}

Since supergravity was developed independently of string theory,
it was natural to try to formulate SUGRA theories in all possible
dimensions. This was indeed done in all dimensions between four
and ten. As it turns out, supersymmetry can only occur with values
${\cal N}=2^n$, where $n$ is an integer. There is a maximum limit,
however, to this that is dimension dependent. For example, in five
dimensions, only theories with ${\cal N}=$2, 4 or 8 SUSY can be
constructed. If one tries to go beyond the maximum, we find that
we need to introduce massless bosonic fields with spins greater
than two, which, in the quantum theory, causes unitarity
violation.

Based on this, it turns out that the maximum number of dimensions
where SUSY can be constructed is eleven, where a single unique
${\cal N}=1$ theory with 32 supercharges (corresponding to the
dimension of its spinors) lives\footnote{If one allows for more
than one time dimension, it turns out that ${\cal N}=1$
supersymmetric theories can be constructed in (10+2) and (6+6)
spacetimes. Extra time dimensions, however, cause the theory to be
unstable. In addition, there is no experimental evidence for the
existence of more than one time dimension. Of course, there is as
yet no evidence of more than three spatial dimensions either, but
those are at least phenomenologically acceptable in that they do
not cause any serious defects in the theory
\cite{Tegmark1,Tegmark2}.}. Since string theory is only consistent
in ten dimensions, there did not appear to be any relation between
strings and $D=11$ SUGRA \cite{CJSh}. It was later realized,
however, that there was indeed a subtle and profound relationship
between the two theories, which we will discuss later. In fact,
$D=11$ SUGRA is now one of the main roads to the ultimate theory.
This is the road along which, in fact, this work proceeds.

The bosonic part of the action is as follows:
\begin{equation} \label{eleven}
    S_{11}  = \int {d^{11} x\sqrt { - G}
    \left[ {R - \frac{1}{{48}}F^2 } \right]}  - \frac{1}{6} \int {A \wedge F \wedge
    F},
\end{equation}
where, for simplicity, we have assumed units where Newton's
constant $\kappa_{11}$ is equal to $1/\sqrt{2}$, otherwise each of
the terms in the action would be multiplied by a factor of
$1/2\kappa^2_{11}$. The theory contains gravity, a single
gravitino field $\psi$, and a 3-form gauge field $A_{MNP}$, where
the indices are eleven dimensional ($M,N=0,\ldots, 10$). The field
strength and its Hodge dual are defined in the usual manner:
\begin{eqnarray}
        A &=& \frac{1}{{3!}}A_{MNP} dx^M  \wedge dx^N  \wedge dx^P  \nonumber \\
        F &=& dA = \frac{1}{{4!}}F_{LMNP} dx^L  \wedge dx^M  \wedge dx^N  \wedge dx^P  \nonumber \\
        F_{LMNP}  &=& \partial _L A_{MNP}  - \partial _M A_{LNP}  + \partial _N A_{LMP}  - \partial _P
        A_{LMN},\nonumber \\
        \star F &=& \frac{1}{{7!}}F_{LMNOPQR} dx^L  \wedge dx^M  \wedge dx^N  \wedge dx^O  \wedge dx^P  \wedge dx^Q  \wedge dx^R  \nonumber \\
        F_{LMNOPQR}  &=& \frac{1}{{4!}}\varepsilon _{LMNOPQRABCD} F^{ABCD}.
\end{eqnarray}

The full action is invariant under the infinitesimal supersymmetry
transformations:
\begin{eqnarray}
    \delta_{\epsilon} e_{\;\;\;M} ^{\hat A }  &=&  - \frac{i}{2}\bar \epsilon \Gamma ^{\hat A } \psi _M    \label{dele}\\
    \delta_{\epsilon} \psi _M   &=& 2\nabla _M  \epsilon  + \frac{i}{{144}}F_{LNPQ } \left( {\Gamma _M  ^{\;\;\;LNPQ }  - 8\delta _M ^L  \Gamma ^{NPQ } } \right)\epsilon  \label{eleventrans} \\
    \delta_{\epsilon} A_{MNP }  &=& \frac{3}{2}\bar \epsilon \Gamma _{\left[ {M N } \right.} \psi _{\left. P
    \right]}\label{dela},
\end{eqnarray}
where $\epsilon$ is the 32-component SUSY spinor, hated indices
are eleven dimensional flat indices in a tangent space (also known
as frame indices), $e$ is the veilbeins defined by (\ref{bein}),
and the covariant derivative is defined by:
\begin{equation}
    \nabla _M \epsilon  = \left( {\partial _M \epsilon } \right) + \
    \frac{1}{4}\omega     _M^{\;\;\;\hat L\hat N} \Gamma _{\hat L\hat N}
    \epsilon,
    \label{covderivative}
\end{equation}
where the $\omega$'s are the spin connections.

The equation of motion for the gauge field is as follows:
\begin{equation}\label{}
    d\star F + \frac{1}{2}F \wedge F = 0,
\end{equation}
giving rise to the conservation of an `electric' charge
\begin{equation}\label{}
    Q_E  = \int\limits_{\partial \mathcal{M}_8 } {\star F + \frac{1}{2}A \wedge
    F},
\end{equation}
while the Bianchi identity $dF=0$ gives rise to the conserved
magnetic charge
\begin{equation}\label{}
    Q_M  = \int\limits_{\partial \mathcal{M}_5 } F.
\end{equation}

The supersymmetry algebra (\ref{SUSYalgebra}) of the theory is
(spinorial indices suppressed):
\begin{equation}\label{}
    \left\{ {Q,Q} \right\} = C\left( {\Gamma ^A P_A  + \Gamma ^{AB}
    Z_{AB}  + \Gamma ^{ABCDE} Z_{ABCDE} } \right).
\end{equation}

The $\Gamma$ matrices in eleven dimensions follow the Clifford
algebra
\begin{equation}\label{}
    \left\{ {\Gamma _M ,\Gamma _N } \right\} = 2G_{MN}.
\end{equation}

They are real $32 \times 32$ matrices acting on the 32-component
Majorana spinors $\epsilon$, the space of which carries the
$Spin(10,1)$-invariant symplectic charge conjugation form $C$,
which may be used to raise or lower spinorial indices.

This sums up the basic properties of eleven dimensional
supergravity. We will postpone further discussion, particularly of
its solutions, to chapter (\ref{Branes}).

\section{A note on compactification}

String theory started out in twenty six spacetime dimensions.
Supersymmetry reduced that number to ten, but we still need to
address the question of where are the extra six spatial dimensions
that we do not observe in our (very) low energy universe. The main
dogma in string theory is that these dimensions are wrapped around
closed cycles of very small manifolds.

For example, one can imagine the simplest possible situation,
where only one extra dimension is wrapped over a circle. Let us
say we can do that for a direction $y$:
\begin{equation}\label{wrap}
    y  = y  + 2\pi R,
\end{equation}
with $R$ obviously being the radius of compactification. As
discovered by Kaluza and Klein almost ninety years ago, every time
a dimension is compactified a new field or more is generated in
the lower dimensional theory. It is as if the fields one observes
in the universe are nothing more than manifestations of the
geometrical and topological properties of the compact manifold
over which the dimensions are wrapped\footnote{This becomes
particularly believable when one observes that the starting theory
in eleven dimensions is purely gravitational with no matter
content. The 3-form gauge field can be thought of as a classical
limit to quantum processes, in the same way tensor and $p$-form
fields in string theory are averages over string states.}.

If we choose to wrap all the extra dimensions over circles, then
the manifold we have wrapped over is nothing more than the six
dimensional torus $T^6$. This is the simplest form of
compactification, more of which will be discussed in the next
section in the context of T-duality. One can make things even more
interesting by further complicating the topology. Six dimensions
allow for a lot of freedom of movement and one can `imagine'
creating a huge number of topologically inequivalent compact
manifolds. So, if nature has chosen to compactify, what type of
manifold has been picked, and why?

One way of answering this question, other than actually
discovering the compactification mechanism, is to note that
supersymmetry is either totally or partially broken by
compactifications\footnote{This is easily seen when noting that
every time a dimension is wrapped, a preferred spatial direction
is being chosen, breaking Poincar\'{e} symmetry. Since
Poincar\'{e} symmetry is a subgroup of supersymmetry, then so is
the later broken.}. Phenomenologically, it is hoped that some
supersymmetry would be preserved below the compactification energy
scale, since it helps in answering many questions in the standard
model; for example the hierarchy problem. Most researchers hope to
discover supersymmetry somewhere in the TeV regime. So one assumes
that whatever compactification mechanism has been chosen by string
theory, the compactification submanifold is of such nature as to
preserve some fraction of supersymmetry.

In order to do so, one needs to show that the compactification
submanifold admits covariantly constant spinors. A six dimensional
manifold that satisfies this property is one that
 admits $SU(3)$ holonomy. This means that a parallel-transported
 vector on a closed path on the manifold would return not to
 itself, but rather rotated by a matrix belonging to the group
 $SU(3)$. If the closed path is contractable to a
 point, this is called a \emph{restricted} holonomy. Such an
 object belongs to a class of complex manifolds known as Calabi-Yau
 manifolds. Appendix \ref{manifolds} contains a more detailed
 discussion of CY manifolds.

 It has also been shown that if one's starting point is not ten
 dimensional string theory, but eleven dimensional
 supergravity/M-theory, the seven dimensional compactification manifold must have restricted $G_2$ holonomy. In chapter (\ref{compact}), we actually give the details of the
 compactification process, $11\rightarrow 5$ over a CY, so
 further discussion of this issue is postponed till then.

\section{\label{Dualities}Dualities}

We have briefly mentioned that all string/SUGRA theories are not
different theories after all. Apparently, there exists a set of
duality transformations that connect the different theories
together, in such a way that they may be thought of as
descriptions of different limits of a single theory.

Some of these dualities connect the perturbative sector of one
string theory with the nonperturbative sector of another. Others
connect the perturbative sector of a theory with itself. This last
is the easiest to check, while the former remains to a certain
extent hypothetical with supporting evidence that does not
generally amount to a rigorous proof.

The simplest type of duality is known as T-duality, which is
manifest even in Bosonic string theory in 26 dimensions. Suppose
one wraps a single dimension of the theory around a circle, as in
(\ref{wrap}), i.e. parameterizing the theory using the metric
\begin{equation}\label{}
    ds^2  = G_{\mu \nu } dx^\mu  dx^\nu   + G_{25,25} \left( {dx^{25}
    + A_\mu  dx^\mu  } \right)^2,
\end{equation}
where $\left( {\mu  = 0, \ldots ,24} \right)$. From the point of
view of the lower dimensional theory, we have generated a scalar
field $\phi \sim G_{25,25}$ known as the dilaton, and a vector
field $A_\mu$. This is the famous Kaluza-Klein mechanism. One
consequence is that the string's momentum in the compactified
direction is quantized: $p_{25}={n \mathord{\left/ {\vphantom {n
R}} \right.
 \kern-\nulldelimiterspace} R}$, where $n$ is an integer, and its mass is given by
\begin{equation}\label{mass2}
    m^2  = \frac{{n^2 }}{{R^2 }}.
\end{equation}

In a theory of closed orientable strings, the string can also wind
over the compactified dimension. The number of times this happens
is clearly another integer $w$; the winding number. The momentum
in the compactified direction hence becomes
\begin{equation}
    p_L  = \frac{n}{R} + \frac{{wR}}{{\alpha '}}, \quad
    p_R  = \frac{n}{R} - \frac{{wR}}{{\alpha '}},
\end{equation}
for left and right windings respectively. The mass spectrum
formula (\ref{mass}) is modified as follows
\begin{equation}\label{wrappedspectrum}
    m^2  = \frac{{n^2 }}{{R^2 }} + \frac{{w^2 R^2 }}{{\alpha '^2 }} +
    \frac{2}{{\alpha '}}\left( {N + \tilde N - 2} \right),
\end{equation}
where the last term corresponds to the contribution to the mass
from oscillations in the large dimensions only. T-duality is the
observation that the spectrum of the theory is invariant under
\begin{equation}\label{}
    R \to \frac{{\alpha '}}{R},\quad n \leftrightarrow w,
\end{equation}
as can be readily seen by checking (\ref{wrappedspectrum}). In
other words, Bosonic string theory is T-dual to itself. In fact,
the $R\to 0$ and $R\to \infty$ limits are physically identical. We
reach the remarkable conclusion that trying to get rid of extra
dimensions by wrapping and contracting to zero does not work,
since the hidden dimensions will simply `pop' back into the
theory. These features are inherent to string theory and are
completely absent in a point particle theory. This discussion can
be extended to superstrings, where we find that type IIA string
theory is T-dual to type IIB, showing that the two theories are,
in a sense, equivalent.

 A closed string wound around the compactified dimension cannot
 then unwind without breaking and violating momentum
 conservation in that direction. What about open strings in the T-dual picture? What prevents an
open string from unwrapping? We find that in order to achieve
this, we must restrict the end points of the string to the
unwrapped dimensions. In other words, the end points must respect
Dirichlet boundary conditions in the compactified direction.
Remarkably, as a consequence of T-duality, one can show that
Neumann and Dirichlet conditions in the wrapped dimension are
T-dual to each other. So, even if we start with Neumann
conditions, we still end up with an open string with end points
fixed in the unwrapped directions. The surfaces upon which the end
points of open strings are constrained have come to be known as
D$p$-branes, where D stands for Dirichlet, and $p$ is the number
of spatial dimensions a brane spans. Historically, this was where
branes first made their debut in string theory.

We have discussed T-duality in some length because it is important
in many respects as we have seen. It is also the simplest. Another
example is S-duality \cite{Sen2} which relates the weak and strong
coupling sectors of either two theories or the same theory $\left(
{g_s \leftrightarrow {1 \mathord{\left/
 {\vphantom {1 {g_s }}} \right. \kern-\nulldelimiterspace} {g_s }}}
\right)$ by rotation under $SL(2,\mathbb{Z})$. The type I theory
is S-dual to the heterotic $SO(32)$ theory, whereas the type IIB
theory is S-dual to itself \cite{Sen}.

\section{M-theory}

In order to complete the picture, at this point we ask the
question: How does the postulated M-theory relate to string
theory? Also, how is it possible that it lives in eleven
dimensions although we have continuously claimed that superstring
theories are consistent only in ten? The answers to these
questions are related. A supersymmetric quantum theory of strings
is mathematically consistent in ten dimensions only, which means
that M-theory, whatever it is, is not a theory of strings. In
fact, as we mentioned earlier, the solitonic solutions in eleven
dimensional supergravity are a (super)membrane and a 5-brane. It
is conjectured that the M2-brane is what gives rise to the
fundamental string in ten dimensions upon wrapping over a circle.
M-theory is also assumed to be completely nonperturbative and that
its perturbative expansion is already known; type IIA string
theory \cite{Witten2}. One way of seeing this is as follows:

Type IIA string theory has a particle-like solitonic solution
known as the D0-brane. Its mass is given by the formula
\begin{equation}\label{}
    m_0  = \frac{1}{{g_s \sqrt {\alpha '} }},
\end{equation}
where $g_s$ is the string coupling constant related to the
strength of the dilaton as discussed earlier. Obviously the
D0-brane is heavy at weak coupling and becomes light at strong
coupling. For any number $n$ of D0-branes, the formula becomes
\begin{equation}\label{}
    M = nm_0  = \frac{n}{{g_s \sqrt {\alpha '} }}\quad  \to \quad M^2
    = \frac{{n^2 }}{{g^2 _s \alpha '}}.
\end{equation}

As the coupling grows stronger, the states become lighter and the
spectrum approaches a continuum. We immediately note that the
above formula is reminiscent of (\ref{mass2}), which arises from
toroidal compactification. This begs the following interpretation:
If we conjecture the existence of an eleventh dimension wrapped
over a circle of radius $R_{10}$, then the string coupling
constant is related to this radius by $R_{10} = g_s \sqrt {\alpha
'} $. The circle becomes larger as the coupling grows stronger. We
reach the remarkable conclusion that an eleven dimensional theory
is the strong coupling limit of weakly coupled type IIA string
theory.

Although M-theory is unknown, its low energy limit, $D=11$ SUGRA
can be used to demonstrate this property. Wrapping one spatial
dimension over a small circle immediately yields type IIA SUGRA.
It is also conjectured that wrapping the eleventh dimension over a
line segment ${{S^1 } \mathord{\left/ {\vphantom {{S^1 } {Z_2 }}}
\right. \kern-\nulldelimiterspace} {Z_2 }}$ gives the $E_8 \times
E_8$ heterotic theory \cite{HoravaWitten}. Other dualities exist,
such as between M-theory on ${{T^5 } \mathord{\left/
 {\vphantom {{T^5 } {Z{}_2}}} \right. \kern-\nulldelimiterspace}
 {Z{}_2}}$ and type IIB string theory on a $K3$ manifold.

Consequently, the solutions of the eleven dimensional theory are
responsible for the various objects that appear in ten dimensions.
So, as we remarked, the M2-brane becomes the fundamental string,
the D2-brane in ten dimensions is a reduced (as opposed to
wrapped) M2-brane, the 4-brane is a M5-brane with one wrapped
direction and the 5-brane (also known as the NS5-brane) is a
reduced M5-brane.

Some proposals as to the explicit form of M-theory have been
presented in the literature, most notable among those is Matrix
theory \cite{Matrix, Matrix2} which proposes that the
aforementioned D0-branes are the fundamental degrees of freedom of
M-theory, and a Hamiltonian describing the D0-branes in ten
dimensions can be rewritten with an eleven dimensional
interpretation such that
\begin{equation}\label{}
M_{11}  = \frac{1}{{g_s ^{{1 \mathord{\left/
 {\vphantom {1 3}} \right.
 \kern-\nulldelimiterspace} 3}} \sqrt {\alpha '} }}
\end{equation}
is the eleven dimensional Planck mass. The Hamiltonian is
dependent on a set of nine matrices $X$ which had the original
interpretation as the components of the nine dimensional position
vector of the D0-branes. We have $n$ D0-branes, so each of these
components become a $(n\times n)$ matrix instead of a scalar.
Aside from Matrix theory, this same behavior appears whenever
brane dynamics is studied. The physical meaning of this (scalar
$\to$ matrix) transformation is ambiguous. It introduces
noncommutative spacetime geometry in string theory; further
deepening the mystery. It may very well be a hint about the nature
of spacetime.

A final remark is that there exists another hypothetical theory
known as F-theory\footnote{Again no agreement exists on what the F
stands for. It could stand for `Further' or even `Father' of all
theories.} \cite{Vafa}. This is a theory in 12 spacetime
dimensions, two of which are time dimensions. It has been proposed
as a strong coupling limit of type IIB string theory in analogy to
M-theory. Whether or not such a theory exists or has any
consistent physical interpretations is yet to be seen.

\section{Various other issues}

We present a quick review of issues of interest in current
research (see \cite{Boer}).

\subsection{Brane world scenarios}

As an alternative to compactification, one may ask what happens if
all or some of the extra dimensions are left unwrapped
\cite{Hamed2,Hamed1,Hamed3,RandallSundrum1,RandallSundrum2}? The
fact that these large extra dimensions are not observed is
explained by assuming that our universe is basically a 3-brane
embedded in a higher dimensional space, and that open string
states (matter particles) are constrained to end on this brane.
Closed string states, however, including the graviton, can
propagate between branes/universes. It has been calculated that
the extra large dimensions transverse to our brane may be of sizes
up to 0.1 mm or so, much larger than the compactification scale,
which is of the order of the Planck length. Hence, the ten
dimensional string scale is much larger than the Planck scale and
strings may be observed in experiments in the not too-distant
future. The hope of experimental verification of string theory has
sparked a lot of interest in these scenarios. For a review, see
\cite{Dick} and the references within.

\subsection{Interest in manifolds with $G_2$ holonomy}

Analogously to string theory compactifications over a six
dimensional Calabi-Yau manifold, M-theory compactified over a
seven dimensional manifold down to four dimensions preserves some
fraction of supersymmetry if the manifold is one with restricted
holonomy belonging to the exceptional group $G_2$ (for a review,
see \cite{Duff}). This type of manifolds is much less understood
than Calabi-Yau's, and it is significantly harder to explicitly
find examples of compact $G_2$ manifolds. Most of the known
examples are non-compact. Upon compactification, we find that
particles can have arbitrarily small momenta in the non-compact
directions, which translates to a continuum of particles in four
dimensions; not a very realistic situation. It is found that
chiral fermions can only arise by compactifying the theory in the
neighborhood of singularities on the manifold, further
complicating the problem. Consequently, this direction of interest
is still in its infancy and there is much work yet to be done. The
first nontrivial examples of $G_2$ compactifications were found in
\cite{Duff3,Duff2,Duff4}.

More compactification scenarios exist, but they only get even more
mysterious. For example, the twelve dimensional theory known as
F-theory seems to require compactification over an eight
dimensional CY manifold. Very little is known about this process.

\subsection{Gauge theory - gravity correspondence}

In 1997, Maldacena conjectured \cite{Maldacena} that there exists
an exact equivalence between type IIB string theory on
AdS$_5\times S^5$ and four dimensional $\mathcal{N}=4$
supersymmetric Yang-Mills theory. The symmetry group of five
dimensional AdS space matches exactly the group of conformal
symmetries of the $\mathcal{N}=4$ theory, hence this
correspondence is also knows as AdS/CFT duality. Other similar
correspondences are conjectured to exist and the search for them
is an active direction of research.

Proving such dualities is hard, but indirect evidence can be found
by looking at the superstring's low energy limits, i.e.
supergravity. The corresponding limit on the CFT side is one where
$N$ and $\left({g_{YM}^2 N}\right)$ become large, where $N$ is the
rank of the $U(N)$ gauge group of the $\mathcal{N}=4$ theory, and
$g_{YM}$ is the Yang-Mills coupling constant. This is related to
't Hooft's original observation \cite{Hooft} that large $N$ gauge
theories are equivalent to a theory of strings.

A more general idea is that of holography \cite{Susskind,Hooft2}.
This is rooted in Hawking and Bekenstein's famous calculation
\cite{Bekenstein,Bekenstein2,Hawking} that the entropy of a black
hole is proportional to the area of its event horizon. So it is
conjectured that gravity in $D$ dimensions is equivalent to a
local field theory in $D-1$ dimensions. This is analogous to a
hologram, where all the information in a 3 dimensional picture is
stored in 2 dimensions; hence the name. Other examples of the
conjecture exist, for a recent review see \cite{Hoker}.

\section{Experimental string physics; an oxymoron?}

Although there is very little possibility of ever directly
observing strings in accelerator experiments, there is still some
hope in indirect verification. For example cosmological
observations, including more detailed measurements of the cosmic
background radiation and the possible detection of gravitational
radiation, may yield conclusions about the very early universe
which can only be explained by string theory. Black hole physics,
as we will discuss in some more detail in the next chapter,
certainly require a final union between quantum mechanics and
gravity. So better observations of actual black holes, like the
one presumably in the center of the Milky Way galaxy, or even the
detection and study of one closer to us, would provide even more
clues as to whether or not string theory describes this universe.

Supersymmetry is a generic prediction of string theory, so even
though it is possible, however unlikely\footnote{At least in this
author's view.}, for SUSY to exist without string theory, the
discovery of even one supersymmetric partner of a known subatomic
particle would lend great support to strings. In addition, the
picture would be complete if, at long last, we manage to exactly
derive the standard model, or a reasonable supersymmetric
extension thereof, from string theory.

Of course, if the brane world scenario is correct, then we may not
be too far off from an actual laboratory observation of strings.
In fact, the fundamental string scale may even be well within the
TeV region, making it accessible to the next generation of
particle accelerators.

So although a final experimental proof, to the complete
satisfaction of the physics community, is probably still decades
away (unless something revolutionary happens), there is still
solace in the hope that strings are not absolutely beyond the
human ability to detect. Even if string theory, or some variation
thereof, turned out not to be a correct theory of nature, the
contributions it has added to our mathematical knowledge are so
vast and so beautiful, that it is very hard to believe that they
will not constitute even a small part of the final theory.

\chapter{Branestorming}\label{Branes}

Historically, branes arose from the requirements of T-duality as
described in \S\ref{Dualities}. In 1995, signalling the beginning
of the so called second string revolution, J. Polchinski showed
that D$p$-branes couple to R-R $(p+1)$-form fields in very much
the same way a point particle couples to a gauge field in ordinary
field theory \cite{Pol3}. In string language, branes emit and
exchange closed strings in the R-R sector. Many of these branes
also arise as stable solitonic solutions to string/supergravity
theories, appearing as clear generalizations of black hole
solutions of ordinary four dimensional GR. This has led to a huge
body of work on classifying the properties of black-brane
spacetimes, for example \cite{Youm}. These solutions satisfy the
Bogomol'nyi-Prasad-Sommerfield (BPS) condition, which states that
the R-R charge density of the brane $Q_{RR}$ and the mass density
of the brane $M$ satisfy:
\begin{equation}\label{BPSCondition}
    M\geq Q_{RR}.
\end{equation}

So a BPS brane cannot decay below the saturation threshold of
(\ref{BPSCondition}) without violating charge conservation. Branes
are hence nonperturbative solutions of string theory; topological
defects that are generalizations of the more familiar Yang-Mills
magnetic monopoles known since 't Hooft and Polyakov's classic
papers \cite{Polyakov,Polyakov2,Hooft3} (for a review of BPS
solitons in YM theory see \cite{Emam}, and more generally
\cite{Riazi}). So even though we do not know how to solve string
theory nonperturbatively, the study of these known solutions,
whether from within string theory or supergravity, provide more
understanding of the spectrum of solutions of the full M-theory.
BPS branes also preserve some portion of supersymmetry, as we will
see.

\section{\label{braneproperties}The basic properties of branes}

One can summarize the dynamics of $p$-branes in $D$ dimensions by
writing down their worldvolume action:
\begin{eqnarray}
    S &=& \int {d^{p + 1} \xi \left[ {\sqrt {-\det  {g_{\mu \nu } } }  - \frac{\bar\varepsilon ^{\mu _1  \cdots \mu _{p + 1} }}{{\left( {p + 1} \right)!}} \left( {\partial _{\mu _1 } X^{M_1 } } \right) \cdots \left( {\partial _{\mu _{p + 1} } X^{M_{p + 1} } } \right)A_{M_1  \cdots M_{p + 1} } } \right]}  \nonumber\\
    M &=& 0, \ldots ,D - 1\quad \quad \mu  = 0, \ldots ,p \label{braneaction}
\end{eqnarray}
where $X^M$ are spacetime coordinates and $\xi_\mu$ are
worldvolume coordinates. The first part of the action is basically
a generalization of the Nambu-Goto string action (\ref{nambu}).
The second part shows the brane's coupling to a R-R $(p+1)$-form
$A$. So for example, the presence of a 3-form field $A$ in the
bosonic action of $D=11$ SUGRA (\ref{eleven}) immediately
indicates the presence of the M2-brane as a BPS solution to the
theory.

The conserved `electric' charge of the brane can be easily
calculated using Gauss' law. In the absence of sources $\left(
{d\star F_{p + 2} =0} \right)$, this can be written as follows
\begin{equation}\label{branecharge}
    Q_p  = \int\limits_{S^{D - p - 2} } {\star F_{p + 2} }.
\end{equation}

In addition, the superalgebra of the theory in question contains
central charge terms that are related to the brane charges. These
are the $Z_{\mu_1 \cdots \mu_{p}}$ forms in (\ref{SUSYalgebra}).
So the structure of the superalgebra indicates how many
fundamental SUSY-preserving branes exist. In the eleven
dimensional case, these are the M2 and M5-branes. The brane
charges $Q_p$ may in fact be identified as the magnitudes of the
forms $Z_{\mu_1 \cdots \mu_{p}}$.

Branes can also form bound states by the exchange of other branes.
They can intersect, overlap, wrap around manifolds and end on each
other. This last is interesting in that it is restricted by Gauss'
law and charge conservation \cite{Callan,Strominger3}. Consider,
as a simple example, an open 1-brane carrying a non-zero charge
$Q_1$ in ten dimensions via (\ref{branecharge}). Now surround the
string by a closed Gaussian `sphere', in this case $S^7$, in order
to calculate its charge. Since the location of this sphere is
arbitrary, one can imagine `sliding' it off the end of the string
and contracting it to a point, giving a charge of zero, which is a
contradiction. This leads to two conclusions; a charged closed
string cannot break without violating charge conservation, and
charged open strings must end on a brane of a particular dimension
that would obstruct the sphere's path and does not allow it to
contract to a point. Furthermore, the end points of the string
would describe charged 0-branes on the brane's worldvolume with
gauge fields restricted to live on the brane. So, the question of
what branes can end on what branes boils down to finding the
brane's worldvolume field content.

Another method, dubbed ``Brane Surgery'', was proposed in
\cite{Town2}. Consider, as an example, our particular case of
interest; $D=11$ SUGRA. The Bianchi identity $dF=0$ tells us right
away that the M5-brane is closed and hence cannot end on another
brane. The M2-brane's charge, on the other hand, receives a
contribution from the Chern-Simons term in the action via the
field equation $d\star F =  - F \wedge F$, or $d\left( {\star F +
F \wedge A} \right) = 0$. Its charge is hence
\begin{equation}\label{}
    Q_2  = \int\limits_{S^7 } {\left( {\star F + F \wedge A} \right)}.
\end{equation}

If we move our Gaussian sphere to the end of the brane, the first
term vanishes and the second term can be broken into two integrals
over $S^7=S^4\times S^3$:
\begin{equation}\label{}
    Q_2  = \int\limits_{S^4 } F  \wedge \int\limits_{S^3 } A.
\end{equation}

The first integral is the charge $Q_5$ of a 5-brane, so the
M2-brane can end on a M5-brane. If we choose $Q_5=1$ and assume
$F=dA=0$ within the brane, which is reasonable in the absence of
sources, then $A$ is a closed form which we may write locally as
$A=dC$, for a 2-form $C$. The charge becomes
\begin{equation}\label{}
    Q_2  = \int\limits_{S^3 } {dC_2 },
\end{equation}
which is the magnetic charge of a string living on the M5-brane's
worldvolume. In short, one can basically answer the question of
what branes can end on what branes by reading it off the
Chern-Simons term in the action.

\section{Branes and black holes}

Ever since the classic work by Hawking and Bekenstein
\cite{Bekenstein,Bekenstein2,Hawking2,Hawking,Hawking4,Hawking3},
the microscopic interpretation of black hole thermodynamics has
been a nagging problem. Initiated by \cite{Maldacena2,Vafa2}, it
has been shown that D-branes provide an interesting answer, at
least in certain special cases \cite{Das}.

The thermodynamics of black holes begins with Bekenstein's
solution to the entropy problem; that the absorption of matter
carrying a certain amount of entropy inside a black hole violates
the second law of thermodynamics, unless the black hole carries
entropy $S_{BH}$ that is directly proportional to the surface area
$A_{H}$ of the event horizon. The formula turns out to be:
\begin{equation}\label{entropyBH}
    S_{BH}  = \frac{{A_H }}{{4 \hbar G_N}};
\end{equation}
$G_N$ being Newton's gravitational constant. \emph{Gedanken}
experiments involving the entropy of regions of space containing a
black hole must involve this formula, as in, for example,
\cite{PelathWald}. Hawking derived, from within classical GR, the
second law of BH thermodynamics stating that the area of a black
hole (hence its entropy) cannot decrease. This led to the writing
of $S_{BH}$ in a Boltzmann-like formula
\begin{equation}\label{boltzmann}
    S_{BH}  \propto \ln \left( \Omega  \right);
\end{equation}
where $\Omega$ is the number of microstates of the system as
usual. The question of the nature of these microstates begs a
microscopic interpretation of the `inside' of a black hole, i.e.
what states are we counting exactly? The mystery deepens if one
calculates the actual number of states using (\ref{entropyBH}) and
(\ref{boltzmann}). For example, for a black hole with one solar
mass, $\Omega = 10^{10^{78} } $! The sheer magnitude of this
number cannot possibly be ignored.

Furthermore, since a BH has entropy, then its temperature must
follow the standard thermodynamic formula
\begin{equation}\label{}
    TdS = dE = dM.
\end{equation}

This implies that black holes must radiate as black bodies.
Classically, this is impossible, but in his famous calculation
\cite{Hawking2}, Hawking showed that the quantum behavior of
fields near the event horizon allows for such radiation and the
gradual loss of mass by the BH. Roughly speaking, vacuum
fluctuations near the horizon allow for pair creation of
particles, which in ordinary space would annihilate each other
very quickly, following Heisenberg's uncertainty principle. In the
extreme environment near a BH, however, one member of the pair can
fall inside the event horizon, while the other propagates away
from it, carrying with it some of the energy of the BH, hence the
radiation and loss of mass.

Hawking's calculation is completely insensitive to the precise
makeup of the matter dropping into the black hole, however the
slow evaporation process follows the same pattern for all black
holes. This implies a certain loss of information, since quantum
mechanics tells us that the final state of any time evolution is
related to the initial state via a unitary evolution operator. In
simple terms, whether one makes a black hole by dropping books,
vases or umbrellas, the modes of radiation that the BH emits will
be the same in all cases. This is known as the ``information
problem''.

All of these issues and questions have evolved from ordinary
classical general relativity and quantum field theory. But the
answers seem to lie only within a theory of quantum gravity, since
an understanding of the structure of a black hole involves very
high gravitational fields on a microscopic scale. String theory
provides such an answer, at least in the case of charged black
holes.

One would expect black holes to appear as solutions of string
theory as some sort of excited quantum states. In 1993, Susskind
\cite{Susskind2} proposed the following idea: Study a particular
quantum string/brane state at weak coupling, then imagine
increasing the coupling gradually (recall that the string coupling
is proportional to the dilaton field and hence variable). The
gravitational coupling also increases until such a time where the
object under study becomes a black hole with a large enough
radius. For this BH we constructed, one can count the microstates,
calculate the entropy, then compare with Bekenstein's formula
(\ref{entropyBH}).

This sort of calculation can only be performed for BPS states,
since (\ref{BPSCondition}) insures that the density of states
would stay the same as the coupling changes. Calculations such as
these have been performed for both extremal, i.e.
(\ref{BPSCondition}) is exactly saturated, and near-extremal black
holes. The first and simplest example of this procedure happened
to be a five dimensional black hole in type IIB string theory
\cite{Vafa2}. The entropy calculated from the counting of string
states accounted \emph{exactly} for the Bekenstein entropy.
Another case of interest is the situation where a group of
$D1$-branes join up to form a longer 1-brane \cite{Das2}.
$D1$-branes bound to $D5$-branes can also be made into black holes
with entropies matching the classical result exactly
\cite{Maldacena3}.

Despite the fact that the calculation requires charged black holes
and is yet to be solved for the case of neutral ones, this indeed
constitutes a huge triumph for string theory and an excellent
confirmation of its consistency with other, more established,
theories of nature.

\section{BPS black holes\label{BH}}

As remarked earlier, brane solutions of supergravity appear as
generalizations of black holes. To see this, we briefly review
solutions of Einstein-Maxwell theory representing static,
electromagnetically charged black holes, i.e. solutions of
\begin{equation}\label{}
S = \int {d^4 x\sqrt { - g} \left( {R - \frac{1}{4}F_{\mu \nu }
F^{\mu \nu } } \right)}.
\end{equation}

In spherical coordinates, the simplest black hole carrying mass
$M$ and an electric charge $Q$ is the Reissner-Nordstr\o m (NR)
solution
\begin{eqnarray}\label{RN}
    ds^2  =  - fdt^2  + f^{ - 1} dr^2  + r^2 d\Omega ^2,\quad A_t  = \frac{Q}{r}
\end{eqnarray}
where $f = \left( {1 - \frac{{2M}}{r} + \frac{{Q^2 }}{{r^2 }}}
\right)$, and $d\Omega ^2  = d\theta ^2  + \sin ^2 \theta d\varphi
^2 $ is the metric of the unit sphere. The case $Q=0$ reduces to
the ordinary Schwarzchild solution.

This BH solution has two event horizons, whose radii are satisfied
by the formula
\begin{equation}\label{}
    r_ \pm   = M \pm \sqrt {M^2  - Q^2 }.
\end{equation}

The two horizons obviously become one at $M^2  = Q^2$. This is the
extremal state of the black hole, since for any value $M^2<Q^2$
the horizon vanishes and the BH becomes a naked singularity. A
naked singularity seems to be an unlikely situation, since
realistic black holes are probably too massive for this to happen,
which is good as a naked singularity is not something that nature
seems to allow (the so called `cosmic censorship', or the
hypothesis that nature abhors a singularity). Interesting physics
arises when one tries to seek scenarios with vanishing event
horizons. For example, it turns out to be impossible to overcharge
a BH by throwing in particles that carry more charge than energy,
since the electrostatic repulsion is always strong enough to repel
the particles away.

The condition $M \geq \left| Q \right|$ that we arrive at with
this argument is exactly the BPS condition (\ref{BPSCondition}).
BPS solutions enjoy interesting properties and are of importance
to string theory for the reasons mentioned earlier in this
chapter.

 An important observation is that metrics of extremal black holes can always be written in terms of a harmonic warp function
 $H$. The RN solution may be written in this form, and, in fact,
 may be generalized to multi-black holes as follows:
\begin{eqnarray}
 ds^2  &=&  - H^{ - 2} dt^2 + H^2 d\vec x  \cdot d\vec x  \nonumber\\
H &=& 1 + \sum\limits_{I = 1}^k {\frac{M_I}{{r_I}}} \quad \quad
r_I    = \left| {\vec x - \vec x_I } \right|;\quad \nabla ^2 H = 0 \nonumber\\
 A_t  &=& H^{-1}.
\end{eqnarray}

This is known as the Majumdar-Papapetrou (MP) result
\cite{MP1,MP2}, where we have redefined the coordinates by $r \to
r + M$ from (\ref{RN}). The solution represents a number $k$
extremal black holes located at $\vec x_I$ that are static with
respect to each other, i.e. gravitational attraction exactly
cancels out electrostatic repulsion for any configuration (the `no
force' condition). The BPS inequality (\ref{BPSCondition}) is
saturated by $M=\left| Q \right|$. It turns out that the harmonic
condition on $H$ is a common feature of BPS solutions. As we will
see, BPS multi-brane solutions have the same structure.

An interesting point is that the `no force' situation arises in
Newtonian physics also, as a consequence of the inverse $r^2$
behavior of both Newtonian gravity and the Coulomb field. One
finds that the force cancellations arise irrespective of the
distribution of the charged masses. Remarkably, this continues in
general relativity.

Finally, we note that BPS solutions in supersymmetric theories
preserve one half of SUSY \cite{GH}. The transition from four
dimensional BPS black holes to multidimensional BPS black branes
should now be obvious.

\section{Constructing brane solutions}

We now specialize to reviewing particular M-brane solutions in
some detail, leading up to our own work in the following
chapter\footnote{A particularly useful review is \cite{Stelle}.}.
Writing down brane solutions to (\ref{eleven}) involves solving
the Einstein equation and the equation of motion of the field
strength.

However, it was shown that a solution satisfying the vanishing of
gravitino variations, i.e. in $D=11$:
\begin{equation}\label{eleventrans2}
    \delta_{\epsilon} \psi _M   = 2\nabla _M  \epsilon  + \frac{i}{{144}}F_{LNPQ } \left( {\Gamma _M  ^{\;\;\;LNPQ }  - 8\delta _M ^L  \Gamma ^{NPQ } }
    \right)\epsilon=0,
\end{equation}
also automatically satisfies Einstein's equation. This is true in
all supergravity theories \cite{Kaya}. The other SUSY variations,
for example (\ref{dele}) and (\ref{dela}), represent variations of
the bosonic fields and these are proportional to the fermions. In
purely bosonic vacua, of interest to us, they are automatically
satisfied.

Since solving a first order differential equation is easier than
solving Einstein's second order equation, we will be using
(\ref{eleventrans2}) instead of $G_{MN}=8 \pi T_{MN}$ in most of
our calculations. In addition, one also gets the projection
conditions that the Killing spinor $\epsilon$ satisfies.

The basic BPS solutions to $D=11$ supergravity are the electric
M2-brane and its magnetically dual M5-brane
\cite{Duff8,Duff7,Duff9,Guven}; respectively:
\begin{eqnarray}{}
    ds_{11}^2 &=& H^{ - \frac{2}{3}} \eta _{ab} dx^a dx^b  + H^{\frac{1}{3}} \delta _{\mu \nu } dx^\mu  dx^\nu,\quad \quad\psi=0 \nonumber\\
    A_{012} &=& \pm H^{-1}, \quad
    a,b = 0,1,2\quad \mu ,\nu  = 3, \ldots ,10
\end{eqnarray}
and
\begin{eqnarray}{}
    ds_{11}^2  &=& H^{ - \frac{1}{3}} \eta _{ab} dx^a dx^b  + H^{\frac{2}{3}} \delta _{\mu \nu } dx^\mu  dx^\nu, \quad\quad \psi=0\nonumber  \\
    A_{012345} &=& \pm H^{-1},\quad a,b = 0, \ldots ,5\quad \mu ,\nu  = 6, \ldots ,10
\end{eqnarray}

The equation of motion of the field strength and the Bianchi
identity show that $H$ is a harmonic function in the coordinates
transverse to the brane's world volume, i.e.
\begin{equation}\label{}
    \nabla ^2 H=\delta ^{\mu \nu } \left( {\partial _\mu  \partial _\nu  H}
    \right) = 0,
\end{equation}
implying the following structure for $H$:
\begin{equation}\label{}
    H = 1 + \frac{M}{{r^m }}\quad \quad r = \left| {\vec x - \vec x_0 }
    \right|;
\end{equation}
$\vec x_0$ being the position vector of the brane in the
transverse space and $M$ its mass density. The power $m$ is equal
to $(D-p-3)$, where $p$ is the dimension of the brane. More
generally, for a set of $k$ parallel branes this becomes:
\begin{equation}\label{}
    H = 1 + \sum\limits_{I = 1}^k {\frac{M_I}{{r_I^m }}} \quad \quad r_I
    = \left| {\vec x - \vec x_I } \right|.
\end{equation}

The gravitino equation also tells us that the supercovariant
Killing spinor can be written in terms of a constant spinor
$\epsilon_0$:
\begin{eqnarray}{}
    {\rm for\;the\;M2-brane:}\quad \quad  \epsilon  &=& H^{-
{1 \mathord{\left/
 {\vphantom {1 6}} \right.
 \kern-\nulldelimiterspace} 6}} \epsilon _0  \nonumber\\
   {\rm for\;the\;M5 -
    brane:}\quad \quad \epsilon  &=& H^{- {1 \mathord{\left/
 {\vphantom {1 12}} \right.
 \kern-\nulldelimiterspace} 12}} \epsilon _0,
\end{eqnarray}
satisfying the projection conditions:
\begin{eqnarray}{}
    \hat\Gamma _{012} \epsilon  &=& \pm \epsilon \quad \quad {\rm for\;the\;M2 - brane} \nonumber\\
    \hat\Gamma _{012345} \epsilon  &=& \pm \epsilon \quad \quad {\rm for\;the\;M5 -
    brane},
\end{eqnarray}
preserving one half of the original supersymmetry. The
$\hat\Gamma$'s are antisymmetrized products of Dirac matrices in
an orthonormal frame.

\subsection{Delocalized intersections}

Now, one may construct new solutions by a variety of means. Two
intersecting M2-branes, for example, have the following form
\cite{Kastor2,Tse}:
\begin{eqnarray}{}
    ds^2  &=&  - \left( {H_1 H_2 } \right)^{ - {2 \mathord{\left/
    {\vphantom {2 3}} \right.
    \kern-\nulldelimiterspace} 3}} dt^2  + H_1^{ - {2 \mathord{\left/
    {\vphantom {2 3}} \right.
    \kern-\nulldelimiterspace} 3}} H_2^{{1 \mathord{\left/
    {\vphantom {1 3}} \right.
    \kern-\nulldelimiterspace} 3}} \left( {dx_1^2  + dx_2^2 } \right)
    + H_1^{{1 \mathord{\left/{\vphantom {1 3}} \right.
    \kern-\nulldelimiterspace} 3}} H_2^{ - {2 \mathord{\left/
    {\vphantom {2 3}} \right.
    \kern-\nulldelimiterspace} 3}} \left( {dx_3^2  + dx_4^2 } \right) \nonumber\\
     & &+ \left( {H_1 H_2 } \right)^{{1 \mathord{\left/
    {\vphantom {1 3}} \right.
    \kern-\nulldelimiterspace} 3}} \delta _{\mu \nu } dx^\mu  dx^\nu  \quad \quad \mu ,\nu  = 5, \ldots ,10 \nonumber\\
    F_{t12\mu }  &=& \pm H_1^{ - 2} \left( {\partial _\mu  H_1 } \right),\quad \quad F_{t34\mu }  = \pm H_2^{ - 2} \left( {\partial _\mu  H_2 }
    \right),
\end{eqnarray}
where the functions $H_i$ are harmonic in the overall transverse
directions $x^\mu$. Obviously, the directions $(x^1,x^2)$ define
the first membrane while $(x^3,x^4)$ define the second. The
intersection is on a point, delocalized in space; one says that
the intersection is `smeared' over the directions of the
membranes. The brane spanning $(x^1,x^2)$ is smeared over
$(x^3,x^4)$ and vice-versa. The harmonic nature of the warp
factors $H_i$ is a manifestation of this smearing. We will see
later that similar factors in localized solutions are no longer
necessarily harmonic. The Killing spinor satisfies:
\begin{eqnarray}{}
    \epsilon  &=& \left( {H_1 H_2 } \right)^{ - {1 \mathord{\left/
    {\vphantom {1 6}} \right.
    \kern-\nulldelimiterspace} 6}} \epsilon _0  \nonumber\\
    \hat\Gamma _{012} \epsilon  &=& \pm \epsilon, \quad
    \hat\Gamma _{034} \epsilon  = \pm \epsilon.
\end{eqnarray}
where the projections commute with each other. The pattern should
be obvious, and the most general solution for a number $N=1,2,3,4$
intersecting membranes can be written as:
\begin{eqnarray}{}
    ds^2  &=&  - \prod\limits_{i = 1}^N {H_i^{ - {2 \mathord{\left/
    {\vphantom {2 3}} \right.
    \kern-\nulldelimiterspace} 3}} } dt^2  + \sum\limits_{i = 1}^N
    {\left[ {H_i^{ - {2 \mathord{\left/
    {\vphantom {2 3}} \right.
    \kern-\nulldelimiterspace} 3}} \prod\limits_{j \ne i} {H_j^{{1
    \mathord{\left/
    {\vphantom {1 3}} \right.
    \kern-\nulldelimiterspace} 3}} } } \right]} \left( {dx_{2i - 1}^2  + dx_{2i}^2 } \right) \nonumber\\
    & &+ \prod\limits_{i = 1}^N {H_i^{{1 \mathord{\left/ {\vphantom {1 3}} \right.
    \kern-\nulldelimiterspace} 3}} } \delta _{\mu \nu } dx^\mu  dx^\nu \quad \mu,\nu  = 2N + 1, \ldots
    ,10\nonumber\\
    F_{t12\mu }  &=& \pm H_1^{ - 2} \left( {\partial _\mu  H_1 } \right),\;\;  \cdots \;\; F_{t,2N - 1,2N,\mu }  = \pm H_N^{ - 2} \left( {\partial _\mu  H_N } \right).
\end{eqnarray}

The solutions preserve $2^{-N}$ of the original supersymmetry,
with Killing spinors of the form $\epsilon  = \prod\nolimits_{i =
1}^N {H_i^{ - {1 \mathord{\left/
 {\vphantom {1 6}} \right.
 \kern-\nulldelimiterspace} 6}} } \epsilon _0 $. Four intersecting membranes of this type fill out all space and hence there can be no $N>4$ solutions.

Following this procedure, one can construct more delocalized
M-brane intersections. The rule is that for self intersecting
$p$-branes to be supersymmetric, they must intersect over
$(p-2)$-branes, such that no spatial direction is shared by any
pair of branes \cite{PapaTown2, Tse2}. For example, M5-branes
intersect over 3-branes \cite{Kastor2}. Dimensional reduction of
these solutions results in electrically and magnetically charged
black hole solutions in four dimensions and so on.

\subsection{Localized intersections and Fayyazuddin-Smith spacetimes}

The solutions of the previous section are delocalized in space,
i.e. the intersection points/strings/branes are smeared out over
the directions of the intersecting branes and cannot be
pinpointed. One may ask how to construct localized brane
intersections, where instead of intersecting on sharply defined
surfaces, the branes are smeared into each other over a smooth
manifold localized in space. The branes are obviously no longer
`flat' everywhere, but rather follow a smooth geometry where the
distinction between them, close to the point of localization,
becomes blurred. This fuzziness also causes the failure of the
warp factors to satisfy the transverse Laplace equation. In
\cite{FS}, Fayyazuddin and Smith (henceforth referred to as FS)
found the following solution representing completely localized
M5-brane intersections:
\begin{eqnarray}{}
    ds^2  &=& H^{ - \frac{1}{3}} \eta _{ab} dx^a dx^b  + 2H^{ - \frac{1}{3}} g_{m\bar n} dz^m dz^{\bar n}  + H^{\frac{2}{3}} \delta _{\mu \nu } dx^\mu  dx^\nu
    \nonumber\\
    H &=& 4g = 4\left( {g_{1\bar 1} g_{2\bar 2}  - g_{1\bar 2} g_{2\bar 1} }
    \right), \quad a,b=0,1,2,3,\quad \mu,\nu=1,2,3\nonumber \\
    F_{m\bar n\mu \nu }  &=& \pm\frac{{i}}{2}\bar \varepsilon _{\mu \nu \rho } \left( {\partial ^\rho  g_{m\bar n} }
    \right), \quad
    F_{m89,10}  =  \mp \frac{{i}}{2}\left( {\partial _m H} \right), \quad  F_{\bar m89,10}  =  \mp
    \frac{{i}}{2}\left( {\partial _{\bar m} H}
    \right) \nonumber\\
    {\rm where}\quad\quad  z^1  &=& x^4  + ix^5, \quad\quad z^2  = x^6  +
    ix^7.\label{complexcoords}
\end{eqnarray}

Both $H$ and the Hermitian metric $g_{m\bar n}$ depend on the
overall transverse coordinates $x^\mu$ and ($z^m,z^{\bar m}$). In
fact, SUSY constrains $g_{m\bar n}$ to be K\"{a}hler (see appendix
\ref{manifolds}). The field equations are:
\begin{eqnarray}{}
    d\star F &=& 0, \quad
    dF = J_{m\bar n} dz^m  \wedge dz^{\bar n}  \wedge dx^8  \wedge dx^9  \wedge dx^{10}  \nonumber\\
    J_{m\bar n}  &=& \pm\frac{{i}}{2}\left[ {4\left( {\partial _m \partial _{\bar n} g} \right) + \left( {\partial _\mu  \partial ^\mu  g_{m\bar n} } \right)}
    \right]; \quad
    g=\left| {g_{m\bar n} } \right| \nonumber \\ \partial _\mu  \ln {g \mathord{\left/
 {\vphantom {g H}} \right.
 \kern-\nulldelimiterspace} H} &=&0, \quad\quad \partial _m \partial _{\bar n} \ln {g \mathord{\left/
 {\vphantom {g H}} \right.
 \kern-\nulldelimiterspace} H} = 0.
\end{eqnarray}

This localized solution reduces to a delocalized one by $g_{m\bar
n}  \to \delta _{m\bar n} $. For a source-free solution, $J_{m\bar
n}=0$ becomes a nonlinear equation for the K\"{a}hler metric.

The FS solutions were extended to describe intersecting M2-branes
\cite{Kastor3}:
\begin{eqnarray}{}
    ds^2  &=&  - H^{ - \frac{2}{3}} dt^2  + 2H^{ - \frac{2}{3}} g_{m\bar n} dz^m dz^{\bar n}  + H^{\frac{1}{3}} \delta _{\mu \nu } dx^\mu  dx^\nu   \nonumber\\
    H &=& 4g = 4\left( {g_{1\bar 1} g_{2\bar 2}  - g_{1\bar 2} g_{2\bar 1} } \right),\quad A_{0m\bar n}  = \pm\frac{{i}}{2}H^{ - 1} g_{m\bar n}  \nonumber\\
    z^1  &=& x^1  + ix^2 ,\quad z^2  = x^3  + ix^4,\quad \mu ,\nu  = 5, \ldots
    ,10.
\end{eqnarray}

The Killing spinor satisfies the projection condition
\begin{equation}\label{projFS}
    \Gamma _{0m\bar n} \epsilon  = \pm\frac{{i}}{2}H^{ - 1} g_{m\bar n}
    \epsilon.
\end{equation}

The significance of (\ref{projFS}) and similar conditions may be
understood by noting the following. Generally, a given $p$-brane
with worldvolume spanning $x_{\mu _1 }  \cdots x_{\mu _p } $
preserves an amount of SUSY given by the number of spinors which
satisfy the equation \cite{Gutowski3}
\begin{equation}\label{SUSY666}
\epsilon  = \frac{1}{{p!}}\varepsilon ^{a_1  \cdots a_p } \Gamma
_{\mu _1  \cdots \mu _p } \left( {\partial _{a_1 } x^{\mu _1 } }
\right) \cdots \left( {\partial _{a_p } x^{\mu _p } }
\right)\epsilon,
\end{equation}
where the $a_i$ are brane worldvolume coordinates. This reduces to
the projection conditions satisfied by the spinor, such as
(\ref{projFS}).

\chapter{Calibrations and FS spacetimes}\label{FS}

The mathematical theory of calibrations developed by Harvey and
Lawson \cite{Harvey} has found extensive application in brane
physics. It has been used to classify all sorts of brane
configurations, for example brane intersections were treated in
\cite{Acharya,Acharya2,Callan,Figueroa,Gauntlett,Papa}, brane
Calabi-Yau wrappings \cite{Becker,BeckerStrominger,Husain},
worldvolume solitons in AdS compactifications \cite{Gutowski} and
more. The calibrated branes in these works have been treated as
test objects only. We note, however, that BPS branes satisfy `no
force' conditions that allow static multi-brane spacetimes
(similar to the MP configurations of \S\ref{BH}). In particular,
we should be able to add static test branes of the type found in
the above references to the solution. This implies that BPS brane
spacetimes should have (generalized) calibrations, as we will see.
In our paper \cite{Emam2}, we took this as an organizing principle
to finding new BPS solutions.

Based on this argument, we propose, and demonstrate, that it is
possible to use calibration techniques to construct charged,
gravitating brane solutions of supergravity. We show that the
class of solutions discovered by Fayyazuddin and Smith admits
calibrations and we propose generalizing the solutions to include
more types of calibrations.

This chapter is structured as follows: In section
\ref{Calibratedmanifolds} we review the calibration technique and
the generalized calibrations we will be using to construct
solutions. In \S\ref{Kahlersolution} we present our solutions
corresponding to M2 and M5-brane intersections over K\"{a}hler
calibrated manifolds. In \S\ref{SpecialSLAG} we discuss a solution
representing M5-branes intersecting over a SLAG calibrated
submanifold.

\section{\label{Calibratedmanifolds}Calibrated manifolds}

The theory of calibrations is a technique used by mathematicians
to find solutions belonging to a certain class of minimal
surfaces. Its usefulness to physics lies in the fact that this
very class admits constant spinors, which means that calibrated
branes preserve some portion of supersymmetry. This is manifest in
the fact that for a calibrated brane, (\ref{SUSY666}) is satisfied
at each point on the surface \cite{Gibbons2}.

We will review the technique in the language of physics rather
than follow Harvey and Lawson's original reasoning. Consider the
action of a neutral static $p$-brane in flat space, i.e. the first
term of (\ref{braneaction}):
\begin{eqnarray}
    S &=& \int {d^{p + 1} \xi \sqrt {-\det  {g_{\mu \nu } } }  }  \nonumber\\
    \mu  &=& 0, \ldots ,p
\end{eqnarray}
with the definitions given earlier in \S\ref{braneproperties}.
Mathematically, this is just the volume of a static
$p$-dimensional manifold, so minimizing the action to find a brane
solution is simply minimizing the volume of the manifold, i.e.
finding a minimal surface.

A $p$-form $\phi$ is called a calibration if:

\begin{enumerate}
    \item It is closed:
            \begin{equation}\label{closeness}
            d\phi  = 0.
            \end{equation}
    \item Its pullback onto any $p$-dimensional
    submanifold $\Sigma$ satisfies the following relation with the
    induced volume on the submanifold:
            \begin{equation}\label{pullback}
            * \phi  \le \varepsilon _\Sigma.
            \end{equation}
\end{enumerate}

If the inequality (\ref{pullback}) is saturated everywhere on
$\Sigma$ then the submanifold minimizes its volume. Pick a closed
$(p-1)$-dimensional $S$ in $\Sigma$, and within $S$ deform
$\Sigma$ into a new submanifold ${\Sigma '}$. Then the following
shows that $Vol\left( \Sigma  \right) \le Vol\left( {\Sigma '}
\right)$:
\begin{equation}\label{calibproof}
    Vol\left( \Sigma  \right) = \int\limits_\Sigma  {\varepsilon
    _\Sigma  }  = \int\limits_\Sigma  { * \phi }  =
    \int\limits_{\Sigma '} { * \phi }  + \int\limits_B {d\phi }  =
    \int\limits_{\Sigma '} { * \phi }  \le \int\limits_{\Sigma '}
    {\varepsilon _{\Sigma '} }  = Vol\left( {\Sigma '} \right),
\end{equation}
where $B$ is the $p$-dimensional region bounded by $\Sigma$ and
$\Sigma '$.

Now, this works well for branes with zero flux, but our interest
lies in BPS branes coupled to $(p+1)$-form fields whose motion
minimizes the full action (\ref{braneaction}). It turns out that
there is a way of relaxing condition (\ref{closeness}) and still
finding a calibration. Basically, instead of minimizing the volume
of the brane, we look for a solution that minimizes the energy
density of the brane, including the contribution from the gauge
field. For a brane coupled to a spacetime gauge field $A$, one may
identify $A \sim \phi$, hence $F=dA=d\phi \neq 0$. This is the so
called `generalized calibration' technique. The new condition can
be shown to yield a minimal surface that satisfies
(\ref{braneaction}) as follows \cite{Gutowski2}: The energy of a
static compact $p$-fold $\Sigma$ coupled to a gauge potential $A$
is
\begin{equation}\label{}
E\left( {\Sigma  } \right) = \int\limits_{\Sigma  } {\left(
{\varepsilon_\Sigma + A } \right)}.
\end{equation}

Now assume that $\Sigma$ is a calibrated manifold and as such
saturates
\begin{equation}\label{}
\int\limits_{\Sigma } {* \phi  }  = Vol\left( {\Sigma  } \right),
\end{equation}
as before. Once again, pick a closed $(p-1)$-dimensional $S$ in
$\Sigma$, and within $S$ deform $\Sigma$ into a new submanifold
${\Sigma '}$, hence
\begin{equation}\label{}
\int\limits_{\Sigma '} * \phi   = \int\limits_\Sigma  * \phi   +
\int\limits_{B} d\phi   = \int\limits_\Sigma * \phi   +
\int\limits_{B} {dA}  = \int\limits_\Sigma \phi +
\int\limits_{\partial B } A  = \int\limits_\Sigma \phi +
\int\limits_\Sigma  A  - \int\limits_{\Sigma '} A,
\end{equation}
demonstrating that
\begin{equation}\label{eee}
\int\limits_{\Sigma '} {\left( {* \phi  + A} \right)}  =
\int\limits_\Sigma  {\left( {* \phi  + A} \right)}.
\end{equation}

Then
\begin{equation}\label{}
\int\limits_\Sigma  {\left( {* \phi  + A} \right)}  = E\left(
\Sigma \right),
\end{equation}
reducing (\ref{eee}) to $E\left( \Sigma  \right) = E\left( {\Sigma
'} \right)$, hence a calibrated manifold minimizes its energy.

Calibrations are classified into three categories:
\begin{enumerate}
    \item K\"{a}hler calibrations: Defining complex coordinates
    $z$ in a $2n$-dimensional space similarly to (\ref{complexcoords}), the K\"{a}hler form is given
    by $\omega  = \frac{i}{2}\sum\limits_i {dz^i  \wedge d\bar z^i }
    $. The calibration $p$-form is hence $\phi  = \frac{1}{{p!}}\omega ^p
    $. The K\"{a}hler form $\omega$, and therefor $\phi$, are
    invariant under $U(n)$ rotations.
    \item Special Lagrangian (SLAG) calibrations: Defining the
    $p$-form $\psi  = dz^1  \wedge  \cdots  \wedge dz^p $, the
    calibration form will be either $\phi=Re \left( \psi  \right)$ or $\phi=Im \left( \psi  \right)$. In
    this case $\phi$ is invariant under $SU(n)$.
    \item The exceptional cases: So called because their group
    invariances follow exceptional Lie algebras. There are three of
    those:
    \begin{enumerate}
        \item The $G_2$ calibrations: a 3-form in seven
        dimensional space known as the `associative' form, invariant under the exceptional group $G_2$.
        \item The $Spin(7)$ calibration: The 4-form dual to the
        associative form, known as the `coassociative' form.
        \item The Cayley calibration: a self-dual 4-form in eight
        dimensional space.
    \end{enumerate}
\end{enumerate}

One notes that the manifolds with restricted holonomies catalogued
in Berger's list (\ref{globalissues}) admit calibrated forms. An
excellent discussion of this and related issues can be found in
\cite{Joyce}. The conditions that a particular form $\phi$ on a
given manifold $\mathcal{M}$ is a calibration can be shown to be
equivalent to the conditions that $\mathcal{M}$ admits a constant
spinor field at every point $x\in \mathcal{M}$ (see
\cite{Gutowski3} and the references within).

\section{\label{Kahlersolution}K\"{a}hler calibrated M-branes}

We now present our partially published results \cite{Emam2}. We
find solutions of $\mathcal{N}=1$ $D=11$ supergravity representing
localized intersecting M-brane configurations using the
calibration technique and show that they are of the FS type. There
are two ways of interpreting these solutions; either as
intersecting branes or as a single brane wrapping calibrated
submanifolds.

The most general FS-type metric ansatz in eleven dimensions is:
\begin{equation}\label{metric}
    ds^2  = G_{MN} dx^M dx^N  = H^{ - 2A} \eta _{ab} dx^a dx^b  + 2H^{ - 2B} g_{m\bar
    n} dz^m dz^{\bar n}  + H^{2C} \delta _{\alpha \beta } dx^\alpha
    dx^\beta.
\end{equation}

The range over which the indices run defines what kind of
intersection we have, in addition to defining the gauge potential
using different K\"{a}hler calibrated forms as was discussed in
the last section. In terms of the K\"{a}hler $(1,1)$-form
$\omega$, the calibration $p$-form $\phi$ can be constructed in
exactly two possible ways that will have an interpretation as
intersecting M-branes. We go through these in turn.

\subsection{$\phi\sim\omega^2$}

The first possibility would be to construct $\phi  =
\frac{1}{2}\omega  \wedge \omega $, with $\omega  = ig_{m\bar n}
dz^m  \wedge dz^{\bar n}$. Since this requires the spatial
dimension of the brane to be at least four, then in the context of
M-theory we will be looking at M5-brane intersections. The gauge
potential is proportional to $\phi$ and hence may be written as
follows\footnote{Note that here we are using the dual six-form
potential $A$ that couples to the M5-brane.}:
\begin{eqnarray}
    A_{abm\bar nr\bar s} &=& \pm H^{ - 1} \bar\varepsilon _{ab} \left( {g_{m\bar
    n}g_{r\bar s}  - g_{m\bar s} g_{r\bar n} } \right)\nonumber\\
    &=& \pm\frac{1}{(N-2)!} H^{ - 1} \bar\varepsilon _{ab} \bar\varepsilon _{\bar n\bar s}
    \bar\varepsilon ^{\bar u\bar v} g_{m\bar u} g_{r\bar v}.
\end{eqnarray}

In the metric (\ref{metric}), we now have $(a,b)$ describing the
worldsheet of a 1-brane, $(m,\bar n=1,\ldots N)$, where
$(N=2,3,4)$ and $(\alpha,\beta=2N+2,\ldots 10)$. The vanishing of
the gravitino variation (\ref{eleventrans}) in dual form is
\begin{equation}\label{eleventransKahler}
    \nabla_A \epsilon  + \left(
    {\frac{1}{{8640}}} \right)F_{AB_1  \cdots B_6 } \Gamma ^{B_1
    \cdots B_6 } \epsilon  - \left( {\frac{1}{{30240}}}
    \right)F_{B_1  \cdots B_7 } \Gamma _A ^{\;\;B_1  \cdots B_7 }
    \epsilon=0,
\end{equation}
where the spinor projections on the compact space are defined by
\begin{eqnarray}
    \Gamma ^{\bar m} \epsilon _ -   &=& \Gamma ^m \epsilon _ +   = 0 \nonumber\\
    \Gamma ^{m\bar n} \epsilon _ \pm   &=&  b G^{m\bar n}
\epsilon _
 \pm,\quad\quad b=\pm 1.
\end{eqnarray}

The nonvanishing components\footnote{For $N=2$, only the third
component exists.} of the field strength are
\begin{eqnarray}
    F_{pabm\bar nr\bar s}  &=&  \mp H^{ - 1} \bar \varepsilon _{ab} \bar \varepsilon _{pmr} \bar \varepsilon ^{qlk} (\partial _q \ln H)g_{{\mathop{\rm l}\nolimits} \bar n} g_{k\bar s}  \nonumber\\
    F_{\bar pabm\bar nr\bar s}  &=&  \mp H^{ - 1} \bar \varepsilon _{ab}
\bar \varepsilon _{\bar n\bar s\bar p} \bar \varepsilon ^{\bar
l\bar k\bar j} (\partial _{\bar l} \ln H)g_{m\bar k} g_{r\bar
j}\nonumber\\
F_{\alpha abm\bar nr\bar s}  &=&  \mp H^{ - 1}\bar \varepsilon
_{ab} \bar \varepsilon _{\bar n\bar s} \bar \varepsilon ^{\bar
u\bar v} \left[ { \left( {\partial _\alpha  \ln H} \right)g_{m\bar
u}g_{r\bar v} -
\partial _\alpha  \left( {g_{m\bar u} g_{r\bar v} } \right)}
\right].
\end{eqnarray}

The supercovariant derivative (\ref{covderivative}) is now
\begin{eqnarray}
 \nabla _a \epsilon _ \pm   &=&  - \frac{A}{2}\left( {\partial _{\bar m} \ln H} \right)\Gamma _a ^{\;\;\bar m} \epsilon _ +   - \frac{A}{2}\left( {\partial _m \ln H} \right)\Gamma _a ^{\;\;m} \epsilon _ -   - \frac{A}{2}\left( {\partial _\alpha  \ln H} \right)\Gamma _a ^{\;\;\alpha}  \epsilon _ \pm   \nonumber\\
 \nabla _m \epsilon _ \pm   &=& \left( {\partial _m \epsilon _ \pm  } \right) - \frac{b}{2}\left( {\partial _m \ln \bar E} \right)\epsilon _ \pm   + \frac{{b B}}{2}\left( {\partial _m \ln H} \right)\epsilon _ \pm   - \frac{B}{2}\left( {\partial _{\bar n} \ln H} \right)\Gamma _m ^{\;\;\bar n} \epsilon _ +   \nonumber\\ & & - \frac{1}{4}H^{ - 2B} \left( {\partial _\alpha  g_{p\bar n} } \right)\Gamma ^{\alpha \bar n} \epsilon _ +   - \frac{B}{2}\left( {\partial _\alpha  \ln H} \right)\Gamma _m ^{\;\;\alpha}  \epsilon _ +  \quad\quad {\rm
 and \;\;c.c.} \nonumber\\
 \nabla _\alpha  \epsilon _ \pm   &=& \left( {\partial _\alpha  \epsilon _ \pm  } \right) - \frac{b}{2}\left( {\partial _\alpha  \ln \bar E} \right)\epsilon _ \pm   + \frac{b}{4}\left( {\partial _m \ln g} \right)\epsilon _ \pm   + \frac{C}{2}\left( {\partial _{\bar n} \ln H} \right)\Gamma _\alpha  ^{\;\;\bar n} \epsilon _ + \nonumber\\& &  + \frac{C}{2}\left( {\partial _n \ln H} \right)\Gamma _\alpha  ^{\;\;n} \epsilon _ -   + \frac{C}{2}\left( {\partial _\beta  \ln H} \right)\Gamma _\alpha  ^{\;\;\beta}  \epsilon _
 \pm,
\end{eqnarray}
where, for $N=3$:
\begin{eqnarray}
        E &=& \frac{1}{3!}\bar\varepsilon ^{mnp} \bar\varepsilon _{\hat r\hat s\hat q} E_{\;\;m}^{\hat r}
        E_{\;\;n}^{\hat s} E_{\;\;p}^{\hat q} \nonumber\\
        \bar E &=& \frac{1}{3!}\bar\varepsilon ^{\bar m\bar n\bar p} \bar\varepsilon _{\hat{\bar
        r}\hat{\bar s}\hat{\bar q}} E_{\;\;\bar m}^{\hat{\bar r}} E_{\;\;\bar n}^{\hat{\bar s}}E_{\;\;\bar p}^{\hat{\bar
        q}}.
\end{eqnarray}

The objects $E_{\;\;\bar n}^{\hat{\bar s}}$ are the beins for the
K\"{a}hler metric which means that the determinant of the metric
is $g = E\bar E$.

For (\ref{eleventransKahler}) to be satisfied, we find, for $N=2$:
\begin{eqnarray}
    ds^2  &=& H^{ - {{1 \mathord{\left/
 {\vphantom {1 3}} \right.
 \kern-\nulldelimiterspace} 3}}} \eta _{ab} dx^a dx^b  + 2H^{ - {{1 \mathord{\left/
 {\vphantom {1 3}} \right.
 \kern-\nulldelimiterspace} 3}}} g_{m\bar
    n} dz^m dz^{\bar n}  + H^{{{2 \mathord{\left/
 {\vphantom {2 3}} \right.
 \kern-\nulldelimiterspace} 3}}} \delta _{\alpha \beta } dx^\alpha
    dx^\beta,  \nonumber \\ a,b&=&0,1 \quad m,\bar n=1,2 \quad \alpha,\beta=1,2,3,4,5.
\nonumber\\
 (\partial _m H) &=& (\partial _{\bar m} H) = 0 \nonumber\\
 (\partial _\alpha  g) &=& (\partial _\alpha  g_{m\bar n} ) = 0
 \nonumber\\
 \epsilon&=&H^{-{{1} \mathord{\left/
 {\vphantom {{1} 12}} \right.
 \kern-\nulldelimiterspace} 12}}\epsilon_0.
\end{eqnarray}
where $\epsilon_0$ is a constant spinor, and the condition on
$g_{m\bar n}$ implies that it is either flat Minkowski or a $K3$
metric. This solution can be thought of as a single M5-brane
wrapping holomorphic 2-cycles (K\"{a}hler) of a Calabi-Yau
manifold.

For $N=3$:
\begin{eqnarray}
    ds^2  &=& H^{ - {{1 \mathord{\left/
 {\vphantom {1 3}} \right.
 \kern-\nulldelimiterspace} 3}}} \eta _{ab} dx^a dx^b  + 2H^{ - {{1 \mathord{\left/
 {\vphantom {1 3}} \right.
 \kern-\nulldelimiterspace} 3}}} g_{m\bar
    n} dz^m dz^{\bar n}  + H^{{{2 \mathord{\left/
 {\vphantom {2 3}} \right.
 \kern-\nulldelimiterspace} 3}}} \delta _{\alpha \beta } dx^\alpha
    dx^\beta,  \nonumber \\ a,b&=&0,1 \quad m,\bar n=1,2,3 \quad
    \alpha,\beta=1,2,3\nonumber\\
    (\partial _\alpha  \ln g) &=& (\partial _\alpha  \ln
    H)\nonumber\\
    \epsilon _ \pm &=& E^{ - \left( {\frac{1}{{12}} \pm \frac{1}{4}}
\right)} \bar E^{ - \left( {\frac{1}{{12}} \mp \frac{1}{4}}
 \right)}\epsilon_0,\label{N3Kahler}
\end{eqnarray}
where we have used $\Gamma _{\hat a\hat b} \epsilon_\pm = \pm\bar
\varepsilon _{\hat a\hat b}\epsilon_\pm $. This is either 3
intersecting M5-branes or a single M5-brane wrapping a K\"{a}hler
3-fold. Finally, for $N=4$:
\begin{eqnarray}
    ds^2  &=& H^{ - 1} \eta _{ab} dx^a dx^b  + 2g_{m\bar
    n} dz^m dz^{\bar n}  + H^2  dy^2,
    \nonumber \\ a,b&=&0,1 \quad m,\bar n=1,2,3,4.
\nonumber\\
 (\partial _\alpha \ln g) &=& (\partial _\alpha  \ln H )  \nonumber\\
    \epsilon _ \pm &=&
E^{ - \frac{1}{4}\left( { \pm 1 + 1} \right)} \bar
E^{\frac{1}{4}\left( { \pm 1 - 1} \right)}  \epsilon_0.
\end{eqnarray}

Using K\"{a}hler calibrations as demonstrated here to find SUGRA
solutions was later extended by the work of
\cite{Husain,Husain4,Husain3,Husain2}. Particularly, it was shown
that our solution presented here is a subset of a more general
solution where it is possible to relax the K\"{a}hler condition on
the metric.

Finally we study the $F$ field equations ($\nabla _A F^{A \cdots }
= 0$). For the trivial case $N=2$, one simply gets the harmonic
condition on the warp factor $H$, i.e. $\delta ^{\mu \nu }
(\partial _\mu  \partial _\nu  H) = 0$. For the other cases, the
conditions are a lot more complex. For $N=3,4$ respectively:
\begin{eqnarray}\label{nonlinear}
2\left( {\partial _m \partial _{\bar n} H} \right) + 2g_{m\bar n}
\delta
                ^{\alpha \beta } \left( {\partial _\alpha  \partial _\beta  \ln g} \right)
                - g_{m\bar s} \delta ^{\alpha \beta } \partial _\alpha  \left[ {g_{p\bar n}
                \left( {\partial                 _\alpha  g^{p\bar s} } \right)} \right]
                &=&
                0,\nonumber\\
4\left( {\partial _m
                \partial_{\bar n} H} \right) - H^{ - 1} \delta ^{\alpha \beta } \left(
                {\partial_\alpha  \ln g} \right)\left( {\partial _\beta  g_{m\bar n} }
                \right)
                - g_{m\bar s} H^{ - 1} \delta ^{\alpha \beta } \partial _\alpha
                \left[ {g_{p\bar n} \left( {\partial _\beta  g^{p\bar s} } \right)} \right]
                &=& 0.\nonumber\\
\end{eqnarray}

The solutions to these complicated differential equations should
specify the explicit metrics $g_{m\bar n}$ on the compact space.

\subsection{$\phi\sim\omega$}

The second, and last, possible nontrivial construct of $A$
involves a single occurrence of the K\"{a}hler form $\omega$
associated with the K\"{a}hler metric in the calibrated
$(1,1)$-form $\phi$. So, the components of the gauge potential
are:
\begin{equation}\label{}
    A_{0m\bar n}  = \pm iH^{ - 1} g_{m\bar n}.
\end{equation}

The coordinates $a,b$ run over the time component only, and the
complex coordinates $m,n = 1, \ldots ,N$ where possible values for
$N$ are 2, 3, 4 and 5. This leaves $\alpha ,\beta  = 1, \ldots ,10
- 2N$ for the transverse coordinates. The solution can be
interpreted as either $N$ M2-branes intersecting smoothly, or a
M2-brane completely wrapping $(1,1)$ cycles of a Calabi-Yau
$N$-fold. In both cases $2^{-N}$ of the supersymmetry is preserved
and we find the following metric and spinors:
\begin{eqnarray}
ds^2  &=& H^{-\frac{2}{3}\left( {N - 1} \right)} dt^2  +
2H^{-\frac{1}{3}\left( {4 - N} \right)} g_{m\bar n} dz^m dz^{\bar
n}  + H^{\frac{1}{3}\left( {N - 1} \right)} \delta _{\alpha \beta
} dx^\alpha  dx^\beta \nonumber\\
    \epsilon_\pm  &=& E^{ - \left( {\frac{{N - 1}}{6} \pm \frac{1}{4}} \right)} \bar E^{ -
    \left({\frac{{N - 1}}{6} \mp \frac{1}{4}} \right)} \epsilon _0.
\end{eqnarray}

\section{\label{SpecialSLAG}A case of SLAG calibrated M-branes}

Another solution we found corresponds to a special Lagrangian
calibrated submanifold with metric (\ref{metric}) for the case
$N=3$. The gauge potential can be constructed in terms of the real
part of the holomorphic 3-form $\Omega$ as follows ($a,b=0,1,2$):
\begin{eqnarray}
        A_{abcmnp}  &=& \pm H^{ - 1} \bar\varepsilon _{abc} \Omega _{mnp}, \quad \quad \Omega
        _{mnp} = E\bar\varepsilon _{mnp}  \nonumber\\
        A_{abc\bar m\bar n\bar p}  &=& \pm H^{ - 1} \bar\varepsilon _{abc} \Omega _{\bar m\bar
        n\bar p}, \quad \quad \Omega _{\bar m\bar n\bar p}  = \bar E\bar\varepsilon _{\bar m\bar
        n\bar p}. \label{specialSLAGA}
\end{eqnarray}

The nonvanishing field strength components are hence
\begin{eqnarray}
 F_{\bar rabcmnp}  &=&  \mp H^{ - 1} \left( {\partial _{\bar r} \ln {H \mathord{\left/
 {\vphantom {H E}} \right.
 \kern-\nulldelimiterspace} E}} \right) \bar \varepsilon _{abc} \Omega _{mnp} \quad {\rm and\;c.c.} \nonumber\\
 F_{\alpha abcmnp} &=&  \mp H^{ - 1} \left( {\partial _\alpha  \ln {H \mathord{\left/
 {\vphantom {H E}} \right.
 \kern-\nulldelimiterspace} E}} \right) \bar
\varepsilon _{abc} \Omega _{mnp} \quad {\rm and\;c.c.}
\end{eqnarray}

The projection conditions for this case are:
\begin{eqnarray}
 \Gamma _{\hat m\hat n\hat p} \epsilon _ +   &=& a_ -  \bar \varepsilon _{\hat m\hat n\hat p} \epsilon _ -   \nonumber\\
 \Gamma _{\hat {\bar m}\hat {\bar n}\hat {\bar p}} \epsilon _ -   &=& a_ +  \bar \varepsilon _{\hat {\bar m}\hat {\bar n}\hat {\bar p}} \epsilon _ +   \nonumber\\
 \Gamma _{\hat a\hat b\hat c} \epsilon _ \pm   &=& d_ \pm  \bar \varepsilon _{\hat a\hat b\hat c} \epsilon _
 \pm,
\end{eqnarray}
where we recall that the hatted indices are in the flat tangent
space. One finds that in order for the SUSY and Einstein equations
to be satisfied, then one must have $d_+ d_- =-1$ and $a_+ a_- =
-8$. This last is a consequence of the following SUSY argument.
Consider the spinorial projections on the compact space. The Dirac
matrices create and annihilate those spinors such that
\begin{equation}
 \Gamma _{\hat 1\hat 2\hat 3} \epsilon _ +   = a_ -  \epsilon _ -,
 \quad\quad
 \Gamma _{\hat {\bar 1}\hat {\bar 2}\hat {\bar 3}} \epsilon _ -   = a_ +  \epsilon _
 +,
\end{equation}
which implies $\Gamma _{\hat {\bar 1}\hat {\bar 2}\hat {\bar 3}}
\Gamma _{\hat 1\hat 2\hat 3} \epsilon _ + \sim\epsilon _ + $. This
can only be satisfied if and only if $a_\pm = \pm \sqrt 8 $. Based
on these projections, the following identities can be derived and
were particularly useful in the calculation
\begin{eqnarray}
 \eta _ \pm   &=& \left( {\epsilon _ +   \pm \epsilon _ -  } \right),\quad \Lambda _ \pm   = \left( {d_ +  \epsilon _ +   \pm d_ -  \epsilon _ -  } \right) \nonumber\\
 \bar \varepsilon _{abc} \Gamma ^{abc} \eta _ \pm   &=&  - 6H^{3A} \Lambda _ \pm   \nonumber\\
 \bar \varepsilon _{abc} \Gamma ^{bc} \eta _ \pm   &=&  - 2H^{3A} \Gamma _a \Lambda _ \pm
 \nonumber\\
 \Omega _{mnp} \Gamma ^{mnp} \eta _ \pm   &=&  \pm 6a_ +  H^{3B} \epsilon _ +   \nonumber\\
 \Omega _{\bar m\bar n\bar p} \Gamma ^{\bar m\bar n\bar p} \eta _ \pm   &=&  + 6a_ -  H^{3B} \epsilon _ -
 \nonumber\\
 \Omega _{mnp} \Gamma ^{np} \eta _ \pm   &=&  \pm a_ +  H^{3B} \Gamma _m \epsilon _ +   \nonumber\\
 \Omega _{\bar m\bar n\bar p} \Gamma ^{\bar n\bar p} \eta _ \pm   &=&  + a_ -  H^{3B} \Gamma _{\bar m} \epsilon _ -
 \nonumber\\
 \Omega _{mnp} \Gamma ^p \eta _ \pm   &=&  \pm \frac{2}{{a_ -  }}H^{3B} \Gamma _{mn} \epsilon _ +   \nonumber\\
 \Omega _{\bar m\bar n\bar p} \Gamma ^{\bar p} \eta _ \pm   &=&  + \frac{2}{{a_ +  }}H^{3B} \Gamma _{\bar m\bar n} \epsilon _
 -.
\end{eqnarray}

The SUSY gravitino equation now acts on the spinor $\eta$, which
is a linear combination of the two chiral spinors $\epsilon$. We
find the metric to be
\begin{eqnarray}
 ds^2  &=& H^{ - \frac{4}{3}} \eta _{ab} dx^a dx^b  + 2H^{\frac{2}{3}} g_{m\bar n} dz^m dz^{\bar n}  + H^{\frac{8}{3}} \delta _{\alpha \beta } dx^\alpha  dx^\beta   \nonumber\\
 a,b &=& 0,1,2\quad \quad m,\bar n = 1,2,3\quad \quad \alpha ,\beta  =
 1,2, \label{specialSLAGmetric}
\end{eqnarray}
which may be interpreted as either M5-branes intersecting over a
SLAG calibrated manifold, or as a single M5-brane wrapped over
SLAG cycles of a Calabi-Yau submanifold as we will discuss in more
detail later.

We also find the conditions $(\partial_{\bar m} \ln E)=(\partial_m
\ln \bar E)=0$ implying that the metric $g_{m\bar n}$ is
Ricci-flat, which is not surprising, since only Ricci-flat, i.e.
Calabi-Yau, metrics allow SLAG calibrations.

Supersymmetry further imposes the following constraints on the
solution
\begin{equation}
 E = \bar E,\quad\quad
 g_{m\bar n}  = f_{m\bar n}\left( {z,\bar z } \right),\quad \quad \left( {\partial _\alpha  \ln g} \right) = \left( {\partial _\alpha  \ln H}
 \right),
\end{equation}
and the SUSY spinors have the form $\epsilon _ \pm   = H^{ -
\frac{1}{3}} \epsilon _0 $.

Now, in constructing the ansatz (\ref{specialSLAGA}), we have only
used the $(3,0)$ forms on the compact space. Later we will discuss
the possibility of a more general solution that involves the CY
$(2,1)$ forms as well. The $(2,1)$ forms are parametrized by the
moduli of the complex structure of the CY 3-fold. The solution we
find here represents the case of constant moduli.

In conclusion, we have proposed the hypothesis that the
calibration technique may be used to find BPS brane solutions to
supergravity. We demonstrated this by finding examples of
localized M-brane intersections in $D=11$. The solutions we found
correspond to a number of M2 and M5 branes intersecting on
K\"{a}hler calibrated submanifolds and the intersection of
M5-branes over a special Lagrangian calibrated surface.

\chapter{Down to five dimensions}\label{compact}

A review of the properties of Calabi-Yau 3-folds (see appendix
\ref{manifolds}) reveals that they have both K\"{a}hler and
special Lagrangian cycles. This hints at the possibility that the
$N=3$ M-brane solutions we have found have the alternate
interpretation as single M-branes wrapping either of these cycles
down to $D=5$. In fact, it has been shown \cite{Kastor4} that the
K\"{a}hler calibrated solution (\ref{N3Kahler}) corresponds to
five dimensional black 1-branes coupled to the $D=5$
$\mathcal{N}=2$ vector multiplet fields. We propose that SLAG
calibrated solutions in $D=11$ correspond to 2-brane solutions
coupled to the hypermultiplets sector of the reduced theory.

In the next chapter, we find 2-brane solutions coupled to the
hypermultiplets, and discuss how they arise from the eleven
dimensional wrapping. This will involve studying $D=5$
$\mathcal{N}=2$ SUGRA theory and how it arises as a dimensional
reduction of $D=11$ SUGRA. The topology of the Calabi-Yau 3-fold
is responsible for the very rich structure of the resulting
theory, particularly of the space of the complex structure moduli,
involving the symplectically invariant special K\"{a}hler
geometry. One might even say that this structure lends a detailed
description of the `internal' mechanics of the $D=11$ solutions.

We first outline the dimensional reduction of eleven dimensional
supergravity over a Calabi-Yau 3-fold down to five dimensions. The
theory we find is ungauged $\mathcal{N}=2$ supergravity
\cite{Aspinwall,Cadavid,Ferrara5,Lukas,PapaTown}\footnote{For a
discussion of the gauged theory, see \cite{ELP}.}, containing, in
addition to the gravity sector\footnote{The graviton $g_{\mu \nu}$
and two gravitini $\psi_\mu^1$ and $\psi_\mu^2$.}, two matter
supermultiplet sectors known as the vector multiplets and the
hypermultiplets (including the $U(1)$ graviphoton $A_\mu$ and the
dilaton $\sigma$ respectively). The scalars of the vector
multiplets arise from the K\"{a}hler moduli of the CY 3-fold,
while the hypermultiplets scalars arise from the complex structure
moduli. M-branes wrapping K\"{a}hler calibrated cycles of the CY
deform the K\"{a}hler moduli while those wrapping SLAG cycles
result in the deformation of the complex structure moduli. The
number of the resulting fields depends on the topology of the CY
manifold. We note that the two sectors naturally decouple from
each other \cite{Witten,BeckerStrominger,Proeyen7}, which makes it
feasible to consistently set one of them to zero. As we are
interested in the hypermultiplets sector only, we will
systematically drop all terms that depend on the K\"{a}hler
$(1,1)$-forms. For reference, similar string theory
compactifications were performed in \cite{Ferrara6,Cadavid2}.
Other interesting studies of $\mathcal{N}=2$ supersymmetry and its
matter content include \cite{Chou,Strominger2,Ketov}. Work within
the hypermultiplets sector, specially the five dimensional case,
is quite rare in the literature. On the other hand, the vector
multiplets sector, specially in four dimensions, is better
understood. Fortunately, as we explain in the appendices, one can
use the technology developed for the latter to discuss the former.

\section{Preliminaries}

We start with the action of eleven dimensional supergravity
\begin{equation}
    S_{11}  = \frac{1}{{2\kappa _{11}^2 }}\int {d^{11} x\sqrt { - G}
    \left( {R - \frac{1}{{48}}F^2 } \right)}  - \frac{1}{12\kappa _{11}^2 } \int {A \wedge F \wedge
    F},
\end{equation}
and discuss only the bosonic part of the calculation. We are
basically following in detail the calculation first performed in
\cite{GS2}. For the eleven dimensional metric, the following
ansatz is used:
\begin{eqnarray}
    ds_{11}^2  &=& G_{MN} dx^M dx^N \nonumber \\
    &=& e^{\frac{2}{3}\sigma } g_{\mu \nu } dx^\mu  dx^\nu
    + e^{ - \frac{\sigma }{3}} k_{ab} dy^a dy^b \nonumber \\
    &=& e^{\frac{2}{3}\sigma } g_{\mu \nu } dx^\mu  dx^\nu
    + e^{ - \frac{\sigma }{3}} \left( {k_{mn} dw^m dw^n  + k_{\bar m\bar n} dw^{\bar m} dw^{\bar n}  + 2k_{m\bar n} dw^m dw^{\bar
    n}} \right),\nonumber \\ \label{expmet}
\end{eqnarray}
where the Greek indices are five dimensional, the capital Latin
indices are eleven dimensional, the small Latin indices from the
beginning of the alphabet are six dimensional and the holomorphic
and antiholomorphic indices $(m,n;\bar m ,\bar n )$ are three
dimensional over the Calabi-Yau space. The metric $g_{\mu\nu}$ is
the five dimensional metric, which will eventually contain our
2-brane solutions. The dilaton $\sigma$ is dependent only on the
$x^\mu$ directions, and the exponents have been chosen such that
the resulting $D=5$ action would have the conventional numerical
coefficients.

Now, generally, the eleven-dimensional 3-form $A_{LMN}$ has two
separable components, the five dimensional part $A_{\mu \nu
\rho}$, with a field strength $F_{\mu \nu \rho \sigma }$ which we
leave alone, and the six dimensional part $A_{abc}$ that lives on
the CY 3-fold. We expand the later in terms of the $H^3$-dual
cohomology forms $\alpha_I$ and $\beta^I$ (\ref{cohbasis}), where
the $I$ and $J$ indices run over $0$ to $h_{2,1}$:
\begin{equation}\label{threeform}
    A = \frac{1}{{3!}}A_{\mu\nu\rho} dx^\mu dx^\nu dx^\rho + \sqrt 2 \left( {\zeta ^I
\alpha _I  + \tilde \zeta _I \beta ^I } \right).
\end{equation}

The coefficients $\zeta^I$ and $\tilde \zeta_I$ appear as axial
scalar fields in the lower dimensional theory. We also note that
the 3-form $A_{\mu \nu \rho}$ in five dimensions is dual to a
scalar field which we will call $a$ (also known as the universal
axion). The set ($a$, $\sigma$, $\zeta^0$, $\tilde \zeta_0$)
comprise the universal hypermultiplet\footnote{So called because
it appears in all Calabi-Yau compactifications, irrespective of
the detailed structure of the CY manifold.}. The rest of the
hypermultiplets are ($z^i$, $z^ {\bar i}$, $\zeta^i$, $\tilde
\zeta_i$ : $i = 1, \ldots ,h_{2,1} $), where we recognize the
$z$'s as the CY's complex structure moduli (as demonstrated in
appendix \ref{specialgeometry}). Note that the total number of
scalar fields in the hypermultiplets sector is $4(h_{2,1}+1)$
(each hypermultiplet has 4 real scalar fields). Later, we will
show that those scalar fields, although exhibiting explicit
special K\"{a}hler geometry, actually parameterize a quaternionic
manifold of $(h_{2,1}+1)$ components (the c-map).

The solutions we are looking for are, as noted earlier, 2-brane
solutions coupled to the hypermultiplets. This makes our solutions
magnetically dual to the instanton solutions of
\cite{GS1,GS2}\footnote{The formula is: a $p$-brane is the
magnetic dual to a $(D-p-4)$-brane in $D$ dimensions. So, a
$(p=2)$-brane is dual to a $(5-2-4=-1)$-brane in 5 dimensions,
i.e. an instanton.}. For easier comparison, we will, with some
exceptions, follow their notation and conventions throughout.

\section{The Einstein-Hilbert term}

\begin{eqnarray}
    S_E  = \frac{1}{{2\kappa _{11}^2 }}\int {d^{11} x\sqrt { - G}R}
\end{eqnarray}

We list the formulae needed for the calculation:

\begin{enumerate}
  \item The metric
        \begin{eqnarray}
            G_{\mu \nu }  &=& e^{\frac{2}{3}\sigma } g_{\mu \nu
            }\quad,\quad G^{\mu \nu }  = e^{-\frac{2}{3}\sigma } g^{\mu \nu } \nonumber\\
            G_{ab}  &=& e^{ - \frac{\sigma }{3}} k_{ab}\quad, \quad G^{ab}  = e^{ \frac{\sigma }{3}} k^{ab}
            \nonumber \\
            G &=& \det  {G_{MN} }  = e^{\frac{4}{3}\sigma } gk, \nonumber \\
            g &=& \det  {g_{\mu \nu } }  \quad, \quad k =  \det k_{ab}.
        \end{eqnarray}
  \item The Christoffel symbols
        \begin{eqnarray}
            \Gamma _{\nu \rho }^\mu   &=& \tilde \Gamma _{\nu \rho }^\mu  \left[ {g} \right] + \frac{1}{3}\left[ {\delta _\rho ^\mu  \left(
            {\partial   _\nu      \sigma } \right) + \delta _\nu ^\mu  \left( {\partial _\rho  \sigma
            }     \right) - \delta_{\nu \rho } \delta^{\mu \kappa} \left( {\partial
            _\kappa      \sigma } \right)} \right]
            \nonumber \\
            \Gamma _{ab}^\mu   &=& \frac{1}{6}e^{ - \sigma } k_{ab} g^{\mu \nu}
            \left( {\partial            _\nu \sigma } \right) - \frac{1}{2}e^{ - \sigma } g^{\mu \nu}
            \left(     {\partial _\nu k_{ab}     } \right) \nonumber \\
            \Gamma _{\mu b}^a  &=& \frac{1}{2}k^{ac} \left( {\partial _\mu  k_{cb} } \right)
            -     \frac{1}{6}\delta _b^a \left( {\partial _\mu  \sigma } \right)
            \nonumber     \\
            \Gamma _{bc}^a  &=& \hat \Gamma _{bc}^a \left[ {k} \right],
        \end{eqnarray}
        where the ( $\tilde{}$ ) and the ( $\hat{}$ ) refer to purely five
        and six dimensional components respectively.
  \item The Ricci scalar
        \begin{eqnarray}
            e^{\frac{2}{3}\sigma } G^{ab} R_{ab}  &=& \tilde \nabla ^2 \sigma
             + \frac{1}{2}\left( {\partial_\mu \ln k} \right)
            \left(     {\partial^\mu \sigma } \right) - \frac{1}{2}\tilde \nabla
            ^2     \ln k - \frac{1}{4}\left( {\partial_\mu \ln k}
            \right)\left( {\partial^\mu \ln k}  \right), \nonumber \\
            e^{\frac{2}{3}\sigma } G^{\mu \nu } R_{\mu \nu }  &=& \tilde
            R\left[{g } \right] - \frac{5}{3}\;\tilde \nabla ^2 \sigma -
            \frac{1}{2}\left( {\partial_\mu \sigma } \right)\left( {\partial^\mu \sigma } \right)  - \frac{1}{2}
            \;\tilde\nabla ^2 \ln k \nonumber \\
            &-& \frac{1}{2}\left( {\partial_\mu \ln k} \right)
            \left(     {\partial^\mu \sigma } \right) + \frac{1}{4}\left( {\partial_\mu k^{ab} }
            \right)    \left( {\partial^\mu k_{ab} } \right), \nonumber \\
            R  &=& \frac{{\sqrt { - gk} }}{{\sqrt { - G} }}\left[ {\tilde
            R\left[  {g } \right] - \frac{2}{3}\;\tilde \nabla ^2 \sigma
            - \frac{1}{2}\left( {\partial_\mu \sigma } \right)\left( {\partial^\mu \sigma } \right)  -
            \tilde     \nabla ^2 \ln k  } \right. \nonumber \\
            &-& \frac{1}{4}\left( {\partial_\mu \ln k} \right)\left( {\partial^\mu \ln k} \right) + \left. {\frac{1}{4}\left( {\partial_\mu k^{ab} }
            \right)  \left(       {\partial^\mu k_{ab} } \right) } \right].
    \end{eqnarray}
\end{enumerate}

Now, the $\tilde \nabla ^2 \sigma$ term is a total derivative and
hence has no effect on the physics, we drop it out\footnote{Also
note that the terms containing ($\ln k$)-dilaton couplings cancel
exactly, which explicitly demonstrates the decoupling of the
vector and hypermultiplets, since this expression includes both
$\delta k_{mn}$ and $\delta k_{m\bar n}$ components.}. Considering
the terms involving the CY metric $k$, we notice that some of
those yield either terms dependent on $k^{mn}$ and $k^{\bar m \bar
n}$, which vanish because of (\ref{Hermitian}), or terms that
contribute to the vector multiplets, which we also drop. The only
surviving terms are of the form $\left[ k^{m\bar n} k^{r\bar p}
\left( {\partial _\mu k_{mr} } \right)\left( {\partial ^\mu
k_{\bar n\bar p} } \right)\right]$. Using the definition
(\ref{Gij}), we write:
\begin{equation}
    \int {d^6 w\sqrt k } \left[ {k^{m\bar n} k^{r\bar p} \left( {\partial _\mu  k_{mr} } \right)\left( {\partial ^\mu  k_{\bar n\bar p} } \right)} \right]
    \sim V_{CY} G_{i\bar j} \;( {\partial _\mu  z^i } )( {\partial
    ^\mu      z^{\bar j} } ), \nonumber
\end{equation}
up to a redefinition of numerical coefficients. Putting things
together, the compactification of the eleven dimensional
Einstein-Hilbert term yields:
\begin{equation}\
    S_E=  \frac{{V_{CY} }}{{2\kappa _{11}^2 }}\int {d^5 x\sqrt { - g }
    \left[     { R_5  - \frac{1}{2}\left( {\partial _\mu  \sigma } \right)\left(
    {\partial     ^\mu  \sigma } \right) - G_{i\bar j} ( {\partial _\mu  z^i }
    )( {\partial ^\mu  z^{\bar j} } )} \right]},
\end{equation}
where the Calabi-Yau volume is defined by:
\begin{equation}\
    V_{CY}=\int {d^6 w\sqrt k }.
\end{equation}

\section{The $F^2$ term}

\begin{eqnarray}
    S_{F^2} &=& - \frac{1}{{2\kappa _{11}^2 }}\int {d^{11} x\sqrt { - G } \frac{1}{{48}}F^{LMNP}
    F_{LMNP}} \label{int1}\nonumber \\
    &=& - \frac{1}{{2\kappa _{11}^2 }}\frac{1}{{48}}\int {d^5 x\sqrt { - g} \int {d^6 w\sqrt k \left[ {e^{ - 2\sigma } F_{\mu \nu \rho \sigma } F^{\mu \nu \rho \sigma }  + e^{\frac{2}{3}\sigma } F_\mu  F^\mu  } \right]} },
\end{eqnarray}
where (compact indices suppressed)
\begin{eqnarray}
    F_\mu &=& \sqrt 2 \left[ {\left( {\partial_\mu \zeta ^I } \right)\alpha _I
    + \left( {\partial_\mu \tilde    \zeta     _I } \right)\beta ^I } \right] \nonumber \\
    \star F_\mu &=& \sqrt 2 \left[ {\left( {\partial_\mu \zeta ^I } \right)\star \alpha _I  + \left(
    {\partial_\mu \tilde    \zeta _I    }     \right)\star \beta ^I } \right]. \label{FdA1}
\end{eqnarray}

Substituting, we get:
\begin{eqnarray}
     S_{F^2} = &-&\frac{{V_{CY} }}{{2\kappa _{11}^2 }}\int {d^5 x\sqrt { - g } \frac{1}{{48}}e^{
    -     2\sigma } F^{\mu \nu \rho \sigma } F_{\mu \nu \rho \sigma } }  \\
    &-&\frac{1}{{2\kappa _{11}^2 }} \int {d^5 x\sqrt { - g } e^\sigma  \left[ {\left( {\partial _\mu  \zeta ^I } \right)\left( {\partial
    ^\mu      \zeta ^J } \right)\int\limits_{CY} {\alpha _I  \wedge \star\alpha _J } } \right. }\nonumber \\
    &+&  ( {\partial   _\mu  \zeta ^I } )( {\partial ^\mu  \tilde \zeta _J }
    )\int\limits_{CY} {\alpha _I  \wedge \star\beta ^J }+( {\partial _\mu  \tilde \zeta _I } )( {\partial ^\mu
    \zeta     ^J } )\int\limits_{CY} {\beta ^I  \wedge \star\alpha _J }  \nonumber \\
    &+& \left. { ( {\partial _\mu \tilde \zeta _I } )( {\partial ^\mu  \tilde \zeta _J } )\int\limits_{CY} {\beta
    ^I      \wedge \star\beta ^J } } \right]. \nonumber
\end{eqnarray}

In order to do the six dimensional integrals, we need explicit
expressions for $\star \alpha$ and $\star \beta$ such that
$\star\star\alpha=-\alpha$ and $\star\star\beta=-\beta$. This has
been done in various sources, for example \cite{Suzuki}, where
they are expressed in terms of the real and imaginary components
of the period matrix. Using matrix notation in the $I,J$ indices,
these are\footnote{From the appendices, recall that ${\mathop{\rm
Re}\nolimits} \mathcal{N} = \theta $ and ${\mathop{\rm
Im}\nolimits} \mathcal{N} =  - \gamma $, where $\mathcal{N}$ is
the period matrix.}:
\begin{eqnarray}
    \star \alpha  &=& \left( {\gamma  + \gamma ^{ - 1} \theta ^2 } \right)\beta  - \gamma
    ^{     -     1} \theta \alpha \nonumber \\
    \star \beta  &=& \gamma ^{ - 1} \theta \beta  - \gamma ^{ - 1} \alpha.
\end{eqnarray}

We find:
\begin{eqnarray}
    S_{F^2}=
    &-& \frac{1}{{2\kappa _{11}^2 }}
    \int{d^5 x\sqrt { - g }} \left\{ {\frac{V_{CY}}{{48}}e^{-2\sigma }F^{\mu \nu \rho \sigma }F_{\mu \nu \rho \sigma}} \right.
     \nonumber \\
    &+& e^\sigma  \left[ {\left( {\gamma  + \gamma ^{ - 1} \theta ^2 } \right)\left(
    {\partial _\mu  \zeta } \right)\left( {\partial ^\mu  \zeta } \right) + \gamma ^{ -
    1}     ( {\partial _\mu  \tilde \zeta } )( {\partial ^\mu  \tilde
    \zeta     } )} \right. \nonumber \\
    &+& \left. {\left. {2\gamma ^{ - 1} \theta ( {\partial _\mu  \zeta } )(
    {\partial ^\mu  \tilde \zeta } )} \right]} \right\}.
\end{eqnarray}

The reader will note that the appearance of the components of the
period matrix in this, seemingly arbitrary, form actually has a
solid reason: the fact that the matrix
\begin{equation}\label{Symplecticmatrix}
\Lambda= \left[ {\begin{array}{*{20}c}
   {\gamma ^{ - 1} \theta } & {\gamma ^{ - 1} }  \\
   {\left( {\gamma  + \gamma ^{ - 1} \theta ^2 } \right)} & { - \gamma ^{ - 1} \theta }  \\
\end{array}} \right]
\end{equation}
is a symplectic matrix as defined by (\ref{sympDef}). It ensures
the symplectic invariance of the action.

\section{The Chern Simons term}

Using the explicit expressions for $A$ and $F$, we write the
Chern-Simons term as follows:
\begin{eqnarray}
    S_{CS}  &=&  - \frac{1}{{12\kappa _{11}^2 }}\int {A \wedge F \wedge F} \nonumber \\ &=&  -\frac{1}{{12\kappa _{11}^2 }} 4\int {d\tilde A \wedge \left[ {\left( {\zeta \alpha  + \tilde \zeta \beta } \right) \wedge \left( {d\zeta \alpha  + d\tilde \zeta \beta } \right)} \right]}   \nonumber \\
    &=&   \frac{-1}{{48\kappa _{11}^2 }}\int {d^5 x \varepsilon _{\mu \nu \rho \sigma \kappa
    }         F^{\mu \nu \rho \sigma } \int\limits_{CY} {\left[ {\zeta ^I (
    {\partial     ^\kappa  \tilde \zeta _J } )\left( {\alpha _I  \wedge \beta ^J
    }         \right) + \tilde \zeta _I ( {\partial ^\kappa  \zeta ^J }
    )\left(     {\beta ^I  \wedge \alpha _J } \right)} \right]}
    }.\nonumber\\
\end{eqnarray}

Now, using (\ref{cohbasis}), we end up with:
\begin{equation}\
    S_{CS}  = -\frac{1}{{2\kappa _{11}^2 }}\frac{1}{{24}}\int {d^5 x\sqrt { -
    g} \bar\varepsilon _{\mu \nu \rho \sigma \kappa } F^{\mu \nu \rho \sigma } \left[
    {\zeta^I \left( {\partial ^\kappa  \tilde \zeta _I } \right) - \tilde \zeta _I \left( {\partial
    ^\kappa  \zeta ^I } \right)} \right]}.
\end{equation}

Summing up, the bosonic part of the action is (choosing $  \kappa
_{11}^2  = \kappa _{5}^2$ and $V_{CY}  = 1$):
\begin{eqnarray}
    S_5  &=& \frac{1}{{2\kappa _{5}^2 }}\int {d^5 x\sqrt { - g} \left[ { R - \frac{1}{2}\left( {\partial
    _\mu \sigma     } \right)\left( {\partial ^\mu  \sigma } \right) - G_{i\bar j} (
    {\partial _\mu  z^i } )( {\partial ^\mu  z^{\bar j} } )}
    \right.}  \nonumber \\
    &-& \left.{ \frac{1}{{48}}e^{ - 2\sigma } F_{\mu \nu \rho \sigma } F^{\mu \nu \rho \sigma }-\frac{1}{{24}}\bar\varepsilon _{\mu \nu \rho \sigma \alpha } F^{\mu \nu
    \rho         \sigma } K^\alpha  \left( {\zeta ,\tilde \zeta } \right) + e^\sigma
    L_\mu     ^\mu  \left( {\zeta ,\tilde \zeta } \right)} \right], \label{d5theory}
\end{eqnarray}
where we have defined the spacetime tensors:
\begin{eqnarray}
    K_\alpha  ( {\zeta ,\tilde \zeta } ) &=& \left[ {\zeta ^I (
    {\partial_\alpha  \tilde \zeta _I } ) - \tilde \zeta _I ( {\partial_\alpha  \zeta
    ^I     } )} \right] \nonumber \\
    L_{\mu \nu}  ( {\zeta ,\tilde \zeta } ) &=&  - \left( {\gamma  + \gamma ^{
    -     1} \theta ^2 } \right)\left(
    {\partial _\mu  \zeta } \right)\left( {\partial _\nu  \zeta } \right)
    - \gamma ^{ - 1}( {\partial _\mu  \tilde \zeta } )( {\partial _\nu
    \tilde \zeta     } ) \nonumber \\
    & & -2\gamma ^{ - 1} \theta \left( {\partial _\mu  \zeta } \right)(
    {\partial _\nu  \tilde \zeta } ). \label{KandL}
\end{eqnarray}

Note that if we write the axion fields in a symplectic vector form
(spacetime indices suppressed)
\begin{equation}\label{}
\Xi  = \left[ {\begin{array}{*{20}c}
   {( {\partial \zeta } )}  \\
   {( {\partial \tilde \zeta } )}  \\
\end{array}} \right],
\end{equation}
then $L_{\mu\nu}$ is nothing more than the norm $\Xi ^T \Lambda
\Xi $ of this vector in symplectic space, with $\Lambda$ defined
by (\ref{Symplecticmatrix}). See appendix \ref{specialgeometry}
for a more detailed discussion.

\section{The supersymmetry transformations}

What are the SUSY transformations that the action
(\ref{d5theory}), after including the fermionic terms, is
invariant under? In this section, we outline the general argument
for acquiring those from the eleven dimensional one \cite{GS2}. We
will also define the pertinent quantities in a way that will be
useful in showing the quaternionic nature of the action. In eleven
dimensions, the only fermionic field is the gravitino, while in
five dimensions, we have two gravitini and a set of hyperini; the
superpartners of the hypermultiplet bosons.

On a CY 3-fold, there are two supercovariantly constant Killing
spinors \cite{WFN}, which may be defined, as usual, in terms of
the Dirac matrices as follows:
\begin{equation}\label{}
    \Gamma ^m \eta_+  = 0\quad ,\quad \Gamma ^{\bar m} \eta_-  =
    0,
\end{equation}
hence
\begin{equation}\label{}
    \Gamma _{\hat m\hat n\hat p} \eta_+  =  \pm \bar \varepsilon
    _{\hat m\hat n\hat p} \eta_-,
\end{equation}
which we use to define the spinors in terms of the $(3,0)$ form:
\begin{eqnarray}
    \Gamma _{mnp} \eta_+  &=& \Omega _{mnp} \eta_-  \nonumber\\
    \eta_-  &=& \frac{1}{{\left| \Omega  \right|^2 }}\Omega ^{mnp}
    \Gamma _{mnp} \eta_+.
\end{eqnarray}

Now, an eleven dimensional spinor $\Lambda $ may be expanded in
terms of the five dimensional spinors $\epsilon_1$ and
$\epsilon_2$ as follows:
\begin{equation}\label{spinorexpansion}
    \Lambda  = \epsilon _1  \otimes \eta_+  + \epsilon _2 \otimes
     \eta_-.
\end{equation}

The strategy is to write the eleven dimensional gravitino
equation, expand in terms of the five dimensional spinors
similarly to (\ref{spinorexpansion}), then identify the terms that
are dependent on the $(2,1)$ and $(1,2)$ forms $\left( {\nabla _i
\Omega } \right)$ and $\left( {\nabla _{\bar i} \bar \Omega }
\right)$, or their components $\chi_i$ and $\chi_{\bar i}$,
defined by (\ref{link}), where the $U(1)$ K\"{a}hler covariant
derivative $\nabla _i$ is defined by (\ref{covderiv}) and
(\ref{covderiv1}). These are taken to represent the hyperini, and
their sum is identified as the hyperino variation equations. The
rest of the terms, dependent on the $(3,0)$ and $(0,3)$ forms,
become the $\mathcal{N}=2$ gravitini equations.

We begin with the $D=11$ gravitino variation:
\begin{equation}
    \delta_{\epsilon} \psi _M = \left( {\nabla _M  \Lambda } \right) -
    \frac{1}{{288}}F_{LNPQ } \left( {\Gamma _M  ^{\;\;\;\;LNPQ }  - 8\delta _M
    ^L  \Gamma ^{NPQ } } \right)\Lambda.\label{eleventranslambda}
\end{equation}

Based on the metric (\ref{expmet}), we collect the relevant
quantities:
\begin{eqnarray}
    e_{\;\;\nu} ^{\hat \mu }  = e^{\frac{\sigma}{3} } N_{\;\;\nu} ^{\hat \mu } \quad \quad &,&\quad
    \quad e_{\;\;b}^{\hat a}  = e^{ - \frac{\sigma}{6} } E_{\;\;b}^{\hat a}  \nonumber \\
    g_{\mu \nu }  = N_{\;\;\mu} ^{\hat \alpha } N_{\;\;\nu} ^{\hat \beta } \eta _{\hat \alpha \hat
    \beta     } \quad \quad &,&\quad \quad k_{ab}  = E_{\;\;a}^{\hat c} E_{\;\;b}^{\hat d} \delta _{\hat c\hat
    d}.
\end{eqnarray}
The non-vanishing components of the spin connections are then:
\begin{eqnarray}
    \omega _\mu ^{\;\;\hat \alpha \hat \beta }&=& \tilde \omega _\mu ^{\;\;\hat \alpha \hat
    \beta     } \left[ {g } \right] - \frac{1}{3}\left( {N^{\hat \alpha \nu } N_{\;\;\mu} ^{\hat \beta }  - N_{\;\;\mu} ^{\hat \alpha } N^{\hat \beta \nu } } \right)\left( {\partial _\nu  \sigma } \right)\nonumber \\
    \omega _\mu ^{\;\;\hat a\hat b}  &=& E^{\hat a d} \left( {\partial _\mu
    E_{\;\;d}^{\hat b} } \right)  - \frac{1}{2}E^{\hat bf} E^{\hat ad} \left( {\partial _\mu  k_{df} } \right) \nonumber \\
    \omega _a^{\;\;\hat \alpha \hat b}  &=&  e^{ - \frac{\sigma }{2}} N^{\hat \alpha \beta
    }     \left[ {\frac{1}{6}E_{\;\;a}^{\hat b} \left( {\partial _\beta  \sigma } \right)-\frac{1}{2}E^{\hat b d} \left( {\partial _\beta  k_{ad} }
    \right)    }    \right]     \nonumber \\
    \omega _c^{\;\;\hat a\hat b}  &=& \hat \omega _c^{\;\;\hat a\hat b}\left[ {k_{ab} }
    \right].
 \end{eqnarray}

We note that $\omega _\mu ^{\;\;\hat a\hat b}$ breaks down into $(
{\hat m\hat {\bar n}} )$, $( {\hat {\bar m}\hat n} )$, $\left(
{\hat m\hat n} \right)$ and $( {\hat {\bar m}\hat {\bar n}} )$
pieces. The last two we can write in terms of the Christoffel
symbol:
\begin{equation}\label{}
    \Gamma _{\mu m}^{\bar n}  = \frac{1}{{2\left| \Omega  \right|^2
    }}\Omega ^{\bar n\bar o\bar p} \chi_{\bar o\bar p m|\bar i} (
    {\partial _\mu  z{}^{\bar i}} ),
\end{equation}
and its complex conjugate. This introduces the complex structure
moduli. The remainder of the $\omega _\mu ^{\;\;\hat a\hat b}$
components are also expanded in terms of the moduli. They
contribute an expression that is identified as the $U(1)$
K\"{a}hler connection. We report on the result below.

To deal with the components $F_{\mu abc}$ of the field strength,
we note that, up to an exact form, one can always expand any three
form, in this case $F_\mu$, in terms of the ($3,0$) and ($2,1$)
forms dual to the homology decomposition $H^3 = H^{3,0} \oplus
H^{2,1} \oplus H^{1,2} \oplus H^{0,3} $ as follows
\cite{Denef,GS2}:
\begin{eqnarray}
    F_\mu &=& ie^{{\mathcal{K} \mathord{\left/ {\vphantom {\mathcal{K} 2}} \right.
    \kern-\nulldelimiterspace} 2}} \bar X\Omega  - ie^{{\mathcal{K}\mathord{\left/{\vphantom {\mathcal{K} 2}} \right.
    \kern-\nulldelimiterspace} 2}} g^{i\bar j} \left( {\nabla _{\bar j} \bar X} \right)\left( {\nabla _i \Omega } \right) + c.c. \nonumber\\
    \bar X &=& \int {F_\mu   \wedge \bar \Omega },
\end{eqnarray}
where the quantities $X$ and $\bar X$ are clearly the coefficients
of the expansion, found in the usual way by making use of:
\begin{eqnarray}
    \left\langle {\Omega }
    \mathrel{\left | {\vphantom {\Omega  {\nabla _i \Omega }}}
    \right. \kern-\nulldelimiterspace}
    {{\nabla _i \Omega }} \right\rangle  &=& \left\langle {\Omega }
    \mathrel{\left | {\vphantom {\Omega  {\nabla _{\bar i} \bar \Omega
    }}} \right. \kern-\nulldelimiterspace}
    {{\nabla _{\bar i} \bar \Omega }} \right\rangle  = 0 \nonumber \\
    \left\langle {\Omega }\mathrel{\left | {\vphantom {\Omega  {\bar \Omega }}}
    \right. \kern-\nulldelimiterspace}{{\bar \Omega }} \right\rangle  &=&  - ie^{ - \mathcal{K}}  \nonumber\\
    \left\langle {{\nabla _i \Omega }} \mathrel{\left | {\vphantom {{\nabla _i \Omega } {\nabla _{\bar j} \bar \Omega }}}
    \right. \kern-\nulldelimiterspace}
    {{\nabla _{\bar j} \bar \Omega }} \right\rangle  &=& iG_{i\bar j} e^{ -
    \mathcal{K}}.
\end{eqnarray}

The field strength form $F_\mu$ then becomes:
\begin{eqnarray}
    F_\mu &=& i\sqrt 2 \left[ {M_I ( {\partial _\mu  \zeta ^I }
    ) + L^I ( {\partial _\mu  \tilde \zeta _I } )}
    \right]\bar \Omega  \nonumber \\
    &-& i\sqrt 2 g^{i\bar j} \left[ {h_{iI} (
    {\partial _\mu  \zeta ^I } ) + f_i^I ( {\partial _\mu
    \tilde \zeta _I } )} \right]\left( {\nabla _{\bar j} \bar
    \Omega } \right) + c.c. \label{Fmu}
\end{eqnarray}
where the quantities $(L,M)$ are the components of the symplectic
section defined by (\ref{symvec}), and $(h,f)$ are their $U(1)$
K\"{a}hler covariant derivatives (\ref{covderiv1}).

Putting everything together, we find that we can write the
resulting equations as follows:

The gravitini equations:
\begin{eqnarray}
    \delta \psi _\mu ^A  &=& \nabla_\mu  \epsilon^A + \left[ {\mathcal{Q}_\mu  } \right]_{\;\;B}^A \epsilon ^B  \nonumber\\
    \left[ {\mathcal{Q}_\mu  } \right] &=& \left[ {\begin{array}{*{20}c}
    {\frac{1}{4}\left( {v_\mu   - \bar v_\mu   - R_\mu  } \right)} & { - \bar
    u_\mu     }  \\
    {u_\mu     } & { - \frac{1}{4}\left( {v_\mu   - \bar v_\mu   - R_\mu  } \right)}
    \\
    \end{array}} \right] \nonumber \\ \label{gravitinotrans}
\end{eqnarray}
where the indices $A$ and $B$ run over $(1,2)$, and
\begin{eqnarray}
    u_\mu   &=&  - ie^{\frac{\sigma }{2}} \left[ {M_I ( {\partial _\mu
    \zeta^I } ) + L^I ( {\partial _\mu  \tilde \zeta _I } )}
    \right] \nonumber \\
    \bar u_\mu   &=& + ie^{\frac{\sigma }{2}} \left[ {\bar M_I ( {\partial
    _\mu \zeta ^I } ) + \bar L^I ( {\partial _\mu  \tilde \zeta _I }
    )} \right] \nonumber \\
    v_\mu   &=& \frac{1}{2}\left( {\partial _\mu  \sigma } \right)
    + \frac{i}{2}e^\sigma  \left[ {\left( {\partial _\mu  a} \right) - K_\mu  } \right]\nonumber \\
    \bar v_\mu   &=& \frac{1}{2}\left( {\partial _\mu  \sigma } \right) -
    \frac{i}{2}e^\sigma  \left[ {\left( {\partial _\mu  a} \right) - K_\mu  }
    \right],\label{eqns5}
\end{eqnarray}
where we have rewritten the 3-form field in terms of its dual; the
universal axion $a$, and $K_\mu$ is defined by (\ref{KandL}). The
fields are grouped together in this manner for a reason that will
become clear shortly. The quantity $R_\mu$ is short for:
\begin{equation}
    R_\mu   = \frac{{\bar Z^I N_{IJ} \left( {\partial _\mu  Z^J } \right) -
    Z^I N_{IJ} \left( {\partial _\mu  \bar Z^J } \right)}}{{\bar Z^I N_{IJ} Z^J
    }},
\end{equation}
where
\begin{equation}
    N_{IJ}  = {\mathop{\rm Im}\nolimits} \left( {\frac{{\partial F_I
    }}{{\partial Z^J }}} \right),
\end{equation}
encoding the dependence of $F_I$ on $Z^I$.

The matrix $\mathcal{Q}$ is the $Sp(1)$ connection of the
quaternionic manifold described by the action\footnote{Recall that
a quaternionic manifold has $Sp(h_{2,1}+1)\otimes Sp(1)$
holonomy.}. One can also see that the quantity $R_\mu$ may be
identified as the K\"{a}hler $U(1)$ connection defined by
(\ref{U1connection}), having used the explicit definition of the
special K\"{a}hler potential (\ref{pot}). One can derive
$\mathcal{Q}$ based on these arguments alone with no reference to
the higher dimensional theory, as was done in \cite{Ferrara4} for
the four dimensional case.

The hyperini equations are:
\begin{eqnarray}
    \delta \xi _1^I  = e_{\;\;\mu} ^{1I} \Gamma ^\mu  \epsilon _1  - \bar e_{\;\;\mu}^{2I}
    \Gamma ^\mu  \epsilon _2  \nonumber \\
    \delta \xi _2^I  = e_{\;\;\mu} ^{2I} \Gamma ^\mu  \epsilon _1  + \bar e_{\;\;\mu}^{1I}
    \Gamma ^\mu  \epsilon _2, \label{hyperinotrans}
\end{eqnarray}
written in terms of the beins:
\begin{eqnarray}
    e_{\;\;\mu} ^{1I}  &=& \left( {\begin{array}{*{20}c}
   {u_\mu  }  \\
   {E_{\;\;\mu} ^{\hat i} }  \\
    \end{array}} \right) \nonumber \\\nonumber \\ e_{\;\;\mu} ^{2I}  &=& \left(
    {\begin{array}{*{20}c}
   {v_\mu  }  \\
   {e_{\;\;\mu} ^{\hat i} }  \\
    \end{array}} \right)
\end{eqnarray}
\begin{eqnarray}
    E_{\;\;\mu} ^{\hat i}  &=&  - ie^{\frac{\sigma }{2}} e^{\hat ij} \left[ {h_{jI} \left(
    {\partial _\mu  \zeta ^I } \right) + f_j^I \left( {\partial _\mu  \tilde \zeta _I }
    \right)} \right] \nonumber \\
    \bar E_{\;\;\mu} ^{\hat i}  &=&  + ie^{\frac{\sigma }{2}} e^{\hat i\bar j} \left[ {h_{\bar
    jI}     \left( {\partial _\mu  \zeta ^I } \right) + f_{\bar j}^I \left( {\partial
    _\mu      \tilde \zeta _I } \right)} \right],
\end{eqnarray}
and the beins of the special K\"{a}hler metric:
\begin{eqnarray}
    e_{\;\;\mu} ^{\hat i}  &=& e_{\;\;j}^{\hat i} \left( {\partial _\mu  z^j } \right)\quad,\quad \quad
    \quad \bar e_{\;\;\mu} ^{\hat i}  = e_{\;\;{\bar j}}^{\hat i} \left( {\partial _\mu  z^{\bar j} }
    \right) \nonumber \\
    G_{i\bar j}  &=& e_{\;\;i}^{\hat k} e_{\;\;{\bar j}}^{\hat l} \delta _{\hat k\hat l}.
\end{eqnarray}

Now, we have intentionally written everything in terms of the
vielbein one forms $e_{\;\;\mu} ^{1I}$, in order to show that we
can write the quaternionic veilbein:
\begin{equation}
V^{\Gamma A}  = \left( {\begin{array}{*{20}c}
   {e_{\;\;\mu} ^{1I} }  \\
   {\bar e_{\;\;\mu} ^{1I} }  \\
   { - e_{\;\;\mu} ^{2I} }  \\
   {\bar e_{\;\;\mu} ^{2I} }  \\
\end{array}} \right),\quad \quad \Gamma  = 1, \ldots ,2h_{2,1} ,\quad A = 1,2
\end{equation}
such that the action (\ref{d5theory}) may be written in the
compact form:
\begin{eqnarray}
    S &=& 2\int {d^5 x\sqrt {-g} \left[ {u_\mu  \bar u_\mu   + v_\mu  \bar v_\mu   +
    \sum\limits_{\hat i} {\left( {e_{\;\;\mu} ^{\hat i} \bar e_{\;\;\mu} ^{\hat i}  + E_{\;\;\mu} ^{\hat
    i}     \bar E_{\;\;\mu} ^{\hat i} } \right)} } \right]}  \nonumber \\
    &=&  2\int {d^5 x\sqrt {-g} \sum\limits_I {\left( {e_{\;\;\mu} ^{1I} \bar e_{\;\;\mu} ^{1I}  +
    e_{\;\;\mu} ^{2I} \bar e_{\;\;\mu} ^{2I} } \right)} } \nonumber \\
    &=& 2\int {d^5 x\sqrt {-g} } \sum\limits_{\scriptstyle A = 1,2 \hfill \atop
    \scriptstyle \Gamma  = 1, \ldots ,2h_{2,1}  \hfill} {\left( {V^{\Gamma A} \bar
    V^{\Gamma     A} } \right)},
\end{eqnarray}
which, we note, is exactly the quaternionic structure discussed
briefly in (\ref{N2D5theory}) if we identify:
\begin{equation}
    V \cdot \bar V \leftrightarrow \left( {\nabla q} \right) \cdot \left( {\nabla q}
    \right),
\end{equation}
and recognize that the real components of the quaternionic degrees
of freedom $q_u$ are linear combinations of the hypermultiplet
scalar fields. This explicitly shows that the special geometric
structure we get is indeed quaternionic, in clear demonstration of
the c-map, as discussed in \S\ref{cmap}.

\section{Symmetries of the theory and the equations of motion}\label{eomfull}

We have discussed how the hypermultiplets define a quaternionic
manifold. There are three isometries of this type of manifold
\cite{Proeyen3}, which physically (i.e. from the Lagrangian's
point of view) correspond to the invariance of the action under
the following infinitesimal shifts:
\begin{eqnarray}
    \zeta _I  \to \zeta _I  &+& \varepsilon _I ,\nonumber\\ \tilde \zeta ^I  \to \tilde \zeta ^I
    &+&     \tilde \varepsilon ^I ,\nonumber\\ a \to a &+& \delta  + \tilde \varepsilon ^I \zeta _I
    -     \varepsilon _I \tilde \zeta ^I,
\end{eqnarray}
where $\varepsilon$, $\tilde \varepsilon$ and $\delta$ are
infinitesimal parameters. Note that these are symmetries of the
dual action, after taking $A_{\mu\nu\rho}\rightarrow a$, since
this is the case where the action can be shown to be quaternionic.

The equations of motion of $\zeta _I$, $\tilde \zeta ^I$ and $a$
are simply the conservation laws of the Noether currents $J_\mu
^I$, $\tilde J_\mu ^I$ and $J_\mu ^5$ associated with these
symmetries. The corresponding charges are:
\begin{eqnarray}
    Q^I_2  &=& \oint {d\Sigma  ^\mu      J_\mu ^I }   \nonumber \\
    \tilde Q^I_2  &=& \oint {d\Sigma ^\mu  \tilde J_\mu ^I }  \nonumber \\
    Q^5  &=& \oint {d\Sigma ^\mu  J_\mu ^5 }, \label{charges1}
\end{eqnarray}
respectively, where $d\Sigma ^\mu$ is an element of the Gaussian
`surface' surrounding the charge. In our case; a 2-brane in five
dimensions, this surface will simply be $S^1$. The charge $Q^5$
corresponds to the wrapping of membranes over the Calabi-Yau
manifold\footnote{Instead of calling it $Q_2$, as would perhaps
seem more appropriate, we choose to keep the notation like that of
\cite{GS1,GS2} for convenience. In their case, the situation is
reversed because they studied the Euclidean dual theory.}. The
other two charges correspond to M5-brane wrappings over the CY. We
note that these charges transform under the given shifts as
follows:
\begin{eqnarray}
    Q^I_2  &\to& Q^I_2  + \tilde \varepsilon ^I Q^5, \nonumber\\ \tilde Q^I_2  &\to& \tilde Q^I_2  -
    \varepsilon ^I Q^5, \nonumber\\ Q^5  &\to& Q^5.
\end{eqnarray}

Now, using the variational principle as usual, we derive the
equations of motion for $\sigma$, $F_{\mu \nu \rho \lambda}$,
$(z,\bar z )$ and $( {\zeta ,\tilde \zeta } )$. These are:
\begin{eqnarray}
    \nabla ^2 \sigma  + e^\sigma  L_\mu ^\mu + \frac{1}{{24}}e^{ - 2\sigma } F_{\mu
    \nu     \rho \sigma }   F^{\mu     \nu \rho \sigma } &=& 0 \label{dilatoneom} \\
    \nabla ^\mu  \left( {e^{ - 2\sigma }  F_{\mu \rho \sigma \lambda}  +
    \bar \varepsilon _{\mu \rho \sigma \lambda \nu} K^\nu  } \right) &=& 0
    \label{Feomgeneral}     \\
    \nabla ^2 z^i  + \Gamma _{jk}^i \left( {\partial _\alpha  z^j } \right)\left(
    {\partial     ^\alpha  z^k } \right) + e^\sigma  \left( {\partial ^i L_\mu ^\mu  }
    \right) &=& 0 \nonumber \\
    \nabla ^2 z^{\bar i}  + \Gamma _{\bar j\bar k}^{\bar i} ( {\partial _\alpha
    z^{\bar j} } )( {\partial ^\alpha  z^{\bar k} } ) + e^\sigma
    (     {\partial ^{\bar i} L_\mu ^\mu  } ) &=& 0
    \label{zeom} \\
    \nabla ^\mu  \left[ {e^\sigma  \left( {\gamma  + \gamma ^{ - 1}
    \theta ^2 } \right)\left( {\partial _\mu  \zeta } \right) + \gamma
    ^{ - 1} \theta e^\sigma  ( {\partial _\mu  \tilde \zeta }
    )}\right. &-& \left.{ \frac{{ 1 }}{{48}}\bar \varepsilon _{\mu \nu \rho \sigma
    \alpha } F^{\mu \nu \rho \sigma } \tilde \zeta } \right] \nonumber\\&=& \frac{1}{{48}}\bar \varepsilon _{\mu\nu \rho \sigma \alpha
    } F^{\mu \nu \rho \sigma } ( {\partial ^\alpha \tilde
    \zeta } ) \nonumber \\
    \nabla ^\mu  \left[ {e^\sigma  \gamma ^{ - 1} \theta \left(
    {\partial _\mu  \zeta } \right) + e^\sigma  \gamma ^{ - 1} (
    {\partial _\mu  \tilde \zeta } )}\right. &+& \left.{ \frac{{e^\sigma
    }}{{48}}\bar \varepsilon _{\mu \nu \rho \sigma \alpha } F^{\mu \nu
    \rho \sigma } \zeta } \right] \nonumber\\&=&  - \frac{1}{{48}}
    \bar \varepsilon _{\mu \nu \rho \sigma \alpha } F^{\mu \nu \rho
    \sigma} \left( {\partial ^\alpha  \zeta } \right),
    \label{xieom}
\end{eqnarray}
with $K^\mu$ and $L_{\mu\nu}$ defined by (\ref{KandL}).

The explicit form of the currents is given by \cite{GS2}:
\begin{eqnarray}
    J_\mu   &=& 2e^\sigma  \left[ {\left( {\gamma  + \gamma ^{ - 1} \theta ^2 } \right)\left( {\partial _\mu  \zeta } \right) + \gamma ^{ - 1} \theta ( {\partial _\mu  \tilde \zeta } )} \right] + e^{2\sigma } \left[ {\left( {\partial _\mu  a} \right) - K_\mu  } \right]\tilde \zeta  \nonumber\\
    \tilde J_\mu   &=& 2e^\sigma  \gamma ^{ - 1} \left[ {\theta \left( {\partial _\mu  \zeta } \right) + ( {\partial _\mu  \tilde \zeta } )} \right] - e^{2\sigma } \left[ {\left( {\partial _\mu  a} \right) - K_\mu  } \right]\zeta  \nonumber\\
    J_\mu ^5  &=& e^{2\sigma } \left[ {\left( {\partial _\mu  a} \right) - K_\mu  }
    \right].
\end{eqnarray}

\section{\label{ExUniv}Exciting the universal hypermultiplet only}

It has been noted in passing that it is possible to find solutions
where only the universal hypermultiplet is excited
\cite{Aspinwall}, setting the rest of the hypermultiplets to zero.
From the compactification point of view, this may be viewed as the
special case ($h_{2,1}=0$). The only hypermultiplet is hence the
set ($a$, $\sigma$, $\zeta^0$, $\tilde \zeta_0$) with their
hyperino superpartners. The universal hypermultiplet is
independent of the details of the structure of the CY 3-fold
(although, globally, it is sensitive to its volume). A consequence
of this is that only the zeroth components of the symplectic
vectors and matrices, so prominent in the full theory, will
feature in this case, contributing constant complex numbers, the
choice of which is a matter of convention, as long as the rules of
symplectic covariance, like (\ref{computational}), are satisfied.
For example, we choose ${\mathcal{N}}_{00}  = -i$, $L^0 = {i
\mathord{\left/ {\vphantom {i {\sqrt 2 }}} \right.
\kern-\nulldelimiterspace} {\sqrt 2 }}$, and $\bar L^0  = {{ - i}
\mathord{\left/ {\vphantom {{ - i} {\sqrt 2 }}} \right.
 \kern-\nulldelimiterspace} {\sqrt 2 }}$. Consequently, we find that $M_0  = \bar
M_0  = L^0 {\mathcal{N}}_{00}  = \bar L^0 \bar {\mathcal{N}}_{00}
= {1 \mathord{\left/ {\vphantom {1 {\sqrt 2 }}} \right.
\kern-\nulldelimiterspace} {\sqrt 2 }}$ and the inner product
(\ref{innerprod}) is satisfied\footnote{Meaning that we  have set
${\mathop{\rm Re}\nolimits} \left( {{\mathcal{N}}_{00} } \right) =
\theta  = 0$ and ${\mathop{\rm Im}\nolimits} \left(
{{\mathcal{N}}_{00} } \right) = -\gamma  = -1$. This simply
amounts to a choice of symplectic gauge and only means that other
actions based on different choices will be related to this one by
a symplectic rotation that has no effect on the physics.}.

Again, following the notation of \cite{GS1}, we choose to redefine
the axions:
\begin{equation}
    \chi  = \zeta ^0  - i\tilde \zeta _0 \quad \quad ,\quad \quad \bar \chi  =  \zeta^0
    +    i\tilde \zeta _0. \label{defchi}
\end{equation}

Based on this notation, we find that the definitions (\ref{KandL})
become
\begin{eqnarray}
    K_\alpha  \left( {\chi ,\bar \chi } \right) &=& \frac{i}{2}\left[ {\chi \left(
    {\partial_\alpha  \bar \chi } \right) - \bar \chi \left( {\partial_\alpha  \chi
    }     \right)} \right] \nonumber \\
    L_{\mu\nu}  \left( {\chi ,\bar \chi } \right) &=& - \left( {\partial _\mu  \chi }
    \right)\left( {\partial_\nu  \bar \chi } \right),
\end{eqnarray}
and the bosonic action reduces to:
\begin{eqnarray}
    S &=& \frac{1}{{2\kappa _5^2 }}\int {d^5 x\sqrt {-g} \left\{ { - R + \frac{1}{2}\left( {\partial
    _\mu   \sigma}   \right)\left( {\partial ^\mu  \sigma } \right) + \frac{1}{{48}}e^{ - 2\sigma }
    F_{\mu     \nu \rho \sigma } F^{\mu \nu \rho \sigma } } \right.}  \nonumber \\
    &+& \left. {{e^\sigma  }\left( {\partial _\mu  \chi } \right)\left(
    {\partial     ^\mu  \bar \chi } \right) + \frac{i}{{48}}\bar\varepsilon _{\mu \nu
    \rho  \sigma \alpha } F^{\mu \nu \rho \sigma } \left[ {\chi \left( {\partial ^\alpha
    \bar  \chi } \right) - \bar \chi \left( {\partial ^\alpha  \chi } \right)}
    \right]}    \right\}.
\end{eqnarray}

Again, we note that the action is invariant under the following
infinitesimal shifts:
\begin{equation}
    \chi  \to \chi  + \varepsilon ,\quad \bar \chi  \to \bar \chi  + \varepsilon ,\quad
    a \to a + \delta  + \frac{i}{4}\left( {\chi \bar \varepsilon  - \bar \chi
    \varepsilon } \right),
\end{equation}
where $\varepsilon$ and $\delta$ are complex and real
infinitesimal parameters respectively, and the arguments of
\S\ref{eomfull} remain valid \cite{GS1}.

The equations of motion of $\sigma$, $F_{\mu \nu \rho \sigma}$ and
$(\chi, \bar \chi)$ are, respectively:
\begin{eqnarray}
    \nabla ^2 \sigma  -  \frac{1}{2}e^\sigma  \left( {\partial _\mu  \chi } \right)\left(
    {\partial     ^\mu  \bar \chi } \right)+ \frac{1}{{24}}e^{ - 2\sigma } F_{\mu \nu
    \rho     \sigma } F^{\mu \nu \rho \sigma }  &=& 0\label{sigmaeom} \\
    \nabla ^\mu  \left( {e^{ - 2\sigma } F_{\mu \nu \rho \sigma}  +
    \bar\varepsilon_{\mu \nu \rho \sigma \alpha} K^\alpha  } \right) &=& 0
    \label{Feom} \\
    \nabla ^\mu  \left[ {e^\sigma  \left( {\partial _\mu   \chi }
    \right) + \frac{i}{{48}}\bar \varepsilon _{\mu \nu \rho \sigma
    \alpha } F^{\nu \rho \sigma \alpha }  \chi } \right] &=& -\frac{i}{{48}}\bar \varepsilon _{\mu \nu \rho \sigma \alpha }
    F^{\mu \nu \rho \sigma } \left( {\partial ^\alpha   \chi }
    \right) \nonumber \\
    \nabla ^\mu  \left[ {e^\sigma  \left( {\partial _\mu  \bar \chi }
    \right) - \frac{i}{{48}}\bar \varepsilon _{\mu \nu \rho \sigma
    \alpha } F^{\nu \rho \sigma \alpha } \bar \chi } \right] &=& +\frac{i}{{48}}\bar \varepsilon _{\mu \nu \rho \sigma \alpha }
    F^{\mu \nu \rho \sigma } \left( {\partial ^\alpha  \bar \chi }
    \right) \nonumber \\ \label{chieom}
\end{eqnarray}

The full action is invariant under a set of supersymmetry
transformations that may be found by directly considering
(\ref{gravitinotrans}) and (\ref{hyperinotrans}), yielding:
\begin{eqnarray}
    \delta \psi _\mu ^1  &=&  \nabla _\mu\epsilon _1  + i \frac{e^{-\sigma}}{{96}} \varepsilon _{\mu
    \nu  \rho \sigma \lambda } F^{\nu \rho \sigma \lambda }  \epsilon
    _1   - \frac{e^{\frac{\sigma }{2}}}{{\sqrt 2 }}\left( {\partial _\mu  \chi } \right)\epsilon _2 \nonumber \\
    \delta \psi _\mu ^2  &=&  \nabla _\mu\epsilon _2  - i\frac{e^{-\sigma}}{{96}} \varepsilon _{\mu
    \nu  \rho \sigma \lambda } F^{\nu \rho \sigma \lambda }  \epsilon
    _2 + \frac{e^{\frac{\sigma }{2}}}{{\sqrt 2 }}\left( {\partial _\mu  \bar\chi } \right)\epsilon _1  \nonumber \\
    \delta \xi _1  &=& \frac{1}{2}\left( {\partial _\mu  \sigma } \right)\Gamma
    ^\mu      \epsilon _1 - \frac{i}{{48}}e^{ - \sigma } \varepsilon _{\mu \nu \rho \sigma \lambda
    }     F^{\mu \nu \rho \sigma } \Gamma ^\lambda  \epsilon _1  +
    \frac{e^{\frac{\sigma }{2}}}{{\sqrt 2 }} \left( {\partial _\mu  \chi } \right)\Gamma
    ^\mu  \epsilon _2  \nonumber \\
    \delta \xi _2  &=& \frac{1}{2}\left( {\partial _\mu  \sigma } \right)\Gamma
    ^\mu      \epsilon _2 + \frac{i}{{48}}e^{ - \sigma } \varepsilon _{\mu \nu \rho \sigma \lambda
    }     F^{\mu \nu \rho \sigma } \Gamma ^\lambda  \epsilon _2  -
    \frac{e^{\frac{\sigma }{2}}}{{\sqrt 2 }} \left( {\partial _\mu  \bar\chi } \right)\Gamma
    ^\mu  \epsilon _1. \label{hyperSUSY}
\end{eqnarray}

\chapter{The five dimensional solutions}\label{5Dsolutions}

We present our 2-brane solutions coupled to the hypermultiplets in
$\mathcal{N}=2$ $D=5$ supergravity. Our objective is the most
general such solution coupled to the full set of hypermultiplets,
which may be used to deduce the general $D=11$ SLAG calibrated
solution. We will, however, begin by finding special case
solutions first and discuss their geometrical interpretation as
wrapped M-branes.

The chapter is structured as follows: In section \ref{Einstein} we
derive the details of the $D=5$ Einstein equation: the components
of the Einstein tensor, as well as the stress tensor. In
\S\ref{univhypersection} we find a 2-brane solution coupled to the
Universal hypermultiplet with vanishing $Q^5$ charge, i.e. the
vanishing of the 3-form gauge field. This solution is interpreted
as the wrapping of a M5-brane over SLAG cycles of a Calabi-Yau
manifold with $h_{2,1}=0$. It corresponds to the special case SLAG
solution found in eleven dimensions in chapter (\ref{FS}). In
\S\ref{Q2equalzero} we find a 2-brane solution coupled to the
Universal hypermultiplet with vanishing $Q_2$ and $\tilde Q_2$
charges, i.e. the vanishing of the axial fields. This has the
interpretation of a reduced M2-brane. In \S\ref{wraped1} and
\ref{11M5to52B} we discuss in more detail the interpretation of
the above solutions as wrapped M-branes. In \S\ref{logHcomment} we
comment on the so called `high brane' behavior of five dimensional
membranes. In \S\ref{general2brane} we present the general 2-brane
solution coupled to the full set of hypermultiplets with vanishing
3-form gauge potential. We study how the solution is completely
satisfied by representing the harmonic functions in terms of the
symplectic sections, as was proposed by Sabra in his work on black
holes coupled to the vector multiplets. In \S\ref{moregeneral},
and for the sake of completeness, we study some of the effects
resulting from switching on the gauge field flux. Finally, in
\S\ref{back}, we review a recently published result proposing a
$D=11$ solution representing the wrapping of a M5-brane over a
SLAG calibrated submanifold. This should dimensionally reduce to
our general 2-brane solution in \S\ref{general2brane}. We argue
that, in fact, it doesn't, and briefly discuss the possibility of
a correction.

The 2-brane ansatz we will use throughout is:
\begin{eqnarray}
    ds^2  &=& e^{2A\sigma } \eta _{ab} dx^a dx^b  + e^{2B\sigma } \delta _{\mu \nu }
    dx^\mu  dx^\nu,\nonumber\\  {a,b}  &=& 0,1,2,\quad\quad
    {\mu ,\nu }  = 3,4,\label{generic}
\end{eqnarray}
where the numbers $A$ and $B$ are to be fixed for each particular
situation. Also, note the change of notation, the Greek indices
now describe the directions transverse to the brane only. In what
follows, we will also redefine the capital Latin indices to be
$M,N=0,\ldots 4$.

\section{\label{Einstein}Einstein's equation}

First, we derive the components of Einstein's equation
$G_{MN}=R_{MN} - \frac{1}{2}g_{MN} R = 8\pi T_{MN}$, for the
metric (\ref{generic}) and the general action (\ref{d5theory}). As
usual, we assume that the fields are dependent only on the
directions transverse to the brane. We later find that the
universal hypermultiplet fields can be expanded in terms of a
single function $H$ harmonic in the transverse directions, i.e $
\delta^{\mu\nu} \left(\partial_\mu
\partial_\nu H \right)=0$. We write all terms of the Einstein equation in terms of this
function for easy reference. For the dilaton, we write $\sigma  =
P\ln H $, where $P$ is a constant to be fixed. This applies to all
solutions, except the most general case, where there will
generally be more than one harmonic function. Hence, we may
rewrite the metric as:
\begin{equation}
    ds^2  = H^{\left( {2AP} \right)} \eta _{ab} dx^a dx^b  + H^{\left( {2BP} \right)} \delta
    _{\alpha \beta }  dx^\alpha  dx^\beta.
\end{equation}

The Ricci tensor is then:
\begin{eqnarray}
    R_{ab}  &=&  - A g_{ab}\left[ {\left( {\partial^\mu  \partial _\mu  \sigma } \right) + 3A \left( {\partial^\mu  \sigma } \right)\left( {\partial _\mu  \sigma } \right)} \right], \nonumber \\
    R_{\mu \nu }  &=& - B g_{\mu \nu } \left( {\partial^\alpha  \partial _\alpha  \sigma } \right) - 3A\left( {\partial _\mu  \partial _\nu  \sigma } \right) \nonumber\\
    & & -3AB g_{\mu \nu } \left( {\partial^\alpha  \sigma } \right)\left( {\partial _\alpha  \sigma } \right) + \left( {6AB - 3A^2 } \right)\left( {\partial _\mu  \sigma } \right)\left( {\partial _\nu  \sigma }
    \right),\nonumber \\
    R_{ab}  &=&  AP\left( {1 - 3AP} \right)g_{ab} \left( {\partial^\mu  \ln H} \right)\left( {\partial _\mu  \ln H} \right), \nonumber \\
    R_{\mu \nu }  &=&  BP\left( {1 - 3AP} \right) g_{\mu \nu } \left( {\partial^\alpha  \ln H} \right)\left( {\partial _\alpha  \ln H} \right) \nonumber\\
    & & +3AP\left( {1 - AP + 2BP} \right)\left( {\partial_\mu  \ln H} \right)\left( {\partial _\nu  \ln H} \right).
\end{eqnarray}

Also:
\begin{equation}
    g^{MN}G_{MN}= g^{MN} \left( {R_{MN}  - \frac{1}{2}g_{MN} R} \right)=-
    \frac{3}{2}R,
\end{equation}
where the Ricci scalar curvature is:
\begin{eqnarray}
 R &=&  - {\left( {6A + 2B} \right)\left( {\partial^\mu  \partial_\mu  \sigma } \right) - 12A^2 \left( {\partial^\mu  \sigma } \right)\left( {\partial_\mu  \sigma } \right)} \nonumber\\
   &=& 2P\left( {3A + B - 6A^2 P} \right)\left( {\partial^\mu  \ln H} \right)\left( {\partial _\mu  \ln H}
   \right).
\end{eqnarray}

The Einstein tensor $G_{MN}$ is:
\begin{eqnarray}
    G_{ab}  &=& g_{ab} \left[ {\left( {2A + B} \right)\left( {\partial ^\mu  \partial _\mu  \sigma } \right) + 3A^2 \left( {\partial ^\mu  \sigma } \right)\left( {\partial _\mu  \sigma } \right)} \right] \nonumber\\
    &=& g_{ab} \left[ {3\left( {AP} \right)^2  - 2\left( {AP} \right) - \left( {BP} \right)} \right]\left( {\partial ^\mu  \ln H} \right)\left( {\partial _\mu  \ln H} \right) \nonumber
\end{eqnarray}
\begin{eqnarray}
    G_{\mu \nu }  &=& 3Ag_{\mu \nu } \left( {\partial ^\alpha  \partial _\alpha  \sigma } \right) + \left( {6A^2  - 3AB} \right)g_{\mu \nu } \left( {\partial ^\alpha  \sigma } \right)\left( {\partial _\alpha  \sigma } \right) \nonumber\\
    & & -3A\left( {\partial _\mu  \partial _\nu  \sigma } \right) + \left( {6AB - 3A^2 } \right)\left( {\partial _\mu  \sigma } \right)\left( {\partial _\nu  \sigma } \right) \nonumber\\
    &=& g_{\mu \nu } \left[ {6\left( {AP} \right)^2  - 3\left( {AP} \right)\left( {BP} \right) - 3\left( {AP} \right)} \right]\left( {\partial ^\alpha  \ln H} \right)\left( {\partial _\alpha  \ln H} \right) \nonumber\\
    & &+ \left[ {6\left( {AP} \right)\left( {BP} \right) - 3\left( {AP} \right)^2  + 3\left( {AP} \right)} \right]\left( {\partial _\mu  \ln H} \right)\left( {\partial _\nu  \ln H}
    \right).
\end{eqnarray}

The energy momentum tensor is found to be:
\begin{eqnarray}
    \left( {8\pi } \right)T_{MN}  &=&  \frac{1}{{\sqrt { - g} }}\frac{{\delta
    S}}{{\delta g^{MN} }} \nonumber \\
    \left( {8\pi } \right)T_{ab} &=&  - \frac{1}{4}g_{ab} \left( {\partial _\alpha  \sigma }
    \right)\left(     {\partial ^\alpha  \sigma } \right) - \frac{1}{2}g_{ab } G_{i\bar j} \left( {\partial _\alpha  z^i }
    \right)\left(     {\partial ^\alpha  z^{\bar j} } \right)  \nonumber\\
    &+& \frac{1}{{12}}e^{ -
    2\sigma     } F_{LMN a} {F^{LMN }} _b-\frac{1}{{96}}e^{ - 2\sigma } g_{ab} F_{LMNP } F^{LMNP }\nonumber \\
    &+& \frac{1}{6}\bar\varepsilon _{LMN a \alpha} {F^{LMN }} _b  K^\alpha   - \frac{1}{{48}}g_{ab}
    \bar\varepsilon _{LMNP \alpha} F^{LMNP } K^\alpha +\frac{1}{2}e^\sigma  g_{ab } L_\alpha ^\alpha \nonumber \\
    \left( {8\pi } \right)T_{\mu \nu}  &=& \frac{1}{2}\left( {\partial _\mu  \sigma } \right)\left( {\partial _\nu  \sigma
    }     \right) - \frac{1}{4}g_{\mu \nu } \left( {\partial _\alpha  \sigma }
    \right)\left(     {\partial ^\alpha  \sigma } \right) \nonumber \\
    &+& G_{i\bar j} \left( {\partial         _\mu  z^i } \right)\left( {\partial _\nu  z^{\bar j} } \right)- \frac{1}{2}g_{\mu \nu } G_{i\bar j} \left( {\partial _\alpha  z^i }
    \right)\left(     {\partial ^\alpha  z^{\bar j} } \right)\nonumber \\
    &+&  \frac{1}{{12}}e^{- 2\sigma     } F_{LMN \mu } {F^{LMN }} _\nu   -  \frac{1}{{96}}e^{ - 2\sigma } g_{\mu \nu } F_{LMNP }
    F^{LMNP}  \nonumber\\
    &+& \frac{1}{{24}}\bar\varepsilon _{LMNP \mu } F^{LMNP} K_\nu   + \frac{1}{6}\bar\varepsilon _{LMN \mu
    \alpha}     {F^{LMN }} _\nu  K^\alpha   \nonumber \\
    &-& \frac{1}{{48}}g_{\mu \nu } \bar\varepsilon _{LMNP \alpha} F^{LMNP } K^\alpha   - e^\sigma  L_{\mu \nu }  + \frac{1}{2}e^\sigma  g_{\mu \nu } L_\alpha ^\alpha.
\end{eqnarray}

Since the only nonvanishing components of the 4-form field
strength are $F_{\mu abc}$, we can rewrite:
\begin{eqnarray}
    \left( {8\pi } \right)T_{ab} &=&  - \frac{1}{4}g_{ab} \left( {\partial _\alpha  \sigma }
    \right)\left(     {\partial ^\alpha  \sigma } \right) - \frac{1}{2}g_{ab } G_{i\bar j} \left( {\partial _\alpha  z^i }
    \right)\left(     {\partial ^\alpha  z^{\bar j} } \right)  \nonumber\\
    &+& \frac{1}{{4}}e^{ - 2\sigma     } F_{\mu dc a} {F^{\mu dc }} _b-\frac{1}{{24}}e^{ - 2\sigma } g_{ab} F_{\mu dce } F^{\mu dce  }
    \nonumber \\
    &+& \frac{1}{2}\bar\varepsilon _{\mu dc a \rho } {F^{\mu dc }} _b  K^\rho   - \frac{1}{{12}}g_{ab}
    \bar\varepsilon _{\mu dce \rho } F^{\mu dce } K^\rho+\frac{1}{2}e^\sigma  g_{ab } L_\alpha ^\alpha \nonumber
\end{eqnarray}
\begin{eqnarray}
    \left( {8\pi } \right)T_{\mu \nu}  &=& \frac{1}{2}\left( {\partial _\mu  \sigma } \right)\left( {\partial _\nu  \sigma
    }     \right) - \frac{1}{4}g_{\mu \nu } \left( {\partial _\alpha  \sigma }
    \right)\left(     {\partial ^\alpha  \sigma } \right)  \nonumber \\
    &+& G_{i\bar j} \left(  {\partial         _\mu  z^i } \right)\left( {\partial _\nu  z^{\bar j} } \right)- \frac{1}{2}g_{\mu \nu } G_{i\bar j} \left( {\partial _\alpha  z^i }
    \right)\left(     {\partial ^\alpha  z^{\bar j} } \right) \nonumber \\
    &+& \frac{1}{{12}}e^{ -  2\sigma     } F_{abc \mu } {F^{abc }} _\nu   -   \frac{1}{{24}}e^{ - 2\sigma } g_{\mu \nu } F_{\alpha abc }
    F^{\alpha abc}  \nonumber \\
    &+&\frac{1}{{6}}\bar\varepsilon _{\alpha abc \mu } F^{\alpha abc} K_\nu   + \frac{1}{6}\bar\varepsilon _{abc \mu \rho
    }     {F^{abc }} _\nu  K^\rho \nonumber \\
    &-& \frac{1}{{12}}g_{\mu \nu } \bar\varepsilon _{\alpha abc \rho } F^{\alpha abc } K^\rho - e^\sigma  L_{\mu \nu }  + \frac{1}{2}e^\sigma  g_{\mu \nu } L_\alpha
    ^\alpha,
\end{eqnarray}
and
\begin{eqnarray}
    \left( {8\pi } \right)T_A^A  =  &-& \frac{3}{4}\left( {\partial _\mu  \sigma } \right)\left( {\partial ^\mu  \sigma } \right) - \frac{3}{2}G_{i\bar j} \left( {\partial _\mu  z^i } \right)\left( {\partial ^\mu  z^{\bar j} } \right) \nonumber\\
    &+& \frac{1}{8}F_{\mu abc} F^{\mu abc}  + \frac{5}{{12}}\bar
    \varepsilon _{\mu abc\alpha } F^{\mu abc} K^\alpha   + \frac{3}{2}e^\sigma  L_\mu^\mu.
\end{eqnarray}

\section{\label{univhypersection}The universal hypermultiplet 2-brane with $\left( {Q^5  = 0} \right)$}

We start by solving the Einstein equation along with the scalar
fields' equations of motion, then show that our solutions satisfy
supersymmetry using the SUSY variation equations. We write the
axions in terms of the harmonic function $H$ as follows:
\begin{eqnarray}
    \left( {\partial _\mu  \chi } \right) &=& ne^{i\varphi } l_\mu ^{\;\;\nu}  \left( {\partial _\nu  H^\eta  } \right) = n\eta e^{i\varphi } H^\eta  l_\mu ^{\;\;\nu}  \left( {\partial _\nu  \ln H} \right) \\
    \left( {\partial _\mu  \bar \chi } \right) &=& ne^{ - i\varphi } l_\mu ^{\;\;\nu}  \left( {\partial _\nu  H^\eta  } \right) = n\eta e^{ - i\varphi } H^\eta  l_\mu ^{\;\;\nu}  \left( {\partial _\nu  \ln H}
    \right),
\end{eqnarray}
where $n$ and $\eta$ are constants to be determined, the constant
phase is put in to make $\chi$ complex, conforming with its
definition (\ref{defchi}), and the tensor $l_\mu ^{\;\;\nu}$ is
either the Kroenicker delta $\delta_\mu ^\nu$ or the Levi-Civita
totally anti-symmetric symbol $\bar\varepsilon_\mu ^{\;\;\;\nu}$.

Plugging our ans\"{a}tze in Einstein's equation and the equations
of motion of the dilaton and axions, we find that we can write
both sides of these equations in terms of $\left( {\partial _\mu
\ln H} \right)\left( {\partial ^\mu  \ln H} \right)$ only, as the
second order terms vanish. We end up with the following algebraic
equations:
\begin{enumerate}
    \item $\left( {2\eta  + P} \right) = 0$ appears in all
    exponents.
    \item $G_{ab}  =   8\pi T_{ab} \quad  \to \quad 3\left( {AP} \right)^2  - 2\left( {AP} \right) - \left( {BP} \right) =  - \frac{{P^2 }}{4}\left( {1 + \frac{{n^2 }}{2}}
    \right)$
    \item $G_{\mu \nu }  =   8\pi T_{\mu \nu } \quad  \to \quad \begin{array}{*{20}c}
   {6\left( {AP} \right)^2  - 3\left( {AP} \right)\left( {BP} \right) - 3\left( {AP} \right) =  - \frac{{P^2 }}{4}\left( {1 + c\frac{{n^2 }}{2}} \right)}  \\
   {\frac{3}{2}\left( {AP} \right)^2  - 3\left( {AP} \right)\left( {BP} \right) - \frac{3}{2}\left( {AP} \right) =  - \frac{{P^2 }}{4}\left( {1 + c\frac{{n^2 }}{2}} \right)}  \\
\end{array}$
    \item \label{item}$g^{\mu \nu } \left( {G_{\mu \nu }  =   8\pi T_{\mu \nu } } \right)\quad  \to \quad 3\left( {AP} \right)^2  = \left( {AP} \right)$
    \item $g^{MN} \left( {G_{MN}  =  8\pi T_{MN} } \right)\quad  \to \quad 6\left( {AP} \right)^2  - 3\left( {AP} \right) - \left( {BP} \right) =  - \frac{{P^2 }}{4}\left( {1 + \frac{{n^2 }}{2}} \right)$
    \item \label{item1}The dilaton's e.o.m$\quad  \to \quad 3\left( {AP} \right) = 1 + \frac{P}{4}n^2 $
    \item The axions' e.o.m$\quad  \to \quad 3\left( {AP} \right) = 1 - \frac{P}{2}$, which is valid iff $l = \delta
    $.
\end{enumerate}
\begin{equation}\label{univhyperalgebra}
\end{equation}

The constant $c$ is either $+1$ or $-1$ depending on whether
$l=\delta$ or $\bar\varepsilon$ respectively. Now, equation
(\ref{item}) has two solutions $\left( {AP} \right) = 0$ and
$\left( {AP} \right)  = {1\mathord{\left/ {\vphantom {1 3}}
\right. \kern-\nulldelimiterspace} 3}$. Using the
${1\mathord{\left/ {\vphantom {1 3}} \right.
\kern-\nulldelimiterspace} 3}$ solution in the rest of the
equations gives $c=+1$, i.e. $l=\delta$. In this case, equation
(\ref{item1}) immediately gives $P=0$, which is a contradiction.
In other words, $\left( {AP} \right)  = {1\mathord{\left/
{\vphantom {1 3}} \right. \kern-\nulldelimiterspace} 3}$ does not
satisfy the field equations. The $\left( {AP} \right) = 0$
solution in turn, branches into two possibilities. One can see
that the equations are satisfied by either value of $c$, except
that the case $c=+1$, corresponding to $l=\delta$, gives the
result $\left( {BP} \right) = 0$. In other words, this is just the
trivial case of flat Minkowski space, expected to satisfy any
theory containing gravity. The remaining possibility is $c=-1$,
which yields:
\begin{equation}
    n^2  = 2,\quad P =  - 2,\quad \eta  = 1,\quad A = 0, \quad
    B =  - 1,\quad l =    \bar\varepsilon.
\end{equation}

The final form of the 2-brane solution coupled to the universal
hypermultiplet with $\left( {Q^5  = 0} \right)$ is:
\begin{eqnarray}
    {\bf Q}^{\bf 5} {\bf  = 0} &;& \left( {{\bf Q}_{\bf 2} {\bf ,\tilde Q}_{\bf 2} }
    \right)  {\bf \ne 0}\nonumber \\
    ds^2  &=& \eta _{ab} dx^a dx^b  + e^{-2\sigma }    \delta _{\alpha \beta } dx^\alpha  dx^\beta \nonumber \\
    &=& \eta _{ab} dx^a dx^b  + H^4  \delta _{\alpha \beta } dx^\alpha  dx^\beta  \label{hypersolQ5=0} \\
    \left( {\partial _\mu  \chi } \right) &=&   \pm \frac{{e^{ - \frac{\sigma }{2}} }}{{\sqrt 2 }}e^{i\varphi } {\bar\varepsilon_\mu}^{\;\;\;\nu}
    \left( {\partial _\nu  \sigma }\right)=\pm \sqrt 2 e^{i\varphi }{\bar\varepsilon_\mu}^{\;\;\;\nu} \left( {\partial _\nu  H }
    \right)\nonumber\\
    \left( {\partial    _\mu  \bar \chi } \right) &=&  \pm \frac{{e^{ - \frac{\sigma }{2}} }}{{\sqrt 2 }}e^{ - i\varphi } {\bar\varepsilon_\mu}^{\;\;\;\nu}
    \left( {\partial _\nu  \sigma } \right)=\pm \sqrt 2 e^{-i\varphi }{\bar\varepsilon_\mu}^{\;\;\;\nu} \left( {\partial _\nu  H }
    \right), \nonumber \\
    H &=& e^{ - {\sigma  \mathord{\left/ {\vphantom {\sigma  2}} \right. \kern-\nulldelimiterspace}
    2}} \quad, \quad \delta^{\mu\nu} \left(\partial_\mu\partial_\nu H    \right)=0.
    \label{summuniv}
\end{eqnarray}

\subsection{Supersymmetry}

We now check that this solution satisfies the supersymmetry
equations (\ref{hyperSUSY}), i.e. whether or not it allows for
supercovariantly constant spinors to exist.

The quantities needed are:
\begin{eqnarray}
    e_{\;\;b}^{\hat a}  &=& e^{A\sigma } \eta _b^{\hat a},\quad \quad \quad \quad
    e_{\;\;\nu}    ^{\hat     \mu }  = e^{B\sigma } \delta _\nu ^{\hat \mu }  \nonumber \\
    \Gamma _{b\alpha }^a  &=& A\delta _b^a \left( {\partial _\alpha  \sigma } \right)
    \nonumber \\
    \Gamma _{ab}^\alpha   &=&  - Ae^{2\left( {A - B} \right)\sigma } \eta _{ab} \delta
    ^{\alpha \beta } \left( {\partial _\beta  \sigma } \right) \nonumber \\
    \Gamma _{\nu \rho }^\mu   &=& B\left[ {\delta _\nu ^\mu  \left( {\partial _\rho  \sigma
    }     \right) + \delta _\rho ^\mu  \left( {\partial _\nu  \sigma } \right) - \delta
    _{\nu \rho } \delta ^{\mu \alpha } \left( {\partial _\alpha  \sigma } \right)}
    \right]     \nonumber \\
\end{eqnarray}
\begin{eqnarray}
    \omega _a^{\;\;\hat b\hat \alpha }  &=& Ae^{\left( {A - B} \right)\sigma } \eta _a^{\hat
    b}     \delta ^{\hat \alpha \beta } \left( {\partial _\beta  \sigma } \right)
    \nonumber     \\
    \omega _\alpha ^{\;\;\hat \beta \hat \gamma }&=& B\left( {\delta _\alpha ^{\hat \beta }
    \delta ^{\hat \gamma \rho }  - \delta ^{\hat \beta \rho } \delta _\alpha ^{\hat \gamma
    }     } \right)\left( {\partial _\rho  \sigma } \right)\nonumber \\
    \nabla _a  &=& \partial _a  + \frac{A}{2}\left( {\partial _\mu  \sigma }
    \right){\Gamma     _a} ^\mu   \nonumber \\
    \nabla _\alpha   &=& \partial _\alpha   + \frac{B}{2}\left( {\partial _\rho  \sigma
    }     \right){\Gamma _\alpha } ^\rho .
\end{eqnarray}

We also use the projection condition\footnote{The Einstein
summation convention is \emph{not} used over the index $r$.}:
\begin{eqnarray}
    \Gamma _{\hat \alpha \hat \beta } \epsilon _r  &=& b_r \bar \varepsilon _{\hat
    \alpha     \hat \beta } \epsilon _r,  \quad\quad r=(1,2),\quad b_r=\pm
    1 \nonumber \\
    \Gamma ^{\mu \nu } \epsilon _r  &=& b_r e^{ - 2B\sigma } \bar \varepsilon ^{\mu \nu
    }     \epsilon _r \nonumber \\
    \Gamma _\mu^{\;\;\;\nu} \epsilon _r  &=& b_r {\bar\varepsilon_\mu}^{\;\;\;\nu}\epsilon
    _r     \nonumber \\
    \Gamma ^\mu  \epsilon _r  &=&  - b_r {\bar\varepsilon_\nu}^{\;\;\;\mu} \Gamma
    ^\nu          \epsilon _r.\label{projection}
\end{eqnarray}

Now, considering equations (\ref{hyperSUSY}); $\left( {\delta \psi
_a^r  = 0} \right)$ vanishes identically and:
\begin{eqnarray}
    \left( {\delta \xi _r  = 0}
    \right)\quad  &\to& \quad \begin{array}{*{20}c}
       \epsilon _1  \pm b_2 e^{i\varphi } \epsilon _2  &=& 0  \nonumber\\
       \epsilon _2  \mp b_1 e^{ - i\varphi } \epsilon _1  &=& 0  \nonumber\\
    \end{array} \\
    \left( {\delta \psi _\mu ^r  = 0} \right)\quad  &\to& \quad \begin{array}{*{20}c}
       \begin{array}{l}
     \left( {\partial _\mu  \epsilon _1 } \right) + \bar \varepsilon _\mu ^{\;\;\;\nu}  \left( {\partial _\nu  \ln H} \right)\left( {b_1 \epsilon _1  \pm e^{i\varphi } \epsilon _2 } \right) =0 \\
     \left( {\partial _\mu  \epsilon _2 } \right) + \bar \varepsilon _\mu ^{\;\;\;\nu}  \left( {\partial _\nu  \ln H} \right)\left( {b_2 \epsilon _1  \mp e^{ - i\varphi } \epsilon _1 } \right) =0 \\
     \end{array}  \\
    \end{array}
\end{eqnarray}\
which can be solved by:
\begin{equation}\label{spinorshyper}
    \epsilon _1  =  \pm e^{i\varphi } \epsilon _2 \quad ,\quad
    \left( {\partial _\mu  \epsilon _1 } \right) = \left( {\partial
    _\mu  \epsilon _2 } \right) = 0.
\end{equation}

\section{\label{Q2equalzero}The universal hypermultiplet 2-brane with $( {Q_2=\tilde Q_2  = 0} )$}

We assume the following ansatz for the 4-form field strength:
\begin{equation}\label{}
    F_{\mu abc}  = \kappa \bar \varepsilon _{abc} l_\mu ^{\;\;\;\nu}  \left(
    {\partial _\nu  H^\eta  } \right) = \kappa \eta H^\eta  \bar
    \varepsilon _{abc} l_\mu ^{\;\;\;\nu}  \left( {\partial _\nu  \ln H}
    \right),
\end{equation}
and calculate the relevant terms of the stress tensor:
\begin{eqnarray}
    F_{\mu abc} F^{\mu abc}  &=&  - 6\kappa ^2 \eta ^2 H^{\left( {2\eta  - 6AP} \right)} \left( {\partial _\mu  \ln H} \right)\left( {\partial ^\mu  \ln H} \right) \nonumber\\
    F_{\mu dca} F^{\mu dc} _{\;\;\;\;\;\;b}  &=&  - 2\kappa ^2 \eta ^2 g_{ab} H^{\left( {2\eta  - 6AP} \right)} \left( {\partial _\mu  \ln H} \right)\left( {\partial ^\mu  \ln H} \right) \nonumber\\
    F_{abc\mu } F^{abc} _{\;\;\;\;\;\;\nu}   &=&  - 6\kappa ^2 \eta ^2 H^{\left( {2\eta  - 6AP} \right)} l_\mu ^{\;\;\alpha}  l_\mu ^{\;\;\beta}  \left( {\partial _\alpha  \ln H} \right)\left( {\partial _\beta  \ln H}
    \right).
\end{eqnarray}

The field equations give:
\begin{enumerate}
    \item $\eta  = \left( {3A + 1} \right)P$ appears in all exponents.
    \item $G_{ab}  =   8\pi T_{ab} \quad  \to \quad 3\left( {AP} \right)^2  - 2\left( {AP} \right) - \left( {BP} \right) =   - \frac{1}{4}\left( {P^2  + \kappa ^2 \eta ^2 } \right)$
    \item $G_{\mu \nu }  =   8\pi T_{\mu \nu } \quad  \to \quad \begin{array}{*{20}c}
   {6\left( {AP} \right)^2  - 3\left( {AP} \right)\left( {BP} \right) - 3\left( {AP} \right) =   - \frac{1}{4}\left( {P^2  - c\kappa ^2 \eta ^2 } \right)}  \\
   {\frac{3}{2}\left( {AP} \right)^2  - 3\left( {AP} \right)\left( {BP} \right) - \frac{3}{2}\left( {AP} \right) =  - \frac{1}{4}\left( {P^2  - c\kappa ^2 \eta ^2 } \right)}  \\
    \end{array}$
    \item $g^{\mu \nu } \left( {G_{\mu \nu }  =   8\pi T_{\mu \nu } } \right)\quad  \to \quad 3\left( {AP} \right)^2  = \left( {AP} \right)$
    \item $g^{MN} \left( {G_{MN}  =   8\pi T_{MN} } \right)\quad  \to \quad 6\left( {AP} \right)^2  - 3\left( {AP} \right) - \left( {BP} \right) =  - \frac{1}{4}\left( {P^2  + \kappa ^2 \eta ^2 } \right)$
    \item The dilaton's e.o.m$\quad  \to \quad 3\left( {AP} \right)P = P  + \kappa ^2 \eta ^2  $
    \item The axions' e.o.m$\quad  \to \quad P=-1$, which is valid iff $l = \delta$.
\end{enumerate}
\begin{equation}\label{}
\end{equation}
Again, the constant $c$ is either $+1$ if $l=\delta$ or $-1$ if
$l=\bar\varepsilon$.

The solution $\left( {AP} \right) = {1 \mathord{\left/ {\vphantom
{1 3}} \right. \kern-\nulldelimiterspace} 3}$ results in
$\kappa^2=0$, and is hence not a solution to the theory. The
solution $\left( {AP} \right) = 0$, as before, results into two
possibilities; one being the trivial case of flat space $\left(
{BP} \right) = 0$, in which case, $l=\bar\varepsilon$. The other
is a 2-brane coupled to a 3-form field, giving:
\begin{eqnarray}
    \kappa^2  &=& 1,\quad P =  - 1,\nonumber\\ \eta  &=& -1,\quad \left( {AP} \right) = A = 0, \\
    \left( {BP} \right) &=& {1 \mathord{\left/ {\vphantom {1 2}} \right.
 \kern-\nulldelimiterspace} 2},\quad B =  - {1 \mathord{\left/ {\vphantom {1 2}} \right.
 \kern-\nulldelimiterspace} 2},\quad l = \delta,
\end{eqnarray}
and the complete solution becomes:
\begin{eqnarray}
    {\bf Q}^{\bf 5} {\bf \ne 0} &;& \left( {{\bf Q}_{\bf 2} {\bf ,\tilde Q}_{\bf 2} }
    \right)   {\bf = 0} \nonumber \\
    ds^2  &=& \eta _{ab} dx^a dx^b  + e^{-\sigma }    \delta _{\alpha \beta } dx^\alpha  dx^\beta \nonumber \\
    &=& \eta _{ab} dx^a dx^b  + H  \delta _{\alpha \beta } dx^\alpha  dx^\beta \nonumber \\
    A_{abc}  &=&  \pm e^\sigma  \bar \varepsilon _{abc} =  \pm H^{ - 1}
    \bar \varepsilon _{abc} \nonumber\\
    F_{\mu abc}  &=&  \pm e^\sigma  \bar \varepsilon _{abc} \left(
    {\partial _\mu  \sigma } \right) =  \mp H^{ - 1} \bar \varepsilon
    _{abc} \left( {\partial _\mu  \ln H} \right) \nonumber \\
    H &=& e^{-\sigma}, \quad\quad \delta^{\mu\nu} \left(\partial_\mu\partial_\nu H    \right)=0.
    \label{Chiequalzero}
\end{eqnarray}

\subsection{Supersymmetry}

We plug into (\ref{hyperSUSY}) to find that both $\left( {\delta
\xi _r  = 0} \right)$ and $\left( {\delta \psi _a^r  = 0} \right)$
are satisfied exactly, where this time, because of the lack of
terms mixing the spinors, we get no conditions defining one spinor
in terms of the other as we did in (\ref{spinorshyper}). Also,
$\left( {\delta \psi _\mu ^r  = 0} \right)$ is trivially satisfied
for $\left( {\partial _\mu  \epsilon _1 } \right)=\left( {\partial
_\mu  \epsilon _2 } \right)=0$.

We will now proceed to discuss the microscopic interpretation of
these solutions in terms of M-branes wrapped over a $T^6$. This
will explicitly demonstrate the relationship between the $\left(
{Q^5 = 0}\right)$ solution with the SLAG calibrated eleven
dimensional solution we found before. Note that the torus $T^6$ is
a degenerate case of a Calabi-Yau 3-fold, preserving all of
supersymmetry.

\section{\label{wraped1}($D=11$) M2-brane $\rightsquigarrow$ ($D=5$) 2-brane}

The M2-brane in eleven dimensions coupled to a three form gauge
field looks like:
\begin{eqnarray}
    ds_{11}^2  &=& H^{ - \frac{2}{3}} dx_{\parallel}^2  + H^{\frac{1}{3}} dx_{\perp}^2
    \nonumber \\
    dx_{\parallel}^2  &=& \left( { - dt^2  + dx_1^2  + dx_2^2 }
    \right),\nonumber \\
    dx_{\perp}^2  &=& \left( {dx_3^2  +  \cdots  + dx_{10}^2 } \right)
    \nonumber \\
    A_{012}  &=&   \pm H^{ - 1} , \nonumber \\ \left( {\nabla _ \bot ^2 H} \right)
    &=&    0.
\end{eqnarray}

We dimensionally reduce this over a $T^6$ in order to demonstrate
the geometrical interpretation of the solution of
\S\ref{Q2equalzero} as a reduced M2-brane. We extract the five
dimensional metric from the above eleven dimensional one by
keeping the 2-brane part (first term in the metric) as it is,
separating six of the eight transverse dimensions, assuming them
compact and contracting to a point. Explicitly:
\begin{eqnarray}
    ds_{11}^2  &=& H^{ - \frac{2}{3}} dx_{\parallel }^2  + H^{\frac{1}{3}}
    dx_\bot ^2  \nonumber\\
    &=& H^{ - \frac{2}{3}} dx_{\parallel }^2  + H^{\frac{1}{3}} \underbrace
    {\left( {dx_3^2  +  \cdots  + dx_8^2 } \right)}_{T^6} +
    H^{\frac{1}{3}}     \left( {dx_9^2  + dx_{10}^2 } \right) \nonumber \\
    &=& \left[ {H^{ - \frac{2}{3}} \left( { - dt^2  + dx_1^2  + dx_2^2 }
    \right) + H^{\frac{1}{3}} \left( {dx_3^2  + dx_4^2 } \right)} \right]+H^{\frac{1}{3}} ds_{T^6}^2 \nonumber \\
    &=& H^{ - \frac{2}{3}} \underbrace {\left( {\eta _{ab} dx^a dx^b  + H\delta _{\alpha \beta }
    dx^\alpha  dx^\beta  } \right)}_{{\rm
    The}\;(5D)\;{\rm metric}} + H^{\frac{1}{3}} ds_{T^6}^2.
    \label{construct}
\end{eqnarray}

Comparing with (\ref{expmet}), we see that $H=e^{-\sigma}$. So the
$D=5$ metric is:
\begin{eqnarray}
    ds^2  &=& {\eta _{ab} dx^a dx^b  + e^{ - \sigma } \delta _{\alpha \beta } dx^\alpha
    dx^\beta  } \nonumber \\
    &=& {\eta _{ab} dx^a dx^b  + H \delta _{\alpha \beta } dx^\alpha
    dx^\beta  },
\end{eqnarray}
coupled to a 3-form field $A_{abc}  =   \pm H^{ - 1}    \bar
\varepsilon _{abc}$, which drops down to five dimensions as is.
This is exactly the case (\ref{Chiequalzero}) and clearly
demonstrates its interpretation as a \emph{reduced} (as opposed to
\emph{wrapped}) M2-brane.

\section{\label{11M5to52B}($D=11$) M5-brane $\rightsquigarrow$ ($D=5$) 2-brane}

A M5-brane in eleven dimensions looks like:
\begin{eqnarray}
    ds_{11}^2  &=& H^{ - \frac{1}{3}} dx_{\parallel}^2  + H^{\frac{2}{3}} dx_{\perp}^2
    \nonumber \\
    dx_{\parallel}^2  &=& \left( { - dt^2  +  \cdots  + dx_5^2  } \right),\quad \quad
    \quad         dx_{\perp}^2  = \left( {dx_6^2  +  \cdots  + dx_{10}^2 } \right)
    \nonumber \\
    A_{012345}  &=&  \pm H^{ - 1}.\label{M5}
\end{eqnarray}

We proceed to show that wrapping this over a $T^6$ yields the
solution of \S\ref{univhypersection}. Separate the terms of the
metric into a 2-brane term, a transverse term and two terms that
will be wrapped on the torus:
\begin{eqnarray}
    ds_{11}^2  &=& H^{ - \frac{1}{3}} dx_{\parallel}^2  + H^{\frac{2}{3}} dx_ \bot ^2  +
    H^{ - \frac{1}{3}} \left( {dx_5^2  + dx_6^2  + dx_7^2 } \right) + H^{\frac{2}{3}}
    \left(     {dx_8^2  + dx_9^2  + dx_{10}^2 } \right) \nonumber \\
    dx_{\parallel}^2  &=& \left( { - dt^2  + dx_1^2  + dx_2^2 } \right) ,\quad\quad dx_
    \bot     ^2  = \left( {dx_3^2  + dx_4^2 } \right),
\end{eqnarray}
then, we pair
\begin{equation}\
    z_1  = x_5  + ix_8 ,\quad\quad z_2  = x_6  + ix_9 ,\quad\quad z_3  = x_7  + ix_{10}.
\end{equation}

Since the warp factors for the (5,6,7) and (8,9,10) directions are
not the same, we do not expect to be able to apply the exact same
technique used in the M2-brane case. Instead, we do the following:
Since we expect this solution to be equivalent to the $D=11$ SLAG
calibrated solution we found earlier (\ref{specialSLAGmetric}), we
consider the SLAG calibrating form:
\begin{eqnarray}
    \phi = {\mathop{\rm Re}\nolimits} \left( {dz_1  \wedge dz_2  \wedge dz_3 } \right)
    &=& {\mathop{\rm Re}\nolimits} \left[ {\left( {dx_5  + idx_8 } \right) \wedge
    \left( {dx_6  + idx_9 } \right) \wedge \left( {dx_7  + idx_{10} } \right)} \right] \nonumber
    \\
    &=& dx_5  \wedge dx_6  \wedge dx_7  - dx_5  \wedge dx_9  \wedge dx_{10}
    \nonumber     \\
    &+& dx_6      \wedge     dx_8  \wedge dx_{10}  - dx_7  \wedge
    dx_8 \wedge dx_9. \label{ReOmega}
\end{eqnarray}

To understand this, recall that $A\sim \phi$, which is simply the
real part of the holomorphic 3-form. Also note that the complex
structure moduli are constant. Other examples of M5-brane
configurations wrapping 3-cycles would deform the complex
structure of the torus.

Now, (\ref{ReOmega}) is analogous to the following configuration
of branes:
\begin{equation}\label{diagram}
\begin{array}{*{20}c}
   t & 1 & 2 &\vline &  \times & \times &\vline &  5 & 6 & 7 &  \times  &  \times  &  \times   \\
   t & 1 & 2 &\vline &  \times & \times &\vline &  5 &  \times  &  \times  &  \times  & 9 & {10}  \\
   t & 1 & 2 &\vline &  \times & \times &\vline &   \times  & 6 &  \times  & 8 &  \times  & {10}  \\
   t & 1 & 2 &\vline &  \times & \times &\vline &   \times  &  \times  & 7 & 8 & 9 &  \times   \\
\end{array}
\end{equation}
giving a metric:
\begin{eqnarray}
    ds_{11}^2  &=& H^{ - \frac{1}{3} \times 4} dx_{\parallel}^2  + H^{\frac{2}{3} \times
    4} dx_ \bot     ^2  + H^{ - \frac{1}{3} \times 2} H^{\frac{2}{3} \times 2} \left(
    {dx_5^2  +  \cdots  + dx_{10}^2 } \right) \nonumber \\
    &=& H^{ - \frac{4}{3}} dx_{||}^2  + H^{\frac{8}{3}} dx_ \bot ^2  + H^{\frac{2}{3}}
    \underbrace {\left( {dx_5^2  +  \cdots  + dx_{10}^2 } \right)}_{T^6}
    \nonumber \\
    &=& H^{ - \frac{4}{3}}\underbrace { \left( {dx_{||}^2 +  H^4 dx_ \bot ^2 }
    \right)}_{{\rm The}\;{\rm (5D)}\;{\rm metric}} + H^{\frac{2}{3}} ds_{CY}^2.
\end{eqnarray}

Note that the middle form of the above equation looks exactly like
the SLAG solution (\ref{specialSLAGmetric}), which is expected
since we began this construction by assuming that an equivalence
exists.

Now, comparing with (\ref{expmet}), we see that $H$ corresponds to
$e^{ - \frac{\sigma }{2}} $, so we rewrite the metric as follows:
\begin{eqnarray}
    ds^2  &=& e^{\frac{2}{3}\sigma } \eta _{ab} dx^a dx^b  + e^{ - \frac{4}{3}\sigma }
    \delta     _{\alpha \beta } dx^\alpha  dx^\beta \nonumber \\
    &=& e^{\frac{2}{3}\sigma } \left( {\eta _{ab} dx^a dx^b  + e^{ - 2\sigma }
    \delta         _{\alpha \beta } dx^\alpha  dx^\beta  } \right).
\end{eqnarray}

To compare with the $D=5$ result, we use (\ref{expmet}) as before.
The five dimensional metric becomes:
\begin{eqnarray}
    ds^2  &=& {\eta _{ab} dx^a dx^b  + e^{- 2 \sigma } \delta _{\alpha \beta } dx^\alpha
    dx^\beta  }\nonumber\\
    &=& {\eta _{ab} dx^a dx^b  + H^4 \delta _{\alpha \beta } dx^\alpha
    dx^\beta  }.\label{UnivHyperSol}
\end{eqnarray}

This is exactly the metric in (\ref{hypersolQ5=0}). We now look at
the way the gauge field wraps down to five dimensions. Writing the
6-form field coupled to the M5-brane in (\ref{M5}) in complex
coordinates yields the 6-form field coupled to the SLAG calibrated
intersecting M5-branes we found before. Explicitly we get:
\begin{equation}\label{}
\begin{array}{*{20}c}
   A_{abcmnp}  = \pm H^{ - 1} \bar \varepsilon _{abc}\bar \varepsilon _{mnp}, \hfill & \quad A_{abc\bar m\bar n\bar p}  = \pm H^{ - 1} \bar \varepsilon _{abc}\bar \varepsilon _{\bar m\bar n\bar p}  \hfill  \\
   F_{\mu abcmnp}  =  \mp H^{ - 2} \left( {\partial _\mu  H} \right)\bar \varepsilon _{abc}\bar \varepsilon
    _{mnp}, \hfill & \quad F_{\mu abc\bar m\bar n\bar p}  =  \mp H^{ - 2} \left( {\partial _\mu  H} \right)\bar \varepsilon _{abc}\bar \varepsilon
    _{\bar m\bar n\bar p}. \hfill  \\
\end{array}
\end{equation}

Since the five dimensional theory was derived using the 4-form
field strength, then in order to compare we derive the Hodge dual
of the result
\begin{equation}\label{wrappedF}
    F_{\mu mnp}  =  \mp \bar \varepsilon _\mu ^{\;\;\;\nu } \left(
    {\partial _\nu  H} \right)\Omega _{mnp} ,\quad\quad F_{\mu \bar m\bar n\bar p}  =  \mp \bar \varepsilon _\mu ^{\;\;\;\nu } \left(
    {\partial _\nu  H} \right)\Omega _{\bar m\bar n\bar p}
\end{equation}
where $\Omega$ is the unique (3,0) form on the CY, which in the
SLAG solution we identified as $\Omega _{mnp}=E \bar
\varepsilon_{mnp}$. Now, using (\ref{threeform}) and (\ref{FdA1})
with $h_{2,1}=0$, we find that the results (\ref{wrappedF}) and
(\ref{summuniv}) match exactly if $e^{i\varphi } \left( {\alpha  +
i\beta } \right) = \Omega $. Using the choice for $L^0$ and $M_0$
we have used before in \S\ref{ExUniv}, we get a trivial constraint
on the value of the constant phase $e^{i\varphi } $ in terms of
the special K\"{a}hler potential which is also a constant in this
case.

This confirms our interpretation of (\ref{hypersolQ5=0},
\ref{summuniv}) as a partially wrapped $M5$-brane, and shows that
the $D=11$ SLAG calibrated solution we found in
\S\ref{SpecialSLAG} corresponds to a 2-brane in five dimensions,
coupled to the universal hypermultiplet with $\left( {Q^5  = 0}
\right)$. The full hypermultiplets solution will be a
generalization of this case, corresponding to a generalization in
eleven dimensions also.

\section{\label{logHcomment}A general comment on $D=5$ membrane solutions}

Let us take a closer look at the properties of a 2-brane in five
dimensions. As we have seen, it is possible to express the dilaton
in terms of a function\footnote{As we will see, we will need to
introduce more harmonic functions for the most general case with
$h_{2,1}>0$, but the discussion in this section will still
basically hold.} $H$ harmonic in the transverse directions, i.e.
$\delta ^{\mu \nu } \left( {\partial _\mu
\partial _\nu  H} \right) = 0$. For a single 2-brane solution, the
harmonic function is thus
\begin{equation}
H\left( {\bf x} \right) = h + q \ln \left| {{\bf x} - {\bf x}_0 }
\right|,
\end{equation}
where ${\bf x}_0 $ represents the position vector of the brane in
transverse space, and $(h,q)$ are constants. In spherical
coordinates centered about ${\bf x}_0 $, this is just $H(r)= h + q
\ln r$.

A familiar analogy is the standard elementary electrostatic
problem of finding the electric potential of a straight,
infinitely long, wire with a homogenous distribution of electric
charge. For a single wire, with 2-dimensional transverse space,
this yields a logarithmic potential.

As an example, let us consider the $\left( {Q^5  = 0} \right)$
solution (\ref{summuniv}). The 2-brane metric looks like:
\begin{equation}
    ds^2  =\eta _{ab} dx^a dx^b  + \left( {h + q\ln r} \right)^4\delta _{\alpha \beta
    } dx^\alpha  dx^\beta.
\end{equation}

Now using the coordinate transformation of derivatives $
r^2\partial _\mu   = r x_\mu \partial _r  - \bar \varepsilon _\mu
^{\;\;\;\nu}  x_\nu\partial
    _\phi$, we find $    \left( {\partial _\mu  H} \right) = x_\mu  {q \mathord{\left/
 {\vphantom {q {r^2 }}} \right.
 \kern-\nulldelimiterspace} {r^2 }}$. This results in:
\begin{eqnarray}
    \left( {\frac{{d\chi }}{{dr}}} \right) &=& \left( {\frac{{d\bar \chi }}{{dr}}} \right) = 0 \nonumber\\
    \left( {\frac{{d\chi }}{{d\phi }}} \right) &=&  \pm \sqrt 2e^{i\varphi } q\quad \quad,\quad \quad \chi  = \chi _0  \pm \sqrt 2 e^{i\varphi } q\phi \nonumber\\
     \left( {\frac{{d\bar \chi }}{{d\phi }}} \right) &=&  \pm \sqrt 2e^{ - i\varphi } q \quad \quad
    ,\quad \quad \bar \chi  = \bar  \chi _0  \pm \sqrt 2 e^{ - i\varphi } q\phi,
\end{eqnarray}
where $\chi _0$ and $\bar \chi _0$ are integration constants. The
solutions describe a `spiral' behavior for the axial fields.

The fact that these, so called `high branes', are not flat at
infinity is interesting and has been studied in various sources,
such as \cite{Panda}. However, as far as our objectives are
concerned, this curious property does not seem to be an
impediment.

\section{\label{general2brane}The general 2-brane solution}

Armed with the experience of the special cases discussed so far,
we now turn to our final objective: the most general 2-brane
solution in the hypermultiplets sector. As remarked earlier,
$\mathcal{N}=2$ $D=5$ solitonic solutions coupled with the
hypermultiplets are quite rare in the literature. On the other
hand, there is an abundance of solutions coupled to the vector
multiplets. $\mathcal{N}=2$ $D=4,5$ black holes coupled to vector
multiplets, for example, have been extensively studied. Due to the
c-map, as discussed before, our solution contains a lot of
similarities to these solutions. See, in particular, the work by
Sabra \emph{et al}
\cite{Sabra1,Sabra2,Sabra9,Sabra8,Sabra4,Sabra7,Sabra6,Sabra5,Sabra3}
and others \cite{Behrndt2,Behrndt1,Ferrara7,Fre2,Kastor}.

We are interested in the case of a vanishing M2-brane charge, i.e.
$Q^5=0$. Later, however, we will briefly discuss the effects of
switching the membrane charge on.

Our ansatz for the metric will be (\ref{generic}) as usual. As for
the scalar fields, we expect that the most general case will be an
expansion in terms of more that one harmonic function, in fact
$2\left({h_{2,1}+1}\right)$ of them. In the previous simple cases,
we looked at situations with either $h_{2,1}=0$ or $Q_2=0$, and we
were able to assume the existence of one type of `electric' charge
$q$, and hence a single function $H=\left( {h+q \ln r} \right)$.
We now need to generalize this by introducing a number
$\left({h_{2,1}+1}\right)$ electric charges $q_I$ and a similar
number of magnetic charges $\tilde q^I$, corresponding to $H_I$
and $\tilde H^I$ harmonic functions associated with each homology
cycle $\in H^3$, hence\footnote{Note that both $(q_I,\tilde q^J)$
and $(H_I,\tilde H^J)$ are symplectic vectors.}:
\begin{equation}\label{harmonicHs}
    H_I  = h_I  + q_I \ln r\quad ,\quad \tilde H^I  = \tilde h^I  + \tilde q^I \ln
    r , \quad I = 0, \ldots ,h_{2,1},
\end{equation}
based on the similar four and five dimensional BH solutions cited
above, as well as the magnetically dual instanton result of
\cite{GS2}.

We find that the ansatz:
\begin{eqnarray}
    \left( {\partial _\mu  \zeta ^I } \right) &=& n e^{a\sigma } {\bar\varepsilon_\mu} ^{\;\;\;\nu}  ( {\partial _\nu
    \tilde H^I }  ) \nonumber \\
    ( {\partial _\mu  \tilde \zeta _I } ) &=& n e^{a\sigma } {\bar\varepsilon_\mu} ^{\;\;\;\nu}
    \left( {\partial _\nu  H_I } \right).
\end{eqnarray}
satisfies the axion equations (\ref{xieom}) exactly. Also, the
Bianchi identity $\bar \varepsilon ^{\mu \nu } (\partial _\mu
\partial _\nu  \zeta ) = 0$ yields the harmonic conditions $\delta^{\mu\nu}(\partial_\mu\partial_\nu H_I)=0$ and $\delta^{\mu\nu}(\partial_\mu\partial_\nu \tilde H^I)=0$, provided that $a=0$, which will be confirmed later by other means.

\subsection{An initial look at supersymmetry}

Due to the expected complexity of these solutions, we make no
ans\"{a}tze concerning the dilaton or the moduli, but rather look
at the SUSY equations first to get clues about their form. The
vanishing hyperino transformations (\ref{hyperinotrans}) give:
\begin{eqnarray}
    \frac{1}{2}\left( {\partial _\mu  \sigma } \right)\Gamma ^\mu  \epsilon _2  +
    e^{\left( {a + {1 \mathord{\left/ {\vphantom {1 2}} \right.
    \kern-\nulldelimiterspace} 2}} \right)\sigma } {\bar\varepsilon_\mu}^{\;\;\;\nu} \left[
    {L^I     \left( {\partial _\nu  H_I } \right) - M_I ( {\partial _\nu  \tilde H^I
    }     )} \right]\Gamma ^\mu  \epsilon _1  &=& 0 \nonumber \\
    \frac{1}{2}\left( {\partial _\mu  \sigma } \right)\Gamma ^\mu  \epsilon _1  -
    e^{\left( {a + {1 \mathord{\left/ {\vphantom {1 2}} \right.
    \kern-\nulldelimiterspace} 2}} \right)\sigma } {\bar\varepsilon_\mu}^{\;\;\;\nu} \left[
    {\bar     L^I \left( {\partial _\nu  H_I } \right) - \bar M_I ( {\partial
    _\nu          \tilde H^I } )} \right]\Gamma ^\mu  \epsilon _2  &=& 0
    \label{hyper1} \\
    ( {\partial _\mu  z^i } )\Gamma ^\mu  \epsilon _1  - e^{\left( {a +
    {1         \mathord{\left/ {\vphantom {1 2}} \right.
    \kern-\nulldelimiterspace} 2}} \right)\sigma } {\bar\varepsilon_\mu}^{\;\;\;\nu}
    G^{i\bar
    j}     \left[ {f_{\bar j}^I \left( {\partial _\nu  H_I } \right) - h_{\bar jI}
    (     {\partial _\nu  \tilde H^I } )} \right]\Gamma ^\mu  \epsilon _2 &=&
    0     \nonumber \\
    ( {\partial _\mu  z^{\bar i} } )\Gamma ^\mu  \epsilon _2  + e^{\left( {a
    +     {1 \mathord{\left/ {\vphantom {1 2}} \right.
    \kern-\nulldelimiterspace} 2}} \right)\sigma } {\bar\varepsilon_\mu}^{\;\;\;\nu}
    G^{\bar
    ij}     \left[ {f_j^I \left( {\partial _\nu  H_I } \right) - h_{jI} (
    {\partial         _\nu  \tilde H^I } )} \right]\Gamma ^\mu  \epsilon _1  &=&
    0,     \label{hyper2}
\end{eqnarray}
implying $\epsilon _1   = \pm \epsilon_2$. Equations
$\left({\delta \psi^r _a=0}\right)$ give $A=0$ as usual, and
equations $\left( {\delta \psi _\mu ^r = 0} \right) $ now give:
\begin{eqnarray}
    \left( {\partial _\mu  \epsilon _1 } \right) - \frac{R_\mu}{4}  \epsilon _1
    +  \frac{b_1 B}{2}\left( {\partial _\nu  \sigma } \right)\bar \varepsilon _\mu ^{\;\;\;\nu}  \epsilon _1  + e^{\left( {a + {1 \mathord{\left/ {\vphantom {1 2}} \right.
    \kern-\nulldelimiterspace} 2}} \right)\sigma } {\bar\varepsilon_\mu}^{\;\;\;\nu} \left[
    {\bar     L^I \left( {\partial _\nu  H_I } \right) - \bar M_I ( {\partial
    _\nu          \tilde H^I } )} \right]\Gamma ^\mu  \epsilon
    _2&=&0 \nonumber\\
    \left( {\partial _\mu  \epsilon _2 } \right) + \frac{R_\mu}{4}  \epsilon
    _2 +  \frac{b_2 B}{2}\left( {\partial _\nu  \sigma } \right)\bar \varepsilon _\mu ^{\;\;\;\nu}  \epsilon _1  + e^{\left( {a + {1 \mathord{\left/ {\vphantom {1 2}} \right.
    \kern-\nulldelimiterspace} 2}} \right)\sigma } {\bar\varepsilon_\mu}^{\;\;\;\nu} \left[
    {\bar     L^I \left( {\partial _\nu  H_I } \right) - \bar M_I ( {\partial
    _\nu          \tilde H^I } )} \right]\Gamma ^\mu
    \epsilon_1&=&0. \nonumber \\\label{gravitino}
\end{eqnarray}

The general form of these equations allows us to construct the
ans\"{a}tze:
\begin{eqnarray}
    \left( {\partial _\mu  \sigma } \right) &=&  \kappa e^{ c \sigma } \left[
    {L^I     \left( {\partial _\mu  H_I } \right) - M_I ( {\partial _\mu  \tilde H^I
    }     )} \right] \nonumber \\
    \left( {\partial _\mu  z^i } \right) &=& r e^{ m\sigma } G^{i\bar j}
    \left[     {f_{\bar j}^I \left( {\partial _\nu  H_I } \right) - h_{\bar jI} (
    {\partial _\nu  \tilde H^I } )} \right] \nonumber \\
    ( {\partial _\mu  z^{\bar i} } ) &=& r e^{ m\sigma } G^{\bar
    ij} \left[ {f_j^I \left( {\partial _\mu  H_I } \right) - h_{jI} ( {\partial _\mu
    \tilde H^I } )} \right], \label{moduliansatz}
\end{eqnarray}
where $\kappa$, $r$, $m$ and $c$ are constants to be fixed.

\subsection{The field equations}

We now go back to the equations of motion. The dilaton equation
(\ref{dilatoneom}) requires calculation of the tensor
$L_{\mu\nu}$. We find that to be:
\begin{eqnarray}
    L_{\mu \nu }  &=& n^2 e^{2a\sigma } \bar\varepsilon_\mu^{\;\;\;\alpha} \bar\varepsilon_\nu^{\;\;\;\beta} \left[ { - \gamma ^{ - 1} \left( {\partial _\alpha  H} \right)\left( {\partial _\beta  H} \right) + \left( {\gamma  + \gamma ^{ - 1} \theta ^2 } \right)( {\partial _\alpha  \tilde H} )( {\partial _\beta  \tilde H} )} \right.\nonumber\\
    &+&\left. { \gamma ^{ - 1} \theta ( {\partial _\alpha  \tilde H} )\left( {\partial _\beta  H} \right) + \gamma ^{ - 1} \theta \left( {\partial _\alpha  H} \right)( {\partial _\beta  \tilde H} )}
    \right].
\end{eqnarray}

The $\left( {\nabla ^2 \sigma } \right)$ term requires that we
calculate $\left({\partial^\mu \partial_\mu \sigma }\right)$,
which in turn, demands knowledge of how to compute the quantities
$\left({\partial_\mu L^I}\right)$ and $\left({\partial_\mu
M_I}\right)$. Since the symplectic sections are functions in the
moduli, which are functions in the transverse spatial coordinates,
we can use the chain rule to find:
\begin{eqnarray}
    \left( {\partial _\mu  L^I } \right) &=&  f_i^I \left( {\partial _\mu  z^i } \right) \nonumber \\
    \left( {\partial _\mu  M_I } \right) &=&  h_{iI} \left( {\partial _\mu  z^i } \right) .
\end{eqnarray}

The dilaton equation becomes:
\begin{eqnarray}
    \left( {\nabla ^2 \sigma } \right) &=& \kappa e^{c\sigma } \left[ {f_i^I \left( {\partial _\mu  z^i } \right) + f_{\bar i}^I ( {\partial _\mu  z^{\bar i} } )} \right]\left( {\partial ^\mu  H_I } \right) + \kappa e^{c\sigma } \left[ {h_{iI} \left( {\partial _\mu  z^i } \right) + h_{\bar iI} ( {\partial _\mu  z^{\bar i} } )} \right]( {\partial ^\mu  \tilde H^I } ) \nonumber\\
    &+& c\kappa ^2 e^{c2\sigma } \left[ {L^I \bar L^J \left( {\partial _\mu  H_I } \right)\left( {\partial ^\mu  H_J } \right) + M_I \bar M_J ( {\partial _\mu  \tilde H^I } )( {\partial ^\mu  \tilde H^J } ) + } \right. \nonumber\\
    &+& \left. {\left( {L^I \bar M_J  + \bar L^J M_I } \right)\left( {\partial _\mu  H_I } \right)( {\partial ^\mu  \tilde H^J } )} \right] = e^\sigma  L_\mu
    ^\mu.
\end{eqnarray}

We now use (\ref{useful}) to evaluate the $\gamma$ and $\theta$
quantities inside $L_\mu^\mu$. We find, in matrix notation:
\begin{eqnarray}
    \gamma ^{ - 1}  &=& 2\left( {G^{i\bar j} f_i f_{\bar j}  + \left| L \right|^2 } \right) \nonumber\\
    \left( {\gamma  + \gamma ^{ - 1} \theta ^2 } \right) &=& 2\left( {G^{i\bar j} h_i h_{\bar j}  + \left| M \right|^2 } \right) \nonumber\\
    \gamma ^{ - 1} \theta  &=& 2\left( {L\bar M + \bar L M}
    \right).
\end{eqnarray}

Plugging in the ans\"{a}tze (\ref{moduliansatz}), cancellations
occur and we end up with the algebraic relations:
\begin{eqnarray}
    \left( {2a + 1} \right) &=& 2c\nonumber \\ &=& c + m \nonumber \\
    2n^2  &=& c\kappa ^2 \nonumber \\ &=& r\kappa,
\end{eqnarray}
where we have used $A=0$ from the SUSY equations.

Analysis of the Einstein equation follows similar arguments,
ending with:
\begin{enumerate}
    \item $\left( {2a  + 1} \right) = 2c$ appears in all
    exponents.
    \item $G_{ab}  =  8\pi T_{ab} \quad  \to \quad B =  - \frac{1}{{4c}} - \frac{1}{2}$
    \item $G_{\mu \nu }  =  8\pi T_{\mu \nu } \quad  \to \quad - \frac{1}{{4c}} + \frac{1}{2}=0$
    \item $g^{MN} \left( {G_{MN}  =  8\pi T_{MN} } \right)\quad  \to \quad B =  - \frac{1}{{4c}} -
    \frac{1}{2}$.
\end{enumerate}
\begin{equation}\label{fullhyperalgebra}
\end{equation}
All the above relations are simultaneously satisfied by:
\begin{eqnarray}
    c &=& m = {1 \mathord{\left/{\vphantom {1 2}} \right.
    \kern-\nulldelimiterspace} 2},\quad n =  \pm 1,\quad \kappa  =  - 2,\quad r =  - 1, \nonumber\\
    A &=& a = 0,\quad B =  - 1.
\end{eqnarray}

Finally, we check that the moduli equations (\ref{zeom}) are
satisfied. For this purpose, we write:
\begin{eqnarray}
    \left( {\nabla ^2 z^i } \right) &=& \left( {\partial _\mu  \partial ^\mu  z^i } \right) - 2\left( {\partial _\mu  \sigma } \right)\left( {\partial ^\mu  z^i } \right) \nonumber\\
    \Gamma _{\bar j\bar k}^{\bar i}  &=& G^{\bar il} \left( {\partial _{\bar j} G_{\bar kl} }
    \right),
\end{eqnarray}
and plug in the expressions for $\left( {\partial _\mu  \sigma }
\right)$ and $\left( {\partial ^\mu  z^i } \right)$. Evaluating
$\left( {\partial^i L^\mu_\mu} \right)$ and $\left( {\partial _\mu
\partial ^\mu  z^i } \right)$ requires using the chain rule again,
but this time on quantities like $\left( {\partial _\mu  f_i^I }
\right)$ which in turn requires the use of (\ref{useful2}). Term
by term cancellations occur and the moduli equations are
satisfied.

Hence, the form of the general solution is, so far:
\begin{eqnarray}
ds^2  &=& \eta _{ab} dx^a dx^b  + e^{ - 2\sigma } \delta _{\mu \nu
} dx^\mu  dx^\nu \label{D11SLAGmetric}\\
    \left( {\partial _\mu  \sigma } \right) &=&  - 2e^{ \frac{\sigma}{2} } \left[
    {L^I     \left( {\partial _\mu  H_I } \right) - M_I ( {\partial _\mu  \tilde H^I
    }     )} \right] \nonumber \\
    &=&  - 2e^{  \frac{\sigma}{2} } \left[ {\bar L^I \left( {\partial _\mu  H_I }
    \right)     - \bar M_I ( {\partial _\mu  \tilde H^I } )}
    \right]\\
    \left( {\partial _\mu  z^i } \right) &=& - e^{ \frac{\sigma}{2} } G^{i\bar j}
    \left[     {f_{\bar j}^I \left( {\partial _\nu  H_I } \right) - h_{\bar jI} (
    {\partial _\nu  \tilde H^I } )} \right] \nonumber \\
    ( {\partial _\mu  z^{\bar i} } ) &=& -e^{ \frac{\sigma}{2} } G^{\bar
    ij}    \left[ {f_j^I \left( {\partial _\mu  H_I } \right) - h_{jI} ( {\partial _\mu
    \tilde H^I } )} \right]\\
    \left( {\partial _\mu  \zeta ^I } \right) &=& \pm {\bar\varepsilon_\mu} ^{\;\;\;\nu}  ( {\partial _\nu
    \tilde H^I }  ) \nonumber \\
    ( {\partial _\mu  \tilde \zeta _I } ) &=& \pm {\bar\varepsilon_\mu} ^{\;\;\;\nu}
    \left( {\partial _\nu  H_I } \right). \label{zetaeom}
\end{eqnarray}

At this point, we run a final check on supersymmetry. Plugging in
the results we have so far in equations (\ref{hyper1}),
(\ref{hyper2}) and (\ref{gravitino}) with the
usual projections immediately gives $\epsilon_1=\pm\epsilon_2$ and%
\begin{equation}\label{}
    \left( {\partial _\mu\epsilon _1 } \right) - \frac{1}{4}R_\mu  \epsilon _1  =
    \left( {\partial _\mu  \epsilon _2 } \right) + \frac{1}{4}R_\mu
    \epsilon _2  = 0.
\end{equation}

This last seems to indicate that $\left( {\partial _\mu\epsilon _1
} \right)=\left( {\partial _\mu\epsilon _2 } \right)=0$ and
$R_\mu=0$, i.e.
\begin{equation}\label{grandsolution}
    \bar Z^I N_{IJ} \left( {\partial _\mu  Z^J } \right) -
    Z^I N_{IJ} \left( {\partial _\mu  \bar Z^J } \right)=0,
\end{equation}
implying the vanishing of the $U(1)$ K\"{a}hler connection
(\ref{U1connection}):
\begin{equation}\label{}
    \left( {\partial _i \mathcal{K}} \right)dz^i  - \left( {\partial _{\bar i}
    \mathcal{K}} \right)dz^{\bar i}  = 0.
\end{equation}

\subsection{Further detail}

To pin down the solution even further, we follow the argument
initiated by Sabra in the case of black hole solutions coupled to
the vector multiplets (see \cite{Sabra1} as an example). The
symplectic invariance of the theory was used to write the harmonic
functions $(H_I, \tilde H^J)$ in terms of the symplectic sections,
more specifically the periods of the 3-form $\Omega$.

We write everything explicitly in terms of the transverse
directions in polar coordinates $\left( {r,\varphi } \right)$,
based on the harmonic functions $H$ and $\tilde H$. In addition,
we recognize the charges $q$ and $\tilde q$ as a c-mapped version
of the electric and magnetic charges defined by (\ref{charges}).
We have hinted in \S\ref{EMdual} at a relationship between these
and the central charge of the theory, which we will discuss now.

As reviewed in \S\ref{SUGRA}, the supersymmetry algebra of any
SUGRA theory has the form (\ref{SUSYalgebra}). The operators
$Z_{\mu \nu \cdots}$ can be diagonalized to yield what are known
as the central charges of the theory, which correspond to,
basically, the charges the solitonic solutions, i.e. the
$p$-branes, carry. For $\mathcal{N}=2$ $D=5$ SUGRA, the central
charge $Z$ carried by the 2-brane is a symplectically invariant
linear combination of the electric and magnetic charges as follows
\cite{WittenOlive}:
\begin{eqnarray}
    Z &=& \left( {L^I q_I  - M_I \tilde q^I } \right) \nonumber \\
    \bar Z &=& \left( {\bar L^I q_I  - \bar M_I \tilde q^I } \right),
\end{eqnarray}
which can be inverted to:
\begin{eqnarray}
    q_I  &=& i\left( {\bar ZM_I  - Z\bar M_I } \right) \nonumber \\
    \tilde q^I  &=& i\left( {\bar ZL^I  - Z\bar L^I } \right).
\end{eqnarray}

Based on these definitions, we can rewrite the field equations as
follows:
\begin{eqnarray}
    \frac{{d\sigma }}{{dr}} &=&  - 2e^{ - \frac{\sigma}{2} } \frac{Z}{r} =  - 2e^{ -
    \frac{\sigma}{2} } \frac{{\bar Z}}{r} \label{eqn4} \\
    \frac{{dz^i }}{{dr}} &=&  - e^{ - \frac{\sigma}{2} } \frac{{\nabla ^i \bar Z}}{r}
    \nonumber \\
    \frac{{dz^{\bar i} }}{{dr}} &=&  - e^{ - \frac{\sigma}{2} } \frac{{\nabla^{\bar i}
    Z}}{r}     \label{zzeom} \\
    \frac{{d\zeta }}{{d\varphi }} &=&  \tilde q \nonumber \\
    \frac{{d\tilde \zeta }}{{d\varphi }} &=&  q,
\end{eqnarray}
where the K\"{a}hler covariant derivative $\nabla^i$ is defined by
(\ref{covderiv}) and (\ref{covderiv1}).

The solution is further specified if we adopt Sabra's ans\"{a}tze:
\begin{eqnarray}
 \sigma  &=&  - \mathcal{K} = \ln \left[ {i\left( {\bar Z^I F_I  - Z^I \bar F_I } \right)} \right] \nonumber\\
 H_I  &=& i\left( {F_I  - \bar F_I } \right) \nonumber\\
 \tilde H^I  &=& i\left( {Z^I  - \bar Z^I } \right). \label{Sabra}
\end{eqnarray}

The calculation showing that this indeed satisfies the field
equations is analogous to Sabra's, detailed in \cite{Sabra4}.
Short of explicitly specifying a metric for the Calabi-Yau
manifold, this is as detailed as one can get the solution to be.

In summary, the general 2-brane solution coupled to the
hypermultiplets with vanishing $Q^5$ charge is:
\begin{eqnarray}
{\bf The}\; {\bf general}\;{\bf (Q}^{\bf 5}&=&{\bf   0)}\; {\bf
solution}\nonumber \\
    ds^2  &=& \eta _{ab} dx^a dx^b  + e^{ - 2\sigma } \delta _{\mu \nu
    } dx^\mu  dx^\nu \nonumber\\
 \sigma  &=&  - \mathcal{K} = \ln \left[ {i\left( {\bar Z^I F_I  - Z^I \bar F_I } \right)} \right] \nonumber\\
 H_I  &=& i\left( {F_I  - \bar F_I } \right), \quad
 \tilde H^I  = i\left( {Z^I  - \bar Z^I } \right) \nonumber\\
    \frac{{dz^i }}{{dr}} &=&  - e^{ - \frac{\sigma}{2} } \frac{{\nabla ^i \bar
    Z}}{r}\quad\quad \frac{{d\zeta }}{{d\varphi }} =  \tilde q
    \nonumber \\
    \frac{{dz^{\bar i} }}{{dr}} &=&  - e^{ - \frac{\sigma}{2} } \frac{{\nabla^{\bar i}
    Z}}{r} \quad\quad     \frac{{d\tilde \zeta }}{{d\varphi }} =
    q.\label{general2branehyper}
\end{eqnarray}

Having set the solution in polar coordinates form, it becomes
easier to check whether or not it reduces to the universal
hypermultiplet solution found in \S\ref{univhypersection}. We set
$\left({h_{2,1}=0}\right)$ and make the gauge choice
$\mathcal{N}_{00}=-i$. The central charge becomes:
\begin{equation}\label{}
    Z = \frac{i}{{\sqrt 2 }}\left( {q + i\tilde q} \right),
\end{equation}
which, by integrating (\ref{eqn4}), gives $\sigma=-2 \ln H$ with
$\left({H=h+Z\ln r}\right)$. Also, by using the definition of the
complex axions $\chi$ and $\bar\chi$, we get exactly the results
found before in equations (\ref{summuniv}).

\section{\label{moregeneral}Switching on the gauge field}

Let us examine the effects a nonvanishing $Q^5$ charge would have
on our solution. This situation may possibly have the geometrical
interpretation of wrapping particular configurations of both M2
and M5-branes \cite{ILPT}. We will only discuss this generally,
without making any ansatz on the explicit structure of the 3-form
$A_{abc}$. Because of their relative simplicity compared to the
equations of motion, we will look at the SUSY equations only.

The only terms affected by a nonvanishing $Q^5$ will be the $v$
and $\bar v$ expressions in (\ref{eqns5}), i.e.
\begin{eqnarray}
    v_\mu   &=& \frac{1}{2}\left( {\partial _\mu  \sigma } \right)
    +  \frac{i}{{2}}e^{-\sigma}  J_\mu ^5   \nonumber \\
    \bar v_\mu   &=& \frac{1}{2}\left( {\partial _\mu  \sigma } \right)
    -    \frac{i}{{2}}e^{-\sigma}  J_\mu ^5 \nonumber\\
    v_\mu - \bar v_\mu &=& i J_\mu ^5.
\end{eqnarray}

Considering the gravitino equations (\ref{gravitinotrans}), the
first thing we note is that, from
$\left({\delta\psi_a^r=0}\right)$, the metric exponent $A$ remains
zero. Equation $\left({\delta\psi_\mu^r=0}\right)$ will be
adjusted by a term proportional to $\left({v_\mu - \bar
v_\mu}\right)$. This new term may be considered to cancel with the
$R_\mu$ term, without affecting the $B$ exponent of the metric,
giving:
\begin{equation}
    R_\mu  = \frac{{\bar Z^I N_{IJ} \left( {\partial _\mu  Z^J } \right) -
    Z^I N_{IJ} \left( {\partial _\mu  \bar Z^J } \right)}}{{\bar Z^I N_{IJ} Z^J
    }}= 2ie^{-\sigma} J_\mu ^5,
\end{equation}
i.e. the $U(1)$ K\"{a}hler connection no longer vanishes.

The significant difference between this solution and the one with
vanishing $Q^5$ becomes apparent when we look at the hyperino
transformations. Equations (\ref{hyper2}) remain unchanged, while
(\ref{hyper1}) become:
\begin{eqnarray}
    \left( {\partial _\mu  \sigma } \right) + 2e^{\frac{\sigma}{2} } \left[ {L^I
    \left(     {\partial _\mu  H_I } \right) - M_I ( {\partial _\mu  \tilde H^I }
    )} \right] &=& \frac{i}{{2}}e^{-\sigma} J_\mu ^5  \nonumber \\
    \left( {\partial _\mu  \sigma } \right) - 2e^{ \frac{\sigma}{2} } \left[ {\bar
    L^I     \left( {\partial _\mu  H_I } \right) - \bar M_I ( {\partial _\mu
    \tilde     H^I } )} \right] &=&   -\frac{i}{{2}}e^{-\sigma}  J_\mu ^5 .
\end{eqnarray}

Now, repeating the arguments of the previous section, we get:
\begin{eqnarray}
    \frac{{d\sigma }}{{dr}} + 2e^{\frac{\sigma}{2}} \frac{Z}{r} &=& \frac{i}{2}\frac{e^{-\sigma}}{r}
     \delta ^{\mu\nu} x_\mu J_\nu ^5    \nonumber \\
    \frac{{d\sigma }}{{dr}} + 2e^{ \frac{\sigma}{2} }\frac{{ \bar Z}}{r} &=&
    -\frac{i}{2}\frac{e^{-\sigma}}{r}     \delta ^{\mu\nu} x_\mu J_\nu ^5 .
\end{eqnarray}

Adding and subtracting, we get the dilaton's new equation:
\begin{equation}\label{newdilaton}
    \frac{{d\sigma }}{{dr}} =  - 2\frac{{e^{ \frac{\sigma}{2}} }}{r}\left( {Z + \bar
    Z}     \right).
\end{equation}
as well as
\begin{equation}\label{}
    \left( {Z - \bar Z} \right) = \frac{i}{2}e^{ - \frac{3}{2}\sigma }
    \delta ^{\mu \nu } J_\mu ^5 x_\nu.
\end{equation}

We note once again that we are by no means exhaustively describing
the case $A \neq 0$, but rather just some of its properties. In
order for this to be considered a proper solution, one needs to
specify the 3-form gauge field, whose dual scalar field $a$ (the
universal axion) will in turn define the current $J_\mu ^5$, then
proceed to solve for the rest of the fields.

We have managed to find the most general 2-brane solution coupled
to the hypermultiplets of five dimensional $\mathcal{N}=2$
supergravity (\ref{general2branehyper}). The special geometry of
the theory and its symplectic invariance provide for a very
interesting and rich structure. We propose that this solution is,
to first order, a dimensional reduction of a M5-brane over SLAG
cycles of a CY 3-fold, deforming the moduli of the CY's complex
structure. This would be a generalization of the solution we found
in \S\ref{SpecialSLAG}.

We have also found special case solutions representing M2 and M5
brane wrappings over SLAG cycles of a CY, with only the universal
hypermultiplet excited - (\ref{Chiequalzero}) and (\ref{summuniv})
respectively. Finally, we briefly outlined some of the major
properties of an even more general 2-brane solution with
hypermultiplets that has the geometrical interpretation of
wrapping both M2 and M5 branes down to $D=5$
(\S\ref{moregeneral}).

\section{\label{back}Back to eleven dimensions?}

The lifting of the five dimensional solution
(\ref{general2branehyper}) is expected to yield a general eleven
dimensional solution representing a M5-brane wrapped over a SLAG
calibrated surface. During the writing of this dissertation, such
a $D=11$ solution was proposed by Martelli and Sparks (henceforth
referred to as MS) using completely different means \cite{MS}.
This solution, however, does not dimensionally reduce to the five
dimensional result (\ref{general2branehyper}).

The MS result has the following metric:
\begin{eqnarray}
 ds^2  &=& e^{2\Delta } \eta _{ab} dx^a dx^b  + e^{2\Delta } k_{ij} dx^i dx^j  + e^{-4\Delta } \delta _{\alpha \beta } dx^\alpha  dx^\beta   \nonumber\\
 a,b &=& 0,1,2\quad \quad i,j = 1, \ldots ,6\quad \quad \alpha ,\beta  =
 1,2. \label{MSmetric}
\end{eqnarray}

On the other hand, the $D=11$ metric we find by the lifting of the
$D=5$ metric (\ref{D11SLAGmetric}) via (\ref{expmet}) is:
\begin{eqnarray}
 ds^2  &=& e^{\frac{2}{3}\sigma } \eta _{ab} dx^a dx^b  + e^{ - \frac{\sigma }{3}} k_{ij} dx^i dx^j  + e^{ - \frac{4}{3}\sigma } \delta _{\alpha \beta } dx^\alpha  dx^\beta   \nonumber\\
 a,b &=& 0,1,2\quad \quad i,j = 1, \ldots ,6\quad \quad \alpha ,\beta  =
 1,2.\label{SLAGMetric}
\end{eqnarray}

Note that with the identification $e^\sigma = H^{-2}$, and
demanding $k_{ij}$ to be Hermitian, (\ref{SLAGMetric}) reduces to
the special case metric (\ref{specialSLAGmetric}) we found
earlier.

Now the MS variable $\Delta$ is not related in any simple way to
our $\sigma$. Hence the MS metric (\ref{MSmetric}) is not related
to ours (\ref{SLAGMetric}) by a conformal scaling, and as such
does not seem to be correct. Initial investigation into
supersymmetry also confirms this.

The MS 4-form field strength
\begin{equation}\label{MSF}
    F =  - e^{ - 6\Delta } d\left( {e^{6\Delta } {\mathop{\rm
    Re}\nolimits} \Omega } \right)
\end{equation}
also does not seem to satisfy the SUSY equations, neither does it
reduce to the correct hypermultiplet fields in five dimensions.

On the other hand, the lifting of the hypermultiplet fields seems
to give the following components of the field strength
\begin{eqnarray}
    F_{\mu mnp}  &=&  \mp \frac{1}{{\sqrt 2 }}e^{ - {\sigma
    \mathord{\left/
    {\vphantom {\sigma  2}} \right.
    \kern-\nulldelimiterspace} 2}}    \bar\varepsilon_\mu^{\;\;\;\nu}\left(
    {\partial _\nu  \sigma } \right)\Omega _{mnp}  \pm \sqrt 2 e^{ -
    {\sigma  \mathord{\left/
     {\vphantom {\sigma  2}} \right.
    \kern-\nulldelimiterspace} 2}}    \bar\varepsilon_\mu^{\;\;\;\nu}\left( {\partial _\nu  \Omega _{mnp} } \right) \nonumber\\
    F_{\mu mn\bar p}  &=&  \mp \frac{1}{{\sqrt 2 }}e^{ - {\sigma
    \mathord{\left/
    {\vphantom {\sigma  2}} \right.
    \kern-\nulldelimiterspace} 2}}    \bar\varepsilon_\mu^{\;\;\;\nu}\left(
    {\partial _\nu  \sigma } \right)\Omega _{mn\bar p}  \pm \sqrt 2 e^{
    - {\sigma  \mathord{\left/ {\vphantom {\sigma  2}} \right.
    \kern-\nulldelimiterspace} 2}}   \bar\varepsilon_\mu^{\;\;\;\nu} \left( {\partial _\nu  \Omega _{mn\bar p} } \right).\label{SLAGfield}
\end{eqnarray}

This reduces to the correct hypermultiplet fields in five
dimensions, while similar components from the MS result
(\ref{MSF}) do not. To show this, recall the compactification
argument of chapter (\ref{compact}) and rewrite (\ref{SLAGfield})
as follows
\begin{eqnarray}
    F_\mu   &=&  \mp \frac{1}{{\sqrt 2 }}e^{ - {\sigma  \mathord{\left/
    {\vphantom {\sigma  2}} \right.
    \kern-\nulldelimiterspace} 2}} \bar\varepsilon_\mu^{\;\;\;\nu}\left( {\partial _\nu  \sigma }
    \right)\Omega  \mp \frac{1}{{\sqrt 2 }}e^{ - {\sigma
    \mathord{\left/
    {\vphantom {\sigma  2}} \right.
    \kern-\nulldelimiterspace} 2}} \bar\varepsilon_\mu^{\;\;\;\nu}\left( {\partial _\nu  \sigma } \right)\bar \Omega  \nonumber\\
    &\quad& \pm \sqrt 2 e^{ - {\sigma  \mathord{\left/
    {\vphantom {\sigma  2}} \right.
    \kern-\nulldelimiterspace} 2}}\bar\varepsilon_\mu^{\;\;\;\nu} \left( {\partial _\nu  z^i }
    \right)\left( {\nabla _i \Omega } \right) \pm \sqrt 2 e^{ -
    {\sigma  \mathord{\left/
    {\vphantom {\sigma  2}} \right.
    \kern-\nulldelimiterspace} 2}} \bar\varepsilon_\mu^{\;\;\;\nu}( {\partial _\nu  z^{\bar i} } )\left( {\nabla _{\bar i} \bar \Omega }
    \right),\label{givencomponents}
\end{eqnarray}
where we have used Kodaira's theorem (\ref{Kodaira}). Comparing
this with the original form of $F_\mu$ (\ref{Fmu}), we can write
the following relations between $\sigma$, $(z,\bar z)$ and the
axions $(\zeta, \tilde \zeta)$:
\begin{eqnarray}
    \bar\varepsilon_\mu^{\;\;\;\nu}\left( {\partial _\nu  \sigma } \right) &=&  \pm i2e^{{\sigma
    \mathord{\left/
    {\vphantom {\sigma  2}} \right.
    \kern-\nulldelimiterspace} 2}} \left[ {( {\partial _\mu  \zeta ^I } )M_I  + ( {\partial _\mu  \tilde \zeta _I } )L^I } \right] \nonumber\\
    \bar\varepsilon_\mu^{\;\;\;\nu}\left( {\partial _\nu  z^i } \right) &=&  \mp ie^{{\sigma
    \mathord{\left/
    {\vphantom {\sigma  2}} \right.
    \kern-\nulldelimiterspace} 2}} G^{i\bar j} \left[ {h_{\bar jI} ( {\partial _\mu  \zeta ^I } ) + f_{\bar j}^I ( {\partial _\mu  \tilde \zeta _I } )}
    \right].
\end{eqnarray}

Using the explicit form of the axions in terms of the harmonic
functions (\ref{zetaeom}), we retrieve the dilaton and moduli
equations (\ref{general2branehyper}) exactly.

The components (\ref{SLAGfield}) do not represent a complete
description of the $D=11$ field strength. Other components, such
as $F_{q\bar m\bar n \bar p}$ are still missing. As far as the
dimensional reduction is concerned, those represent second order
effects and are ignored. The full eleven dimensional formulation
is required to specify the missing components as well as to verify
(\ref{givencomponents}). A guiding principle is that, based on
(\ref{specialSLAGA}), one would expect the six form gauge
potential to have a form similar to
\begin{eqnarray}
    A_{abcmnp}  &=&  \pm \sqrt 2 e^{  {\sigma  \mathord{\left/
    {\vphantom {\sigma  2}} \right.
    \kern-\nulldelimiterspace} 2}} \bar\varepsilon _{abc}\Omega _{mnp}  \nonumber\\
    A_{abcmn\bar p}  &=&  \pm \sqrt 2 e^{  {\sigma  \mathord{\left/
    {\vphantom {\sigma  2}} \right.
    \kern-\nulldelimiterspace} 2}} \bar\varepsilon _{abc}\Omega _{mn\bar
    p}.
\end{eqnarray}

Further investigation is deferred to future research.

\unnumberedchapter{Conclusion}

The theory of calibrations was used to find BPS solutions to
$D=11$ supergravity representing localized M-brane intersections.
Various special cases corresponding to K\"{a}hler calibrated
submanifolds were found and shown to be in the form of the
Fayyazuddin-Smith metric \cite{FS}. We also found a certain
special Lagrangian calibrated case, in the FS form as well,
corresponding to intersecting M5-branes (\S\ref{SpecialSLAG}).

The solutions have the alternate interpretation as M-branes
wrapping supersymmetric cycles of a Calabi-Yau 3-fold. This is
demonstrated by studying the dimensional reduction of $D=11$ SUGRA
and finding solutions representing the wrapping of M-branes over
SLAG cycles of a CY 3-fold with constant complex structure moduli.
These are five dimensional 2-branes coupled to the $\mathcal{N}=2$
universal hypermultiplet. A more general 2-brane solution coupled
to the full set of hypermultiplets was calculated and shown to be
magnetically dual to the instanton solutions of \cite{GS1,GS2}.
The reduced theory has a rich structure involving the
symplectically invariant special K\"{a}hler geometry, explained in
terms of the topology of the Calabi-Yau manifold.

Several directions of research can follow from this work. For
example, one can proceed to further specify the general SLAG
calibrated result and compare to the seemingly different MS
solution \cite{MS}. This may involve understanding the new
technique they used and comparing it to our methods.

The MS paper also discusses a $D=11$ solution representing
M-branes wrapped over a $G_2$ calibrated submanifold. It would be
interesting to verify this, then find and study the corresponding
reduced solutions. This case is of particular importance due to
the current interest in the literature in studying M-theory
compactifications over manifolds with $G_2$ holonomy. The
properties of such manifolds, particularly the lack of an analogue
to Yau's theorem guaranteeing the existence of a metric, make it a
considerably more difficult problem to do generally. Particular
choices of a $G_2$ metric must be used.

Finally, as remarked earlier, BPS solutions coupled to the
$\mathcal{N}=2$ hypermultiplets are lacking in the literature,
compared to their vector multiplets counterparts, which have been
extensively studied. This is due to the notorious difficulty of
dealing with actions parametrized by quaternionic manifolds. Now,
due to the c-map, it has been shown that, at least in $D=5$, these
can be cast into the special geometric structure traditionally
associated with the vector multiplets sector \cite{GS2}. This
correspondence allowed us to use the well understood structure of
special geometry to find solutions in the hypermultiplets sector.
We propose further study in this direction, perhaps even expanding
the search into theories with higher supersymmetry.

\appendix
\chapter{Manifolds, from Riemann to Yau}\label{manifolds}

We review the properties of the various classes of complex
manifolds referred to in the body of the text. Starting with
elementary definitions, we write down the various properties with
minimal mathematics, reporting only on those formulae which were
essential for our particular problem. For more detail, the reader
is referred to the literature (for example
\cite{GaumeFreedman,Ferrara1,CandelasOssa,Fre,Nakahara} and the
references therein).

\section{K\"{a}hler Manifolds}

We define the notion of a real $2k$-dimensional manifold
$\mathcal{M}$ as a set of points that behaves locally like
$\mathbb{R}^{2k}$, such that $2k$ real parameters ($x^1, \ldots,
x^\alpha, \ldots, x^{2k}$) are coordinates on $\mathcal{M}$
\cite{Wald}\footnote{The number of dimensions is chosen to be even
because they will be complexified in a moment, however, generally
the dimensions of a manifold may also be odd.}. Similarly, a
complex $q$-dimensional manifold may be defined as a set of points
that behaves locally like $\mathbb{C}^q$.

A Riemannian manifold is a manifold on which a smooth symmetric
positive-definite metric tensor $g_{\mu \nu } \left( {x^\alpha  }
\right)$ can be defined, describing a line element on the
manifold. A manifold is called Lorentzian if its metric is
Riemannian with a Lorentzian signature\footnote{Which we take to
be $\left( { -  +  +  \cdots + } \right)$ throughout.}. From the
metric, a Levi-Civita connection (a.k.a. Christoffel symbols), the
Riemann and Ricci tensors and the Ricci scalar may be defined in
the usual way:
\begin{eqnarray}
    ds^2  &=& g_{\mu \nu } dx^\mu  dx^\nu   \nonumber \\
    \Gamma _{\mu \nu }^\lambda   &=& \frac{1}{2}g^{\lambda \kappa } \left[ {\left(
    {\partial_\mu  g_{\nu \kappa } } \right) + \left( {\partial _\nu  g_{\mu \kappa } }
    \right) - \left( {\partial _\kappa  g_{\mu \nu } } \right)} \right] \nonumber \\
    {R_{\mu \nu \rho } }^\sigma   &=& \left( {\partial _\nu  \Gamma _{\mu \rho }^\sigma
    }\right) - \left( {\partial _\mu  \Gamma _{\nu \rho }^\sigma  } \right) + \Gamma
    _{\mu \rho }^\alpha  \Gamma _{\alpha \nu }^\sigma   - \Gamma _{\nu \rho }^\alpha
    \Gamma_{\alpha \mu
    }^\sigma \nonumber \\
    R_{\mu \nu }  &=& {R_{\mu \rho\nu  }} ^\rho  ,\quad \quad R = R_\mu ^\mu.
\end{eqnarray}

Locally, one can complexify $\mathcal{M}$ as follows:
\begin{eqnarray}\label{complex}
    w^\alpha  &=& x^\alpha   + \tau _{(\alpha ,\alpha  + k)} x^{\alpha  + k}  \nonumber \\
    \bar w^ \alpha   &=& x^\alpha   + \bar \tau _{(\alpha ,\alpha  + k)} x^{\alpha  + k}
    = w^{\bar \alpha},
\end{eqnarray}
i.e. $\mathbb{C}^q \sim \mathbb{R}^{2k}$, where the $\tau$'s are
complex parameters that specify a complex structure on the
manifold (more on that later). We define $w^m$ to be a set of $2k$
complex coordinates where the index runs through the $k$ unbarred
(holomorphic) indices, then through the barred (antiholomorphic)
indices. Reality of the line element is insured by the conditions
\begin{eqnarray}
    g_{mn}=g_{\bar m \bar n} \nonumber \\
    g_{m \bar n}=g_{n \bar m}.
\end{eqnarray}

A Hermitian manifold is defined as a complex manifold where there
is a preferred class of coordinate systems such that
\begin{equation}\label{Hermitian}
    g_{mn}=g_{\bar m \bar n}=0.
\end{equation}

The line element becomes
\begin{equation}\label{}
    ds^2  = 2g_{m\bar n} dw^m dw^{\bar n}.
\end{equation}

On any Hermitian manifold, a real 2-form can be defined as a (1,1)
form as follows:
\begin{equation}\label{form}
    \omega  =  ig_{m\bar n} dw^m  \wedge dw^{\bar n}.
\end{equation}

A K\"{a}hler manifold is a Hermitian manifold whose 2-form is
closed, i.e. $d\omega=0$, henceforth we will call $\omega$ the
K\"{a}hler form. This leads to the `curl-free' condition:
\begin{equation}\label{kahler}
    \partial _m g_{n\bar p}  - \partial _n g_{m\bar p} = 0,
\end{equation}
which may equivalently be used as the definition of a K\"{a}hler
manifold. This implies that, locally, the K\"{a}hler metric can be
determined in terms of a scalar function, known as the K\"{a}hler
potential ${\mathcal{K}}(w,\bar w)$, as follows:
\begin{equation}
    g_{m\bar n}  = \partial _m \partial _{\bar n} \mathcal{K}.
\end{equation}

Obviously, the metric is invariant under changes of the K\"{a}hler
potential of the form $\mathcal{K}\left( {w,\bar w} \right) \to
\mathcal{K}\left( {w,\bar w} \right) + f\left( w \right) + h\left(
{\bar w} \right)$, known as the K\"{a}hler gauge transformations.

The condition (\ref{kahler}) simplifies the properties of the
manifold considerably, for example one finds that
\begin{equation}
    \Gamma _{mn}^r  = g^{r\bar l} \left( {\partial _m g_{n\bar l} }
    \right)\quad,\quad  \Gamma _{\bar m\bar n}^{\bar r}  = g^{l\bar r} \left( {\partial _{\bar m} g_{\bar nl} }
    \right),
\end{equation}
are the only non-vanishing Christoffel symbols, indicating that
parallel transport does not mix the holomorphic with the
antiholomorphic components of a vector.

And the non-vanishing components of the Ricci tensor are
\begin{equation}\label{ricci}
    R_{m\bar n}  = \partial _m \partial _{\bar n} \ln g, \quad {\rm where\;\;} g = \det  {g_{m\bar n} }.
\end{equation}

\subsection{\label{globalissues}Global issues}

Technically, the assumption that any real $2n$-dimensional
manifold can be made into a complex manifold is only valid
locally. Global considerations must be included in order to
properly decide if a given manifold is truly complex everywhere.

A key element to such considerations is the so called complex
structure of the manifold. Intuitively, it is nothing more than
the formalization of multiplication by $i$ smoothly over the
manifold, i.e. an operation on geometrical objects whose square is
minus the identity. A tensor $J$ is called an \emph{almost}
complex structure if it satisfies the condition:
\begin{eqnarray}
    J^2 \sim - 1\quad :\quad
    J_\mu ^\rho  \left( x \right)J_\rho ^\nu  \left( x \right) =  -
    \delta _\mu ^\nu.
\end{eqnarray}
If a manifold $\mathcal{M}$ has a smooth almost complex structure,
it is called an almost complex manifold. An almost complex
structure becomes a complex structure when its so called Nijenhuis
tensor
\begin{equation}
    N_{\mu \nu }^\rho   = J_\mu ^\alpha  \left[ {\left( {\partial
    _\alpha  J_\nu ^\rho  } \right) - \left( {\partial _\nu  J_\alpha
    ^\rho  } \right)} \right] - J_\nu ^\alpha  \left[ {\left(
    {\partial _\alpha  J_\mu ^\rho  } \right) - \left( {\partial _\mu
    J_\alpha ^\rho  } \right)} \right]
\end{equation}
vanishes everywhere. This condition is achieved by demanding that
different complex structures on a manifold smoothly patch
together.

So, any $2n$-dimensional real manifold is locally complex (almost
complex manifold), but only globally so (complex manifold) when it
admits a complex structure with vanishing Nijenhuis tensor. This
is analogous to the concept that any Riemannian manifold is
locally flat, but only globally so when the Riemann tensor
vanishes everywhere.

Now, an almost complex manifold is called almost Hermitian if
there exists a Riemannian metric $g_{\mu \nu}$ which satisfies the
invariance:
\begin{equation}
    J_\mu ^\lambda  g_{\lambda \kappa } J_\nu ^\kappa   = g_{\mu
    \nu}.
\end{equation}

And finally, an almost Hermitian manifold becomes almost
K\"{a}hler if the K\"{a}hler form is closed. All of these
definitions lose the `almost' part of their titles once the
Nijenhuis tensor is found to vanish everywhere.

Another point of global importance is the question of holonomy
groups on a K\"{a}hler manifold \cite{Joyce}. If we consider a
vector $V^\mu$ on a $n$-fold and parallel transport it around a
closed loop, generally the vector will not return to itself, but
rather rotated by an element of $GL(n,\mathbb{R})$. The subset of
$GL(n,\mathbb{R})$ defined in this way forms the holonomy group of
the manifold. The \emph{restricted} holonomy group would be the
subset defined by paths which may be smoothly shrunk to a point
(contractable loops). The classification of the restricted
holonomy groups of all Riemannian manifolds has been performed by
Berger \cite{Berger}, which we list for completeness:

\begin{description}
    \item[Berger's theorem] Suppose $\mathcal{M}$ is a simply-connected
    manifold of dimension $n$, and that $g$ is a Riemannian metric on
    $\mathcal{M}$, then exactly seven restricted, or special, holonomy cases are possible:
        \begin{enumerate}
            \item Generic Riemannian manifolds, $Hol(g) = SO(n)$.
            \item K\"{a}hler manifolds, where $n = 2m$ with $m \ge
            2$ and $Hol(g) = U(m) \subset SO(2m)$.
            \item Calabi-Yau manifolds, where $n = 2m$ with $m \ge
            2$ and $Hol(g) = SU(m) \subset SO(2m)$. These are also Ricci-flat.
            \item HyperK\"{a}hler manifolds, where $n = 4m$ with $m \ge
            2$ and $Hol(g) = Sp(m) \subset SO(4m)$.
            \item Quaternionic K\"{a}hler manifolds, where $n = 4m$ with $m \ge
            2$ and $Hol(g) = Sp(m)\otimes Sp(1) \subset SO(4m)$.
            \item Manifolds with $n = 7$ and $Hol(g) = G_2 \subset SO(7)$.
            \item Manifolds with $n = 8$ and $Hol(g) = Spin(7) \subset
            SO(8)$. The groups $G_2$ and $Spin(7)$ are exceptional holonomy groups.
        \end{enumerate}
\end{description}

The list may also be understood in terms of the four division
algebras in the following way: It is well-known that one can
define exactly four algebras where, for two quantities $w_1$ and
$w_2$, the property $\left| {w_1 w_2 } \right| = \left| {w_1 }
\right|\left| {w_2 } \right|$ is satisfied. These are the real
numbers $\mathbb{R}$, the complex numbers $\mathbb{C}$, the
quaternions $\mathbb{H}$ and the octonions, or Cayley numbers,
$\mathbb{O}$. The Berger list fits into this classification by
noting that:
\begin{itemize}
  \item $SO(n)$ is a group of automorphisms of ${\mathbb{R}}^n$.
  \item $U(n)$ and $SU(n)$ are groups of automorphisms of ${\mathbb{C}}^n$.
  \item $Sp(n)$ and $Sp(n)\otimes Sp(1)$ are groups of automorphisms of ${\mathbb{H}}^n$.
  \item $G_2$ is the group of automorphisms of ${\mathop{\rm Im}\nolimits} \mathbb{O} \approx \mathbb{R}^7 $.
  \item $Spin(7)$ is a group of automorphisms of $\mathbb{O} \approx \mathbb{R}^8 $.
\end{itemize}

It is interesting to note that all of the manifolds on Berger's
list have found applications in string theory. In addition to the
usual use of Riemann manifolds with Lorentzian signature to
describe spacetime, one finds that supersymmetric theories
constrain certain fields to parameterize K\"{a}hler,
hyperK\"{a}hler and quaternionic manifolds. Calabi-Yau, $G_2$ and
$Spin(7)$-holonomy manifolds are all considered as possible
candidates for string/M-theory compactifications.

\section{A bit of differential geometry}

We recall that, given a Riemannian manifold $\mathcal{M}$, with a
metric $g_{\mu \nu}$, one can define the veilbeins $e$, the
connection 1-form $\omega$ (a.k.a. spin connection) in the
following way:
\begin{eqnarray}
    ds^2  &=& g_{\mu \nu } dx^\mu  dx^\nu   =  e^{\hat a} \eta _{\hat a\hat c} e^{\hat c} \quad ; \quad e^{\hat a}  = e_{\;\; \mu} ^{\hat a} dx^\mu   \label{bein} \\
    \omega _\mu ^{\;\;\hat a\hat c}  &=& e^{\hat a\lambda } \left[ {\left({\partial _\mu  e_{\;\; \lambda} ^{\hat c} } \right) - \Gamma _{\mu
    \lambda }^\nu  e_{\;\; \nu} ^{\hat c} } \right]\quad; \quad\omega ^{\hat a\hat b}  = \omega_\mu ^{\;\;\hat a\hat b}
    dx^\mu,
\end{eqnarray}
such that the so called Cartan structure equations define the
torsion and curvature 2-forms:
\begin{eqnarray}
    T^{\hat a}  &=& de^{\hat a}  + \omega _{\;\; \hat c}^{\hat a}  \wedge e^{\hat
    c}=0 \nonumber \\
    \Omega _{\;\; \hat c}^{\hat a}  &=& d\omega _{\;\; \hat c}^{\hat a}  + \omega_{\;\;
    \hat  b}^{\hat a}  \wedge \omega _{\;\; \hat c}^{\hat b},
\end{eqnarray}
where the hated indices are raised and lowered by the Minkowski
metric $\eta_{\hat a\hat c}$, describing a flat space tangent to
each point on the curved spacetime manifold. They are also
sometimes referred to as `frame' indices, as opposed to the
spacetime `world' indices.

Using this structure, one can define the so called total Chern
form \cite{Riazi}, which is a polynomial in the curvature as
follows:
\begin{equation}
    C\left( \Omega  \right) = \det \left( {1 + \frac{i}{{2\pi }}\Omega
    } \right) = 1 + c_1 \left( \Omega  \right) + c_2 \left( \Omega
    \right) +  \cdots.
\end{equation}

The terms $c_r$ are the Chern classes. They belong to
topologically distinct cohomology classes. Integrals such as
\begin{equation}
    \int\limits_M {c_2 \left( \Omega  \right)} \quad {\rm and}\quad
    \int\limits_M {c_1 \left( \Omega  \right) \wedge c_1 \left(
    \Omega  \right)} \nonumber
\end{equation}
are invariant integers, usually called the Chern numbers.

The Chern classes are a very general and efficient way of
classifying and distinguishing topologically inequivalent
manifolds and fiber bundles. Although we have not defined what a
fiber bundle is, some important examples may be worth mentioning
in this context since they represent familiar physical quantities.
For instance, the Chern classes reduce to the well-known
Pontrjagin classes in the special case of vector bundles, which
are of main interest for the classification of topological
solutions in the standard model. The simplest examples are the
first and second Pontrjagin classes for a pure electromagnetic
field which give the energy density and Poynting vector. Another
familiar example is the case of a $U(1)$ bundle of the Dirac
monopole over $S^2$. Here the first Chern class is $c_1 \left(
\Omega \right) = - {F \mathord{\left/ {\vphantom {F {2\pi }}}
\right. \kern-\nulldelimiterspace} {2\pi }}$, where $F$ is the
electromagnetic field strength. The corresponding Chern number
exactly yields the monopole's conserved charge. For a $SU(N)$
bundle,the second Chern number is the invariant
\begin{equation}
    S =  - \frac{1}{2}\int {Tr\left( {F \wedge *F} \right)}  =  \mp
    \frac{1}{2}\int {Tr\left( {F \wedge F} \right)}  = 4\pi \left|
    {C_2 } \right|,
\end{equation}
where the case $\left|{C_2 } \right|=1$ corresponds to the 't
Hooft instanton.

\section{Hodge-K\"{a}hler manifolds}

The Chern classes can be used to topologically distinguish various
types of manifolds. Given the Ricci tensor $R_{m \bar n}$ of a
K\"{a}hler manifold, we define the $(1,1)$ Ricci form
\begin{equation}
    \mathcal{R} = R_{m\bar n} dw^m \wedge dw^{\bar n}.
\end{equation}

Since the Ricci form is necessarily closed; $d\mathcal{R}=0$, then
it defines an equivalence class in the homology group $H^{1,1}$.

The first Chern class is simply:
\begin{equation}
    c_1  = \frac{\mathcal{R}}{{2\pi }}.
\end{equation}

Consider a line bundle $\mathcal{L}$ over a K\"{a}hler manifold.
By definition, this is a holomorphic vector bundle of rank $r=1$.
If its first Chern class, defined in terms of a metric on the
bundle, equals the cohomology class of the manifold's K\"{a}hler
form;
\begin{equation}\label{hodge-k}
    c_1 \left( \mathcal{L} \right) = \left[ \omega \right],
\end{equation}
then we call this a Hodge-K\"{a}hler manifold \cite{Ferrara1}. It
is particularly important to us, since this is the type of
manifold described by the complex structure moduli in $D=5$ SUGRA.

An equivalent definition is that the exponential of the K\"{a}hler
potential of the manifold is equal to the metric of the line
bundle $\mathcal{L}$. So, if a one-component real function
$h\left( {w,\bar w} \right)$ is the hermitian fiber metric on the
line bundle $\mathcal{L}$, then the manifold is Hodge-K\"{a}hler
if we can write:
\begin{equation}\label{hodge}
    h\left( {w,\bar w} \right) = e^{\mathcal{K}\left( {w,\bar w}
    \right)},
\end{equation}
which clearly enables us to write:
\begin{equation}
    c_1 \left( \mathcal{L} \right) = \frac{i}{{2\pi }}\partial \bar \partial \ln
    h.
\end{equation}

Since $\mathcal{L}$ is a line bundle, its connection (Christoffel
symbol) is a 1-form defined by $h$ as follows
\begin{equation}
    \vartheta  \equiv h^{ - 1} \left( {\partial h} \right) = h^{ - 1}
    \left( {\partial _m h} \right)dw^m \quad ,\quad \bar \vartheta
    \equiv h^{ - 1} \left( {\bar \partial h} \right) = h^{ - 1} \left(
    {\partial _{\bar n} h} \right)dw^{\bar n},
\end{equation}
which, by virtue of (\ref{hodge}), becomes
\begin{equation}\label{conn}
    \vartheta  = \left( {\partial \mathcal{K}} \right) = \left( {\partial _m \mathcal{K}}
    \right)dw^m \quad ,\quad \bar \vartheta  = \left( {\bar \partial
    \mathcal{K}} \right) = \left( {\partial _{\bar n} \mathcal{K}} \right)dw^{\bar n}.
\end{equation}

Now, it is known that there exists a correspondence between line
bundles and $U(1)$ bundles. At the level of connections this
reduces to
\begin{equation}\label{U1connection}
    U(1)\; {\rm connection} \equiv \mathcal{P} = {\mathop{\rm Im}\nolimits} \vartheta = -
    \frac{i}{2}\left( {\vartheta  - \bar \vartheta } \right),
\end{equation}
which, in our case, becomes
\begin{equation}
    \mathcal{P} =  - \frac{i}{2}\left[ {\left( {\partial _m \mathcal{K}} \right)dw^m  -
    \left( {\partial _{\bar n} \mathcal{K}} \right)dw^{\bar n} } \right],
\end{equation}
and the covariant derivatives can be constructed accordingly.

\section{Calabi-Yau manifolds}

In 1954 Calabi proposed the following conjecture: If $\mathcal{M}$
is a complex manifold with a K\"{a}hler metric and vanishing first
Chern class, then there exists a unique Ricci flat metric for each
K\"{a}hler class on $\mathcal{M}$. In 1976, Calabi's conjecture
was proven by Yau, also showing that a Ricci flat metric
necessarily has $SU(m)$ holonomy; $m$ being the number of complex
dimensions. Henceforth, we will define Calabi-Yau manifolds as
K\"{a}hler manifolds with Ricci flat ($c_1=0$) metrics.

From its general properties, it turns out that a large number of
different CY manifolds exist. It also turns out that defining them
explicitly is a nontrivial task. Indeed, very few explicit CY
metrics have ever been written down. However, the properties of CY
manifolds make it possible to work with them without explicit
knowledge of the metric, as far as string theory compactifications
are concerned. We will restrict ourselves to six dimensional CY
manifolds admitting $SU(3)$ holonomy, since this is the type of
interest to string theory in general and to this work in
particular.

The importance of this class of manifolds to physics lies in the
fact that they admit covariantly constant spinors. As a
consequence, it can be shown \cite{CHSW} that string theory
compactifications over CY 3-folds preserve some supersymmetry
(also see \cite{PapaTown} and the references therein). Such
compactifications have indeed yielded rich, physically
interesting, theories in lower dimensions. For example, the scalar
fields in the compactified theory correspond to the parameters
that describe possible deformations of the CY 3-fold. This
parameters' space factorizes, at least locally, into a product
manifold ${\mathcal{M}}_C \otimes {\mathcal{M}}_K$, with
${\mathcal{M}}_C$ being the manifold of complex structures and
${\mathcal{M}}_K$ being a complexification of the parameters of
the K\"{a}hler class. These moduli spaces turn out to be
K\"{a}hler manifolds of a restricted type, which we will describe
in detail later. In addition, there exists a symmetry in the
structures of ${\mathcal{M}}_C$ and ${\mathcal{M}}_K$ which lends
support to the so called mirror symmetry hypothesis of CY 3-folds.

Calabi-Yau 3-folds have a cohomology groups structure that may be
summed up by the so called Hodge diamond:
\begin{equation}\label{diamond}
    \begin{array}{*{20}c}
       {} & {} & {} & 1 & {} & {} & {}  \\
       {} & {} & 0 & {} & 0 & {} & {}  \\
       {} & 0 & {} & {h_{1,1} } & {} & 0 & {}  \\
       1 & {} & {h_{1,2} } & {} & {h_{2,1} } & {} & 1  \\
       {} & 0 & {} & {h_{1,1} } & {} & 0 & {}  \\
       {} & {} & 0 & {} & 0 & {} & {}  \\
       {} & {} & {} & 1 & {} & {} & {}
    \end{array}
\end{equation}
where the Hodge numbers $h_{p,q}$ are the dimensions of the
respective cohomology groups the manifold admits\footnote{The
equivalent to the Betti numbers for a real manifold.}, so the
diamond tells us that CY 3-folds have a single (3,0) cohomology
form; $h_{3,0}=\dim \left( {H^{3,0} } \right) = 1$, which we will
call $\Omega$ (the holomorphic volume form) and an arbitrary
number of (1,1) and (2,1) forms determined by the corresponding
$h$'s \footnote{Whose values depend on the particular choice of CY
manifold.}. The Hodge number $h_{2,1}$ determines the dimensions
of ${\mathcal{M}}_C$, while $h_{1,1}$ determines the dimensions of
${\mathcal{M}}_K$. The K\"{a}hler form $\omega$ of a CY 3-fold
$\mathcal{M}$ is defined by (\ref{form}). The pair
($\mathcal{M},\omega$) can be deformed by either deforming the
complex structure of $\mathcal{M}$ or by deforming the K\"{a}hler
form $\omega$. In string compactifications, each of these two
choices yields a different set of fields in the lower dimensional
theory.

\subsection{The space $\mathcal{M}_C$ of complex structure moduli}\label{21forms}

The way the (2,1) forms $\chi$ are linked to the complex structure
deformations $\delta g_{m n}$ and $\delta g_{\bar m\bar n}$ is
defined via the unique (3,0) form $\Omega$ as follows
\cite{CandelasOssa}:
\begin{equation}
    \delta g_{\bar p\bar r}  =  - \frac{1}{{\left| \Omega  \right|^2
    }}{\Omega _{\bar p}} ^{mn} \chi _{i|mn\bar r} \delta z^i \quad ; \quad
    \left| \Omega  \right|^2  \equiv \frac{1}{{3!}}\Omega _{mnp} \bar
    \Omega ^{mnp},
\end{equation}
with the inverse relation
\begin{equation}\label{link}
    \chi _{i|mn\bar p}  =  - \frac{1}{2}{\Omega _{mn}} ^{\bar r} \left(
    {\frac{{\partial g_{\bar p\bar r} }}{{\partial z^i }}}
    \right)\quad ; \quad \chi _i  = \frac{1}{2}\chi _{i|mn\bar p} dw^m
    \wedge dw^n  \wedge dw^{\bar p},
\end{equation}
where $\left( {z^i :i = 1, \ldots ,h_{2,1} } \right)$ are the
parameters, or moduli, of the complex structure. Each $\chi _i $
defines a (2,1) cohomology class. An important observation is that
the moduli can be treated as complex coordinates that define a
K\"{a}hler manifold $\mathcal{M}_C$ of their own, with metric
$G_{i\bar j}$ defined by
\begin{equation}\label{Gij}
    2G_{i\bar j} \left( {\delta z^i } \right)\left( {\delta z^{\bar j}
    } \right) \equiv \frac{1}{{2V_{CY}}}\int {d^6 x\sqrt g g^{m\bar n} g^{r\bar
    p} \left( {\delta g_{mr} } \right)\left( {\delta g_{\bar n\bar p}
    } \right)},
\end{equation}
where $V_{CY}$ is the volume of the real space CY 3-fold. In
differential geometric notation, we can write
\begin{equation}\label{G}
 G_{i\bar j}  =  - \frac{{\int {\chi _i  \wedge \bar \chi _{\bar j}
    } }}{{\int {\Omega  \wedge \bar \Omega } }} = \partial _i
    \partial _{\bar j} \mathcal{K}_C=  - \partial _i \partial _{\bar j}
    \ln \left( {i\int {\Omega  \wedge \bar \Omega } } \right)
\end{equation}
which also defines its K\"{a}hler potential $\mathcal{K}_C$
\cite{CandelasOssa}. A particularly useful theorem, attributed to
Kodaira (see \cite{Kodaira}), states that
\begin{equation}\label{Kodaira}
\left( {\partial _i \Omega } \right) = k_i \Omega  + \chi _i,
\end{equation}
where the coefficients $k_i$ may depend on the moduli.

The space $\mathcal{M}_C$ of complex structures may be described
in terms of the periods of the holomorphic 3-form $\Omega$. Let
$\left( {A^I ,B_I } \right)$, where $I,J,K = 0, \ldots ,h_{2,1} $,
be a canonical homology basis for $H^3$ and let $\left( {\alpha _I
,\beta ^I } \right)$ be the dual cohomology basis forms such that
\begin{eqnarray}
    \int {\alpha _I  \wedge } \beta ^J  &=&  \int\limits_{A^J } {\alpha _I } = \delta
    _I^J,\quad \int {\beta ^I  \wedge \alpha _J } = \int\limits_{B_J } {\beta ^I } =  -
    \delta _J^I , \nonumber \\
    \int {\alpha _I  \wedge } \alpha _J  &=& \int {\beta ^I  \wedge \beta ^J }  = 0.
    \label{cohbasis}
\end{eqnarray}

The periods of $\Omega$ are defined by
\begin{equation}\label{periods}
    Z^I  = \int\limits_{A^I } {\Omega },\quad\quad F_I  = \int\limits_{B_I
    } \Omega.
\end{equation}

Based on the periods, we also define the matrix:
\begin{equation}\label{}
    \varpi \equiv \left[\; {\int\limits_{A^J } {\left( {\partial _I \Omega }
    \right)} ,\int\limits_{B_J } {\left( {\partial _I \Omega }
    \right)} } \right].
\end{equation}

Now, given a particular choice of basis, it can be shown that this
matrix may be written as
\begin{equation}\label{periodmat}
    \varpi = \left({\mathbf{1},\mathcal{N}_{IJ}}\right).
\end{equation}

In the early literature, the matrix $\varpi$ was known as the
period matrix. However, it is $\mathcal{N}_{IJ}$ that appears in
supersymmetric actions, therefore the title `period matrix' has
been transferred to $\mathcal{N}_{IJ}$ in the more recent
literature, and we will follow this convention here.

Now, it can be shown that, locally in the moduli space, the
complex structure is entirely determined by $Z^I$, so one can
write $F_I = F_I \left( Z^J \right)$ \cite{BryantGriffiths}. Also,
a rescaling $Z^I  \to \lambda Z^I $, where $\lambda$ is a nonzero
constant, corresponds to a rescaling of $\Omega$ that does not
change the complex structure, which implies that the $Z$'s are
projective coordinates for the complex structure. In fact, we can
choose a set of independent `special coordinates' $z$ as follows:
\begin{equation}
    z^I  = \frac{{Z^I }}{{Z^0 }},
\end{equation}
which are identified with the complex structure moduli.

So, given the cohomology basis defined above, one can invert
(\ref{periods}) as follows\footnote{The significance of the minus
sign will become apparent when we discuss the symplectic
invariance behind these expressions.}
\begin{equation}\label{defomega}
    \Omega  = Z^I \alpha _I  - F_I \beta ^I,
\end{equation}
and the K\"{a}hler potential of $\mathcal{M}_C$ becomes
\begin{equation}\label{pot}
    \mathcal{K}_C =  - \ln \left[ {i\left( {\bar Z^I F_I  - Z^I \bar F_I }
    \right)} \right].
\end{equation}

One final remark is that it is usually possible to define $F_I$ as
being the derivative with respect to $Z^I$ of a scalar function
$F$ known as the prepotential. In the early days of string theory,
this would have been the starting point of any CY
compactifications calculation, but since it is not always possible
to write down this $F$, it is more convenient and general to work
without this assumption.

\subsubsection{A simple example}

To get a more intuitive understanding of the subject of moduli
spaces, we consider the simplest example of a Calabi-Yau manifold:
the ordinary torus $T^2$ \cite{Bellido}. In this case, there are
two real periodic degrees of freedom $x$ and $y$, such that:
\begin{equation}
    x = x + R_1  , \quad y = y + R_2,
\end{equation}
corresponding to the $H^1$ homology cycles $A$ and $B$
respectively. The cohomology basis forms would then be:
\begin{eqnarray}
    \alpha  = \frac{{dx}}{{R_1 }}\quad \quad &,&\quad \quad \beta  =  - \frac{{dy}}{{R_2 }}
    \nonumber \\
    \int\limits_A \alpha   =  - \int\limits_B \beta   = 1\quad \quad &,&\quad \quad
    \int\limits_{T^2 } {\alpha  \wedge \beta }  = 1,
 \end{eqnarray}
such that the volume form would be the holomorphic $(1,0)$-form:
\begin{eqnarray}
    \Omega _T  &=& dx + idy = R_1 \alpha  - iR_2 \beta = Z\alpha  - F\beta  \nonumber \\
    Z &=& \int\limits_A {\Omega _T }  = R_1 \quad \quad ,\quad \quad F = \int\limits_B
    {\Omega     _T }  = iR_2,
\end{eqnarray}
and the period matrix is trivially:
\begin{equation}\label{}
    \varpi  = \left( {1,0} \right),\quad \mathcal{N} = 0.
\end{equation}

\subsection{The space $\mathcal{M}_K$ of K\"{a}hler moduli}

The structure of the space defined by the parameters of the
K\"{a}hler deformations class is quite analogous to the one
already discussed in the last section. This is presumably the
result of the hypothesized mirror symmetry mentioned earlier
\cite{FerraraMirror}, where it is believed that for any given CY
3-fold, there exists a mirror manifold where the properties of the
complex structure and K\"{a}hler classes switch
positions\footnote{Basically, in a mirror symmetric manifold, an
exchange of rows and columns in the Hodge diamond (\ref{diamond})
would occur.}. A detailed understanding of this symmetry is yet to
be accomplished. One expected consequence is the decoupling of the
vector multiplets and hypermultiplets sectors in $\mathcal{N}=2$
supersymmetric theories. In chapter (\ref{compact}), we explicitly
demonstrate that this indeed does occur.

On $\mathcal{M}_K$, the inner product of two (1,1) forms $\rho$
and $\sigma$ can be defined as follows
\begin{equation}
    G\left( {\rho , \sigma} \right) = \frac{1}{{2V_{CY}}}\int {\rho \wedge
    \star\sigma},
\end{equation}
in clear analogy with (\ref{G}). As observed in \cite{Strominger},
$G\left( {\rho , \sigma} \right)$ may be written entirely in terms
of the cubic form
\begin{equation}
    d\left( {\rho , \sigma , \tau} \right) \equiv \int {\rho \wedge \sigma  \wedge \tau
    },
\end{equation}
which defines the volume of the manifold in terms of the
K\"{a}hler form $\omega$
\begin{equation}
    V_{CY} = \frac{1}{{3!}}d\left( {\omega ,\omega ,\omega } \right).
\end{equation}

As we did in the case of the (2,1) forms, let us choose a basis
$e_i$ of the homology class $H^2$, where now $i,j,k = 1, \ldots
,h_{1,1} $. One can then expand the forms in terms of this basis,
where the coefficients will play the role of the moduli $z^i$. We
will use the same symbols to denote analogous parameters.

The metric of the space of the K\"{a}hler moduli $z^i$ is defined
in terms of the inner product of the basis $e_i$ as follows
\begin{eqnarray}
    G_{i\bar j}  &=& \frac{1}{2}G\left( {e_i ,e_{\bar j} } \right) \nonumber\\&=&  -
    \partial _i \partial _{\bar j} \ln d\left( {\omega ,\omega ,\omega
    } \right) =  - \partial _i \partial _{\bar j} \mathcal{K}_K,
\end{eqnarray}
meaning that it also parameterizes a Hodge-K\"{a}hler manifold
described by the coordinates $z^i$ with the given K\"{a}hler
potential, which is basically the logarithm of the volume of the
CY manifold.

Once again, traditionally most authors on this subject would, at
this point, define the prepotential $F$ in terms of the components
of the cubic form $d$, otherwise known as the classical
intersection numbers of the CY manifold. As we have noted before,
a prepotential does not always exist\footnote{Although it is
worthwhile noting that a theory with no prepotential can, at least
in principle, be smoothly mapped to a theory with a prepotential
via a symplectic rotation, which we will define in appendix
\ref{specialgeometry}.}, hence we will skip this step and instead
make the connection with physics (in chapter \ref{compact}) via
the special geometric structures that arise in CY compactified
theories, discussed in appendix \ref{specialgeometry}. Meanwhile,
it is perhaps worthwhile repeating the important observation that
the restricted K\"{a}hler manifolds both deformation classes
describe exhibit very much the same structure, despite the fact
that they represent different objects.

\chapter{Special Geometry}\label{specialgeometry}

If one begins from ten dimensional type IIA string/supergravity
theory or eleven dimensional M/supergravity theory and
compactifies over a six dimensional Calabi-Yau manifold, we end up
having just about the same structure; a $\mathcal{N}=2$
supersymmetric theory in four or five dimensions, exhibiting the
so called special geometry. Historically, though, the constraints
of special geometry have been derived based on the requirements of
supersymmetry in $D=4,5$ without reference to higher dimensions.

Reducing the higher dimensional theories over a CY manifold,
gives, in addition to the gravity sector, two sets of fields: the
vector multiplets and the hypermultiplets. As noted elsewhere,
these two sectors completely decouple from each other and one may
set one of them to zero consistently. The theory as a whole
exhibits a symplectic structure. We will begin this appendix by
defining what that is, then proceed to understand the special
geometry that arises. The connection with the CY moduli spaces
defined in appendix \ref{manifolds} will become more apparent as
we go along. Basically, the scalar moduli of the CY define a
geometry of a restricted type; special geometry, that depends on
the topological properties of the CY manifold. In turn, special
geometry determines the dynamics of the massless fields in the
dimensionally reduced theory \cite{CandelasOssa}.

Although we are mainly interested in the five dimensional theory
with hypermultiplets, we will spend some time developing the
terminology of the four dimensional vector multiplet theory, since
they are related by a smooth transformation known as the c-map, as
discussed elsewhere in this appendix.

The literature on special geometry is extensive. In addition to
cited papers, other reviews include
\cite{Ferrara3,Ferrara2,Proeyen2,Proeyen1,Vandoren,Proeyen6,Proeyen3,Proeyen4,Proeyen5}.

\section{\label{EMdual}Electric-magnetic duality and symplectic covariance}

The symplectic structure characteristic of special geometry first
arose in the study of a supersymmetric collection of abelian
vector fields \cite{GaillardZumino}. Consider a general four
dimensional abelian theory of vector and scalar fields exhibiting
covariance under a group of duality rotations; a generalization of
the electromagnetic duality of ordinary Maxwell theory. Generally,
we can have a set of $n$ 1-form $U(1)$ gauge fields, and a similar
number of scalar fields $\phi$, such that we can write, as usual
\begin{eqnarray}\label{SU}
    A^I  &\equiv& A_\mu ^I dx^\mu  \quad , \quad F^I  = dA^I  = {\mathcal{F}}_{\mu \nu }^I dx^\mu   \wedge dx^\nu   \nonumber \\
    {\mathcal{F}}_{\mu \nu }^I  &\equiv& \frac{1}{2}\left( {\partial _\mu  A_\nu ^I  - \partial _\nu  A_\mu ^I }
    \right) \nonumber \\
    \star F^I  &\equiv& \tilde {\mathcal{F}}_{\mu \nu }^I dx^\mu   \wedge dx^\nu  \quad ; \quad \tilde {\mathcal{F}}^I  \equiv \frac{1}{2}\varepsilon _{\mu \nu \rho \sigma } {\mathcal{F}}^{I|\rho \sigma
    },
\end{eqnarray}
where the indices ($I,J,K$) range over $\left( {1, \ldots ,n}
\right)$. The Lagrangian has the form
\begin{equation}\label{lag}
    \mathcal{L} = \frac{1}{2}\left[ {\gamma _{IJ} \left( \phi  \right)F^I  \wedge \star F^J  +
    \theta _{IJ} \left( \phi  \right)F^I  \wedge F^J } \right] + \frac{1}{2}
    G_{IJ} \left( \phi  \right)\left( {\partial _\mu  \phi ^I }
    \right)\left( {\partial ^\mu  \phi ^J } \right),
\end{equation}
where the symmetric matrix ${\gamma _{IJ}(\phi) }$ is a
generalization of the inverse of the square of the coupling
constant (${1 \mathord{\left/ {\vphantom {1 {g^2 }}} \right.
\kern-\nulldelimiterspace} {g^2 }}$) in ordinary gauge theories
and the symmetric matrix ${\theta _{IJ}(\phi) }$ is a
generalization of the theta-angle of QCD. The scalar fields $\phi$
define a $n$-dimensional manifold $\mathcal{M}_\phi$ with metric
$G_{IJ}$. This is the kind of bosonic structure (we will neglect
the fermionic terms for brevity) that the vector multiplets sector
of $\mathcal{N}=2$ supersymmetry theory exhibits.

Now, if we define a formal operator $j$ that has the effect of
mapping a field strength into its dual;
\begin{equation}
    \left( {j{\mathcal{F}}^I } \right)_{\mu \nu }  = \tilde {\mathcal{F}}_{\mu \nu }^I  =
    \frac{1}{2}\varepsilon _{\mu \nu \rho \sigma } {\mathcal{F}}^{I|\rho \sigma
    },
\end{equation}
then the Lagrangian can be recast in the matrix form ($I$,$J$
indices suppressed):
\begin{equation}
    \mathcal{L} = {\mathcal{F}}^T \left( { - \gamma  \otimes \mathbf{1} + \theta  \otimes j} \right){\mathcal{F}}
    + \frac{1}{2}G\left( \phi  \right)\left( {\partial_\mu  \phi }
    \right)\left( {\partial^\mu   \phi } \right).
\end{equation}

Since the operator $j$ satisfies $j^2=-\mathbf{1}$, then its
eigenvalues are $\pm i$ and we can define the self-dual and
antiself-dual matrices
\begin{eqnarray}
    {\mathcal{F}}^ \pm   &=& \frac{1}{2}\left( {{\mathcal{F}} \pm ij{\mathcal{F}}} \right) \nonumber \\
    {\rm s}{\rm .t}{\rm .} \quad j\mathcal{F}^ \pm   &=&  \mp i\mathcal{F}^ \pm,
\end{eqnarray}
and the $\phi$-dependent symmetric matrix
\begin{eqnarray}
    \mathcal{N} = \theta  - i\gamma  \nonumber \\
    {\bar{\mathcal{N}}} = \theta  + i\gamma
\end{eqnarray}
which is either referred to as the kinetic matrix for obvious
reasons, or the period matrix because it is identified with
(\ref{periodmat}). We can again rewrite the vector part of the
Lagrangian in the matrix form
\begin{equation}
    \mathcal{L} = i\left( {{\mathcal{F}}^{ - T} \bar{\mathcal{N}}{\mathcal{F}}^ -   - {\mathcal{F}}^{ + T} \mathcal{N}{\mathcal{F}}^ +  }
    \right).
\end{equation}

Finally, we define the new `conjugate momenta' tensor
\begin{equation}\label{new}
    \tilde {\mathcal{G}}_{I | \mu \nu }  \equiv \frac{1}{2}\left( {\frac{{\partial
    \mathcal{L}}}{{\partial {\mathcal{F}}_{\mu \nu }^I }}}
    \right),
\end{equation}
and note that its self-dual and antiself-dual forms can be written
as
\begin{equation}
    \mathcal{G}^ +   = \mathcal{N}{\mathcal{F}}^ +  ,\quad \mathcal{G}^ -   = \bar {\mathcal{N}}{\mathcal{F}}^
    -.
\end{equation}

Now the Bianchi identities and field equations of the Lagrangian
$\mathcal{L}$ yield
\begin{eqnarray}\label{eom}
    \left( {\partial ^\mu  \tilde {\mathcal{F}}_{\mu \nu }^I } \right) &=& 0 \nonumber \\
    \left( {\partial ^\mu  \tilde {\mathcal{G}}_{I | \mu \nu } } \right) &=&
    0.
\end{eqnarray}

In the presence of electric and magnetic sources, Gauss' law gives
the corresponding charges
\begin{equation}\label{charges}
    \int\limits_{S^2 } {\mathcal{F}}  = \tilde q \;({\rm magnetic})\quad ,\quad
    \int\limits_{S^2 } {\mathcal{G} }  = q \;({\rm electric}).
\end{equation}

We note that these charges can be linked to the central charge of
the supersymmetry algebra as coefficients of the later's
symplectic expansion \cite{WittenOlive}. We discuss this in more
detail in chapter (\ref{5Dsolutions}).

The equations of motion (\ref{eom}) suggest that if we introduce a
$2n$ column vector
\begin{equation}
    V \equiv \left( {\begin{array}{*{20}c}
       {j{\mathcal{F}}}  \\
       {j\mathcal{G}}  \\
    \end{array}} \right)
\end{equation}
we find that for any ($2n \times 2n$) matrix $ \in GL\left(
{2n,\mathbb{R}} \right)$, the general linear transformation
\begin{equation}\label{rot}
    \left( {\begin{array}{*{20}c}
       {j{\mathcal{F}}}  \\
       {j\mathcal{G}}  \\
    \end{array}} \right)^\prime   = \left( {\begin{array}{*{20}c}
       A & B  \\
       C & D  \\
    \end{array}} \right)\left( {\begin{array}{*{20}c}
       {j{\mathcal{F}}}  \\
       {j\mathcal{G}}  \\
    \end{array}} \right)
\end{equation}
produces a new vector that still satisfies (\ref{eom}), i.e.
\begin{equation}
    \left( {\partial V} \right) = 0\quad  \Leftrightarrow \quad \left(
    {\partial V'} \right) = 0.
\end{equation}

The reader may recognize this as a generalization of
electric-magnetic duality in Maxwell-Dirac theory (Maxwell theory
with magnetic currents).

Rewriting in terms of the self-dual and antiself-dual formulation,
the duality rotation (\ref{rot}) is recast into
\begin{eqnarray}
    \left( {\begin{array}{*{20}c}
       {{\mathcal{F}}^ +  }  \\
       {\mathcal{G}^ +  }  \\
    \end{array}} \right)^\prime   &=& \left( {\begin{array}{*{20}c}
       A & B  \\
       C & D  \\
    \end{array}} \right)\left( {\begin{array}{*{20}c}
       {{\mathcal{F}}^ +  }  \\
       {\mathcal{N}{\mathcal{F}}^ +  }  \\
    \end{array}} \right) \nonumber \\ \left( {\begin{array}{*{20}c}
       {{\mathcal{F}}^ -  }  \\
       {\mathcal{G}^ -  }  \\
    \end{array}} \right)^\prime   &=& \left( {\begin{array}{*{20}c}
       A & B  \\
       C & D  \\
    \end{array}} \right)\left( {\begin{array}{*{20}c}
       {{\mathcal{F}}^ -  }  \\
       {\bar {\mathcal{N}}F^ -  }  \\
    \end{array}} \right) \label{trans}
\end{eqnarray}

Now, we demand that the transformation rule (\ref{trans}) of
${\mathcal{G}}^ \pm$ be consistent with the definition
(\ref{new}). This constraint restricts the form of the
transformation matrix $\Lambda$, and a straightforward calculation
shows that it cannot be any general $GL$ matrix, but must belong
to the so called symplectic group
\begin{equation}\label{sympDef}
    \Lambda  \equiv \left( {\begin{array}{*{20}c}
       A & B  \\
       C & D  \\
    \end{array}} \right) \in Sp(2n,\mathbb{R}) \subset
    GL(2n,\mathbb{R}),
\end{equation}
defined as the group of ($2n \times 2n$) matrices that leave the
symplectic matrix
\begin{equation}
    \mathbf{\Omega}  = \left( {\begin{array}{*{20}c}
       0 & \mathbf{1}  \\
       { - \mathbf{1}} & 0  \\
    \end{array}} \right)
\end{equation}
invariant, i.e.
\begin{equation}
    \Lambda ^T \mathbf{\Omega} \Lambda  = \mathbf{\Omega}.
\end{equation}

So when we say a theory has symplectic structure, we mean that a
family of Lagrangians exist such that they differ from each other
by the symplectic duality rotation (\ref{trans}).

Finally, we note that in general we can relax the condition that
the matrix $\Lambda$ should be real and find that it must then
belong to the complex symplectic group $Sp(2n,\mathbb{C})$.
Furthermore, it is a known group-theoretic result that the
following isomorphism is true
\begin{eqnarray}
    Sp(2n,\mathbb{R}) &\sim& Usp(n,n) \nonumber \\
    Usp(n,n) &\equiv& Sp(2n,\mathbb{C}) \cap U(n,n),
\end{eqnarray}
where $Usp(n,n)$ is the group of unitary symplectic matrices
$\mathcal{S}$ that satisfy \emph{both} the following relations
\begin{equation}
    \mathcal{S}^T \mathbf{\Omega} \mathcal{S} = \mathbf{\Omega} \quad ,\quad \mathcal{S}^\dag \mathbb{H}\mathcal{S} = \mathbb{H}
\end{equation}
simultaneously, where
\begin{equation}
    \mathbb{H} = \left( {\begin{array}{*{20}c}
       \mathbf{1} & 0  \\
       0 & { - \mathbf{1}}  \\
    \end{array}} \right).
\end{equation}

\section{Special K\"{a}hler geometry; a general discussion}

At this point, we present a general discussion of what a special
K\"{a}hler manifold is. Recall that a Hodge-K\"{a}hler manifold is
defined as a K\"{a}hler manifold with a line bundle $\mathcal{L}$
whose first Chern class equals the cohomology class of the
K\"{a}hler 2-form (\ref{hodge-k}). Now consider a Hodge-K\"{a}hler
manifold with an additional vector bundle $\mathcal{V}$, as
demanded by $\mathcal{N}=2$ supersymmetry. The bundle
$\mathcal{V}$ is constrained to have the symplectic structure
discussed in the previous section. One says that a
(local)\footnote{We are generally ignoring the rigid SUSY case,
which would result in the so called rigid special K\"{a}hler
manifold.} special K\"{a}hler manifold is a K\"{a}hler manifold
with a tensor bundle $\mathcal{H}=\mathcal{V} \otimes
\mathcal{L}$.

In 1979, Zumino discovered that the self-interaction of
supersymmetric multiplets in the Wess-Zumino model is governed by
K\"{a}hler geometry \cite{Zumino}, which means that one can define
a K\"{a}hler metric that appears in the Lagrangian in the same way
$G_{I J}$ does in (\ref{lag}). This was found to be a property of
all $\mathcal{N}=2$ supersymmetric theories, and, in fact, may be
generalized to $\mathcal{N}>2$ by adding more fields. Such
Lagrangians are the result of compactifications over Calabi-Yau
3-folds (or can be thought of as such, since this is not how they
were originally discovered) resulting into two sets of bosonic
fields, corresponding to the two deformation classes, as discussed
in the first appendix.

So, to make the connection to CY 3-folds clearer, we note that we
can identify the scalar fields $\phi$ of the last section with the
moduli $z$ and $\bar z$ from appendix \ref{manifolds}, which can
be either the complex structure moduli or the K\"{a}hler moduli.
The symplectic structure is the same even though the scalar field
is now complex. Comparing equations (\ref{hodge}) and (\ref{pot}),
we make the following identification
\begin{equation}\label{ident}
    h\left( {w,\bar w} \right) = - {i\left( {\bar Z^I F_I  - Z^I \bar F_I }
    \right)},
\end{equation}
and proceed to define the rest of the parameters in view of their
symplectic properties.

Since there are two classes of CY 3-fold deformations, we will
discuss the one that results in the set of fields known as the
vector multiplets first, associated with the special K\"{a}hler
geometry. Following that, we will discuss the hypermultiplets
sector, associated with quaternionic geometry. We also note that
the discussion so far has been restricted to the $\mathcal{N}=2$
theory in four spacetime dimensions. The reason is that the five
dimensional theory is related to the four dimensional theory via
compactification over a circle $S^1$ (the c-map). This maps the
quaternionic hypermultiplet sector in $D=5$ to the special
K\"{a}hler vector multiplet sector in $D=4$. Hence one can write
the $D=5$ theory with hypermultiplets using special K\"{a}hler
parameters, specifically the components of the period matrix
$\mathcal{N}$.

\section{Special K\"{a}hler geometry; the details}

A special K\"{a}hler manifold is a K\"{a}hler manifold with a
tensor bundle $\mathcal{H}$. There are, in fact, two types of
special K\"{a}hler manifolds; the local and the rigid. The rigid
manifold arises in globally supersymmetric Yang-Mills theories,
defined by the moduli of certain Riemannian manifolds, while the
local manifold arises in the locally supersymmetric gravity
theories (i.e. supergravity), defined by the moduli of Calabi-Yau
manifolds. Since the later is our main interest, we will not
discuss the rigid type in any detail. Suffice to note that in the
rigid case, the first Chern class of the manifold actually
vanishes, hence the line bundle is flat. At the level of the
Lagrangian this reflects into a different behavior of the
fermionic fields. A special K\"{a}hler manifold is local if the
line bundle $\mathcal{L}$ is non-flat and can be identified with
$U(1)$ as discussed earlier.

We find that the number of vector multiplets in the theory,
previously denoted by $n$, is actually equal to $2(h_{1,1}+1)$. We
proceed by defining a typical holomorphic section of the
$\mathcal{H} = U(1) \otimes Sp(2h_{1,1}+2,\mathbb{R})$ bundle:
\begin{equation}
    \Lambda  = \left( {\begin{array}{*{20}c}
       {Z^I }  \\
       {F_J }  \\
    \end{array}} \right).
\end{equation}

The choice of notation is no accident, these are exactly the
periods of the unique CY (3,0) volume form $\Omega$ defined in
appendix \ref{manifolds}, now written as a section of
$\mathcal{H}$. That is why, in the literature, the section
$\Lambda$ is sometimes called the period vector. We define the
hermitian metric on $\mathcal{H}$:
\begin{equation}
    i\left\langle \Lambda  \right|\left. {\bar \Lambda } \right\rangle \equiv
    - i\Lambda ^T \left( {\begin{array}{*{20}c}
       0 & 1  \\
       { - 1} & 0  \\
    \end{array}} \right)\bar \Lambda
\end{equation}
which appears explicitly in (\ref{ident}). Hence the resulting
K\"{a}hler form is
\begin{equation}
    \omega  = \frac{i}{{2\pi }}\partial \bar \partial \ln \left(
    {i\left\langle \Lambda  \right|\left. {\bar \Lambda } \right\rangle }
    \right).
\end{equation}

Furthermore, we define the symplectic vector $V$ by properly
normalizing $\Lambda$:
\begin{equation}\label{symvec}
    V = \left( {\begin{array}{*{20}c}
       {L^I }  \\
       {M_J }  \\
    \end{array}} \right) \equiv e^{{\mathcal{K} \mathord{\left/
     {\vphantom {\mathcal{K} 2}} \right.
     \kern-\nulldelimiterspace} 2}} \left( {\begin{array}{*{20}c}
       {Z^I }  \\
       {F_J }  \\
    \end{array}} \right)
\end{equation}
such that we immediately find that $L$ and $M$ define a complete
symplectic basis with an inner product
\begin{equation}\label{innerprod}
    i\left\langle V \right|\left. {\bar V} \right\rangle  = i\left(
    {\bar L^I M_I  - L^I \bar M_I } \right) = 1.
\end{equation}

It also follows that we can define a covariant derivative, using
the K\"{a}hler connection (\ref{conn}) such that
\begin{equation}\label{covderiv}
    \nabla _{\bar i} V = \left[ {\partial _{\bar i}  -
    \frac{1}{2}\left( {\partial _{\bar i} \mathcal{K}} \right)} \right]V =
    0,
\end{equation}
whereas an orthogonal quantity may be defined by
\begin{equation}\label{covderiv1}
    \nabla _i V = \left[ {\partial _i  - \frac{1}{2}\left( {\partial
    _i \mathcal{K}} \right)} \right]V = U_i  \equiv \left(
    {\begin{array}{*{20}c}
       {f_i^I }  \\
       {h_{J|i} }  \\
    \end{array}} \right) \ne 0,
\end{equation}
such that
\begin{equation}\label{}
    \left\langle V \right|\left. {U_i } \right\rangle  = \left\langle V \right|\left. {U_{\bar i} } \right\rangle  = 0.
\end{equation}

Using these definitions, one can reintroduce the period matrix via
\begin{equation}
    \mathcal{N}_{IJ} L^J  = M_I \quad ,\quad \mathcal{N}_{IJ} f_i^J  = h_{I|i}.
\end{equation}

The following relations can be derived given the above structure
and are particularly useful in calculations:
\begin{eqnarray}
    \left( {{\mathop{\rm Im}\nolimits} \mathcal{N}^{IJ} } \right)^{ - 1}  &=& -\gamma ^{IJ}  = -2\left( {G^{i\bar j} f_i^I f_{\bar j}^J  + L^I \bar L^J }
    \right) \label{useful}\\
    \left( {\nabla _{\bar j} f_i^I } \right) &=& G_{i\bar j} L^I ,\quad \left( {\nabla _{\bar j} h_{iI} } \right) = G_{i\bar j} M_I
    \label{useful2} \\
    {\mathop{\rm Im}\nolimits} \mathcal{N}_{IJ} L^I \bar L^J  &=&  - \gamma _{IJ} L^I \bar L^J  =  - \frac{1}{2} \nonumber \\
    G_{i\bar j}  &=&   - 2f_i^I {\mathop{\rm Im}\nolimits} \mathcal{N}_{IJ} f_{\bar j}^J  = 2f_i^I \gamma _{IJ} f_{\bar
    j}^J,\label{computational}
\end{eqnarray}
where the last formula in particular implies that the imaginary
part of the period matrix (${\mathop{\rm Im}\nolimits} \mathcal{N}
=  - \gamma $) acts as a metric to raise and lower the ($I,J,K$)
indices, and that $f$ is the veilbein that relates it to the
special K\"{a}hler metric $G$ similar to (\ref{bein}).

In summary, it appears that the periods $F$ and $Z$ of the
Calabi-Yau 3-fold's homology cycles define a special K\"{a}hler
manifold with a symplectically covariant structure. And if one
recognizes that the $F$'s are dependent on the $Z$'s and that
$Z^0$ is not linearly independent of the rest, we can define the
special K\"{a}hler manifold using the moduli $z$ as special
coordinates. The fact that, via the c-map, this four dimensional
structure corresponds to the quaternionic structure of the five
dimensional theory, allows us to write the $D=5$ hypermultiplets
sector in terms of the special K\"{a}hler parameters of the $D=4$
vector multiplets sector.

\section{Hypergeometry}\label{hypergeometry}

We have actually finished developing the technology needed for our
purposes, but for the sake of completeness, we now turn to the
four dimensional hypermultiplets sector, which is related to the
five dimensional vector multiplets sector via the so called r-map.
The geometry described by the Hodge-K\"{a}hler manifold in this
case is hypergeometric, which in the rigid case exhibits the
geometry of the so called hyperK\"{a}hler manifolds, which we will
not discuss. In the local case, we find that it describes a
quaternionic manifold. The dimension of the manifold is $4n$,
where $n$ is now the number of hypermultiplets and is equal to
$\left( {h_{2,1}  + 1} \right)$.

We will discuss the general structure of a quaternionic manifold
in this section, and we will end up by writing down the bosonic
action of a theory that exhibits both vector and hypermultiplets.

Define a $4n$-dimensional quaternionic real manifold with the
metric
\begin{eqnarray}
    ds^2  &=& h_{uv} \left( q \right)dq^u dq^v ,\nonumber\\ u,v &=& 1,
    \ldots ,4n.
\end{eqnarray}

It turns out that such a manifold has \emph{three} complex
structures on it, i.e. three different ways of defining an
operation whose square is $-1$. We denote those by $J^x$ where
$x=1,2,3$. The complex structures must satisfy the quaternionic
algebra
\begin{equation}
    J^x J^y  =  - \delta ^{xy} \mathbf{1} + \bar\varepsilon ^{xyz}
    J^z.
\end{equation}

It follows that we can construct three 2-forms which we call the
hyperK\"{a}hler forms
\begin{eqnarray}
    \omega ^x  &=& \omega _{uv}^x dq^u \wedge dq^v, \nonumber\\ \omega _{uv}^x
    &=& h_{uw} \left( {J^x } \right)_v^w,
\end{eqnarray}
generalizing the concept of a K\"{a}hler form. The hyperK\"{a}hler
forms follow a $SU(2)$ Lie-algebra, in the same way the ordinary
K\"{a}hler form follows a $U(1)$ Lie-algebra.

In the special K\"{a}hler case, we managed to identify the
K\"{a}hler 2-form with the curvature of a line bundle on a
Hodge-K\"{a}hler manifold, here we can follow similar steps that
leads us into defining the quaternionic manifold as a
Hodge-K\"{a}hler manifold with a non-flat $SU(2)$ line bundle on
it. If the $SU$-bundle is flat, we get a HyperK\"{a}hler (rigid)
manifold.

We end this section by writing down the general form of the
bosonic SUGRA Lagrangian of the (ungauged) $\mathcal{N}=2$
supersymmetric theory:
\begin{eqnarray}
    \mathcal{L}_4 &=& \sqrt { - g} \left[ {  R \left( g \right) - h_{uv} \left( q \right)\left( {\nabla _\mu  q^u } \right)\left( {\nabla ^\mu  q^v } \right) - G_{i\bar j} \left( {z,\bar z} \right)( {\nabla _\mu  z^i } )( {\nabla ^\mu  z^{\bar j} } ) } \right. \nonumber \\
    & & - \left. { i\left( {\bar {\mathcal{N}}_{IJ} F_{\mu \nu }^{ - I} F^{ - J\mu \nu }  - \mathcal{N}_{IJ} F_{\mu \nu }^{ + I} F^{ + J\mu \nu } } \right)}
    \right], \label{N2D5theory}
\end{eqnarray}
where the complex fields $z$ are the K\"{a}hler moduli spanning a
special K\"{a}hler manifold and the real fields $q$ include the
complex structure moduli spanning a quaternionic manifold. $G$ and
$h$ are the metrics of these manifolds respectively. The complete
Lagrangian with fermionic fields and terms due to gauging with the
full set of SUSY transformations may be found in \cite{Ferrara1}.

\section{\label{cmap}The c-map}

The matter fields in supergravity theory are constrained by
supersymmetry to describe certain manifolds. For example, in $D=3$
$\mathcal{N}=4$ SUGRA, the multiplets decouple into two
quaternionic manifolds. In $D=4$ $\mathcal{N}=2$ SUGRA (which we
discussed in detail in this appendix) we find that the scalars of
the vector multiplets sector describe a special K\"{a}hler
manifold, while those of the hypermultiplets describe a
quaternionic manifold. Now since one can get the three dimensional
theory by wrapping the four dimensional one over a circle, we find
that both the $D=4$ special K\"{a}hler and quaternionic manifolds
map into $D=3$ quaternionic manifolds.

Our case of interest is the five dimensional theory. In this case,
the vector multiplets are described by the so called very special
real manifold, which, upon compactification, maps to the $D=4$
quaternionic manifold. This is the r-map:
\begin{equation}
    \mathbb{R}  \stackrel{\bf r}{\longrightarrow}  \mathbb{C}.
\end{equation}

The hypermultiplets sector is defined by a quaternionic manifold,
which, upon compactification, maps to the $D=4$ special K\"{a}hler
manifold. This is the c-map:
\begin{equation}
    \mathbb{C}  \stackrel{\bf c}{\longrightarrow}  \mathbb{H}.
\end{equation}

For more detail on this subject, one is referred to, for example,
\cite{Vandoren,Proeyen3,Proeyen5}, and for an explicit calculation
see \cite{Ferrara4}. For our purposes, we show in chapter
(\ref{compact}) that direct compactification of $D=11$ SUGRA over
a CY 3-fold, ignoring the vector multiplets, yields the set of
scalar fields known as the hypermultiplets,which can be naturally
written in terms of the symplectic special K\"{a}hler parameters
defined for the four dimensional vector multiplets. We also show
that writing the fields in a certain form immediately yields a
quaternionic structure analogous to the one discussed in
\S\ref{hypergeometry}.

\backmatter

\bibliographystyle{abbrv}
\bibliography{thesis}

\end{document}